\documentclass[preprint]{aastex}
\input epsf
\begin{document}
\title{Cosmic Infrared Background and Early Galaxy Evolution}

\author{A. Kashlinsky\\Laboratory for Astronomy and Solar Physics, Goddard Space Flight Center,
Greenbelt, MD 20771, and \\ SSAI\\
e-mail address: kashlinsky@stars.gsfc.nasa.gov}



\def\plotone#1{\centering \leavevmode
\epsfxsize=\columnwidth \epsfbox{#1}}

\def\wisk#1{\ifmmode{#1}\else{$#1$}\fi}

\def\wm2sr {Wm$^{-2}$sr$^{-1}$ }        
\def\nw2m4sr2 {nW$^2$m$^{-4}$sr$^{-2}$\ }       
\def\nwm2sr {nWm$^{-2}$sr$^{-1}$\ }     
\def\nw2m4sr {nW$^2$m$^{-4}$sr$^{-1}$\ }
\def\Ncut {$N_{\rm cut}$\ }
\def\lt     {\wisk{<}}
\def\gt     {\wisk{>}}
\def\le     {\wisk{_<\atop^=}}
\def\ge     {\wisk{_>\atop^=}}
\def\lsim   {\wisk{_<\atop^{\sim}}}
\def\gsim   {\wisk{_>\atop^{\sim}}}
\def\kms    {\wisk{{\rm ~km~s^{-1}}}}
\def\Lsun   {\wisk{{\rm L_\odot}}}
\def\Msun   {\wisk{{\rm M_\odot}}}
\def\um     { $\mu$m\ }
\def\sig    {\wisk{\sigma}}
\def\etal   {{\sl et~al.\ }}
\def\eg     {{\it e.g.\ }}
\def\ie     {{\it i.e.\ }}
\def\bsl    {\wisk{\backslash}}
\def\by     {\wisk{\times}}
\def\cosec {\wisk{\rm cosec}}
\def\mic {\wisk{ \mu{\rm m }}}

\def\amin   {\wisk{^\prime\ }}
\def\asec   {\wisk{^{\prime\prime}\ }}
\def\cc     {\wisk{{\rm cm^{-3}\ }}}
\def\deg     {\wisk{^\circ}}
\def\ddeg   {\wisk{{\rlap.}^\circ}}
\def\damin  {\wisk{{\rlap.}^\prime}}
\def\dasec  {\wisk{{\rlap.}^{\prime\prime}}}
\def\approxeq{$\sim \over =$}
\def\abouteq{$\sim \over -$}
\def\percm{cm$^{-1}$}
\def\percmsq{cm$^{-2}$}
\def\percmcub{cm$^{-3}$}
\def\perhz{Hz$^{-1}$}
\def\perpc{$\rm pc^{-1}$}
\def\persec{s$^{-1}$}
\def\peryr{yr$^{-1}$}
\def\te{$\rm T_e$}
\def\tenup#1{10$^{#1}$}
\def\to{\wisk{\rightarrow}}
\def\thin{\thinspace}
\def\uk{$\rm \mu K$}
\def\p{\vskip 13pt}


\begin{abstract}
The Cosmic Infrared Background (CIB) reflects the sum total of
galactic luminosities integrated over the entire age of the
universe. From its measurement the red-shifted starlight and
dust-absorbed and re-radiated starlight of the CIB can be used to
determine (or constrain) the rates of star formation and metal
production as a function of time and deduce information about
objects at epochs currently inaccessible to telescopic studies.
This review discusses the state of current CIB  measurements and
the (mostly space-based) instruments with which these measurements
have been made, the obstacles (the various foreground emissions)
and the physics behind the CIB and its structure. Theoretical
discussion of the CIB levels can now be normalized to the standard
cosmological model narrowing down theoretical uncertainties. We
review the information behind and theoretical modeling of both the
mean (isotropic) levels of the CIB and their fluctuations. The CIB
is divided into three broad bands: near-IR, mid-IR and far-IR. For
each of the bands we review the main contributors to the CIB flux
and the epochs at which the bulk of the flux originates. We also
discuss the data on the various quantities relevant for correct
interpretation of the CIB levels: the star-formation history, the
present-day luminosity function measurements, resolving the
various galaxy contributors to the CIB, etc. The integrated light
of all galaxies in the deepest near-IR galaxy counts to date fails
to match the observed mean level of the CIB, probably indicating a
significant high-redshift contribution to the CIB. Additionally,
Population III stars should have left a strong and measurable
signature via their contribution to the cosmic infrared background
(CIB) anisotropies for a wide range of their formation scenarios,
and  measuring the excess CIB anisotropies coming from high $z$
would provide direct information on the epoch of the first stars.
\end{abstract}


\newpage

\section*{Table of contents}
{\bf Abstract}\\
{\bf 1. Introduction} \\
{\bf 2. Miscellaneous: definitions, units, magnitudes, etc}\\
{\bf 3. Theoretical preliminaries}

3.1 Mean level

3.2 CIB anisotropies\\
{\it 3.2.1 CIB anisotropies from galaxy clustering}\\
{\it 3.2.2 Shot  noise CIB fluctuations from galaxies}\\
{\it 3.2.3 Cosmic variance}\\
{\it 3.2.4 CIB dipole component}

3.3 Structure formation: cosmological paradigms

3.4 From cosmological paradigm to galaxies \\
{\bf 4. Obstacles to measurement: confusion and foregrounds}

4.1 Atmospheric emission

4.2 Galactic stars

4.3 Zodiacal emission

4.4 Galactic cirrus

4.5 Cosmic microwave background\\
{\bf 5. Current CIB measurements}

5.1 COBE/DIRBE

5.2 COBE/FIRAS

5.3 IRTS

5.4 ISO

5.5 2MASS


5.6 Results:\\
{\it 5.6.1 Near-IR}\\
{\it 5.6.2 Mid-IR}\\
{\it 5.6.3 Far-IR}\\
{\it 5.6.4 Bolometric CIB flux}\\
{\bf 6. 'Ordinary' contributors to CIB}

6.1 IMF and star formation history: from UV to far-IR

6.2 Normal stellar populations

6.3 Dust emission from galaxies: mid-IR to sub-mm

6.4 Contribution from quasars/AGNs

6.5 Present-day luminosity density

6.6 Deep galaxy counts\\
{\it 6.6.1 Near-IR}\\
{\it 6.6.2 Mid-IR}\\
{\it 6.6.3 Far-IR and sub-mm}

6.7 CIB fluctuations from clustering of ordinary galaxies

6.9 Cumulative flux from galaxy counts vs CIB measurements\\
{\bf 8. Population III}

8.1 What were the first stars?

8.2 Isotropic component of CIB

8.3 Contribution to anisotropies in CIB

8.4 Can CIB anisotropies from Population III be measured?\\
{\bf 9. Snapshot of the future}\\
{\bf 10. Concluding remarks}\\

\newpage

\section{Introduction}

Diffuse backgrounds contain important information about the
history of the early Universe, when discrete objects either did
not exist or are not accessible to telescopic studies. The cosmic
infrared background (CIB) arises from accumulated emissions from
early galaxy populations spanning a large range of redshifts. The
earliest epoch for the production of this background occurred when
star formation first began, and contributions to the CIB continued
through the present epoch. The CIB is thus an integrated summary
of the collective star forming events, star-burst activity and
other luminous events in cosmic history to the present time. As
photons move to observer they lose energy to cosmic expansion and
any stellar emission from high redshift populations will now be
seen in the infrared (mostly near-IR with $\lambda \lsim$ a few
\um, unless it comes from very cold stars at very high $z$).
Emission from galactic dust will be shifted to still longer IR
wavelengths. CIB thus probes the physics in he Universe between
the present epoch and the last scattering surface and is
complementary to its more famous cousin, the cosmic microwave
background radiation (CMB) which probes mainly the physics at the
last scattering.

Considerable effort has now gone into studying the luminosity
sources during the most recent history of the universe ($z <
2-3$), but the period from recombination to the redshift of the
Hubble Deep Field (HDF) remains largely an unexplored era because
of the difficulty of detecting many distant galaxies over large
areas of the sky. Significant progress in finding high redshift
galaxies has been achieved using the Lyman dropout technique
(Steidel et al 1996), but uncovering substantial populations of
galaxies at $z\gsim 5$ is extremely difficult and the total sample
is still very small. It is there that the CIB measurements can
provide critical information about the early history of the
Universe largely inaccessible to telescopic studies. At more
recent epochs galaxy evolution is also constrained by the
measurements of the visible part of the extragalactic background
light or EBL (Bernstein et al 2002a,b; cf. Mattila 2003).

Chronologically, the importance of the CIB and early predictions
about its levels (Partridge \& Peebles 1968) followed the
discovery of the CMB which observationally established the Big
Bang model for the origin and evolution of the Universe. Because
of the Earth's atmosphere is was clear from the start that the
measurement of the CIB must be done from space and even there the
Solar system and Galactic foregrounds presented a formidable
challenge. It took a while for technology to reach the required
sensitivity and, when the first rocket and space based CIB
measurements were conducted, the results were inconclusive
(Matsumoto, Akiba \& Murakami 1988, Noda et al 1992). Following
its launch in 1983 the IRAS satellite was the first to conduct
all-sky infrared measurements of point sources between 12 and 100
\um (Soifer, Neugebauer \& Houck 1987). It revealed that galaxies
were efficient infrared emitters, but its design was not optimized
for diffuse background measurements. The COBE DIRBE instrument,
launched in 1990, operated between 1.25 and 240 \um and was the
first to be devoted specifically to CIB measurements (Hauser \&
Dwek 2001). It led to the first reliable measurements of and
limits on the CIB over a wide range of infrared bands and
literally began the observational CIB era.

The status of the CIB measurements in the immediate post-DIRBE era
has been reviewed extensively by Hauser \& Dwek (2001) and a
slightly earlier review by Leinert et al (1998) has provided a
detailed discussion of the issues involved in measuring the
extragalactic background light (EBL) levels. Since then several
important developments happened, most notably the establishment,
through CMB measurements, of the standard cosmological model for
the overall evolution of the Universe, particularly the existence
of the so-called dark energy, and of the standard paradigm
($\Lambda$CDM) for structure formation. CMB polarization
measurements have identified the epoch of re-ionization,
presumably due to first collapsed objects in the Universe. This
now allows narrowing of CIB predictions and puts refined searches
for the CIB on a much firmer basis. Also, in the CIB context,
several new measurements have now been accomplished and several
old ones are now more firmly confirmed. Scientific exploration is
never static and the very recently (August 2003) launched NASA
Spitzer satellite already led to several important findings for
CIB and many more are expected to follow as Spitzer observational
programs get implemented. New space instruments in the IR and
sub-mm bands are planned by NASA and ESA which will shed new light
on the early Universe and the interconnection between the CIB and
early luminous systems. This thus seems like an opportune moment
for a new attempt to review the CIB measurements and their
interpretation.

Indeed, following the WMAP and balloon experiments we now know
that the Universe is flat and dominated by the vacuum energy
(cosmological constant) and/or even more exotic quintessence
field. The recent WMAP results (Bennett et al 2003, Spergel et al
2003) imply that the Universe had a large optical depth since last
scattering ($\tau \sim 0.2$) indicating unexpectedly early star
formation ($z \sim 20$). Direct deep studies of the universe at
$z>$2-3 show that much of the luminosity of the universe at that
time was probably involved in the process of galaxy formation, and
the birth of the first generations of stars. Many systems, such as
elliptical galaxies, already seem well-formed by the epoch of the
HDF, and the metallicities for many systems are already near
solar. Somewhere between $z\sim 20-30$ and $z\sim 6$ reside the
systems made up of the first population of stars, the so-called
zero-metallicity Population III, about which little is known
observationally. All these galaxies should have left their imprint
in the CIB and its structure.

Observationally, the CIB is difficult to distinguish from the
generally brighter foregrounds contributed by the local matter
within the solar system, and the stars and ISM of the Galaxy. A
number of investigations have attempted to extract the isotropic
component (mean level) of the CIB from ground- and satellite-based
data as described below. This has in nearly all instances been a
complicated task due to a lack of detailed knowledge of the
absolute brightness levels and spatial variations across the sky
for the many foregrounds that overlay the CIB signal. It is thus
very important to understand and estimate to high accuracy the
various foreground emissions which need to subtracted or removed
before uncovering the CIB.

The outline of this review is as follows: in Sec. 2 we start with
a summary of the units, the various relevant astronomical
magnitude systems and the list of frequent abbreviations used
throughout the review. Sec. 3 discusses theoretical basis for CIB
and its structure and how these are related to information on the
early galaxies and stellar systems. Sec. 4 reviews the foregrounds
that inhibit the CIB measurements and Sec. 5 reviews the status of
the current CIB measurements. In Sec. 6 we discuss the topics
related to CIB measurements and their interpretation such as the
contribution from ordinary galaxies composed of Population I and
II stars, the role of star formation, dust, the current status of
the present day galaxy luminosity density measurements, and the
deep galaxy counts from near-IR to far-IR. Sec. 7 is devoted to
the signature of Population III in the CIB and its structure and
the almost unique role CIB plays in uncovering the Population III
era. We end up with a review of the currently planned instruments
and space missions that will or can have direct bearing on the CIB
in Sec. 8 followed by a summary section.

\section{Miscellaneous: definitions, units, magnitudes, abbreviations, etc.}

We start with definitions. The surface brightness of the CIB per
unit wavelength will be denoted as $I_\lambda$, per unit frequency
as $I_\nu$, and per logarithmic wavelength interval $F=\lambda
I_\lambda = \nu I_\nu$, and we call them all ``flux.'' Throughout
the paper $B_\nu$ will denote the Planck black-body function per
unit frequency.

Fluxes of astronomical sources are often measured in narrow band
filters. The flux units commonly used in astronomy  for $I_\nu$
are Jy=$10^{-26}$ W/m$^2$/Hz. The surface brightness of the CIB is
usually given in units of either MJy/sr or \nwm2sr . The
conversion between the two is:
\begin{equation}
1 \;\;\; \frac{\rm nW}{{\rm m}^2 {\rm sr}}=
\frac{3000}{\lambda(\mu {\rm m})} \;\;\; \frac{\rm MJy}{\rm sr}
\end{equation}

The range of wavelengths used in this review is divided into the
following groups: near-infrared (NIR) covers 1 \um $\lsim \lambda
\lsim $ 5--10 \um. Mid-infrared (MIR) is defined to lie in  5--10
\um $\lsim \lambda \lsim $ 50--100 \um range. We call the
Far-infrared (FIR) the region corresponding to 50--100 \um $\lsim
\lambda \lsim $ 500 \um, and beyond that will be sometimes
referred to as sub-mm. These definitions are neither exact nor
unique and, although the band ranges cover (largely) different
physical processes, this division is used for convenience only.

\begin{deluxetable}{c | c c c c c c c }
\tabletypesize{\scriptsize}
\tablecaption{Magnitude system conversion \label{magtable} }
\startdata
Filter &  J & H & K & IRAC-1 & IRAC-2 & IRAC-3 & IRAC-4 \\
\hline
$\lambda$ (\um) & 1.25 & 1.65 & 2.2 & 3.6 & 4.5 & 5.8 & 8\\
\hline
m$_{\rm AB}$--m$_{\rm Vega}$& 0.90 & 1.37 & 1.84 & 2.79 & 3.26 & 3.73  & 4.40 \\
\enddata
\end{deluxetable}

Throughout the review both Vega magnitudes and AB magnitudes are
used. By definition, Vega's magnitudes are zero in all filters.
However, because of uncertainties in the absolute flux calibration
of Vega, magnitudes of this star have been slightly corrected over
time. The zero point of AB magnitude system is constant flux of
3631 Jy for apparent brightness (Oke \& Gunn 1983). Thus in AB
magnitude system an object with $I_\nu=$constant (flat energy
distribution) has the same magnitude in all bands, and all colors
are zero. For the wavelengths used frequently in the paper the
conversion is given in Table \ref{magtable}. The numbers were
adopted in the J,H,K photometric bands from the 2MASS Explanatory
Supplement
\footnote{http://www.ipac.caltech.edu/2mass/releases/allsky/doc/explsup.html}
and in the four Spitzer IRAC channels from the Spitzer Observer's
Manual \footnote{http://ssc.spitzer.caltech.edu/documents}.

Table \ref{acronyms} lists the abbreviations and acronyms that
will appear below.

\begin{deluxetable}{c c}
\tabletypesize{\scriptsize}
\tablecaption{Summary of frequently used acronyms and
abbreviations.}
 \startdata
2dF & 2 degree Field \\
2MASS & 2 Micron All Sky Survey \\
AGN & Active Galactic Nucleus\\
CDM & Cold Dark Matter \\
CIB & Cosmic Infrared Background\\
CMB & Cosmic Microwave Background\\
COBE & COsmic Background Explorer\\
DIRBE & Diffuse InfraRed Background Experiment \\
EBL & Extragalactic Background Light\\
ELAIS & European Large Area ISO Survey\\
ESA & European Space Agency\\
FIR & Far IR \\
FIRAS & Far InfraRed Absolute Spectrometer \\
FOV & Field-Of-View \\
FSM & Faint Source Model \\
HDF & Hubble Deep Field \\
IGM & InterGalactic Medium \\
IMF & Initial Mass Function \\
IN & Instrument Noise \\
IPD & InterPlanetary Dust \\
IR & InfraRed \\
IRAC & InfraRed Array Camera\\
IRAS & InfraRed Astronomical Satellite \\
IRTS & InfraRed Telescope in Space \\
ISM & InterStellar Matter \\
ISO & Infrared Space Observatory \\
JWST & James Webb Space Telescope \\
MIPS & Multiband Imaging Photometer System for Spitzer\\
MIR & Mid IR \\
NASA & National Aeronautics \& Space Agency \\
NGP/SGP & North/South Galactic Pole \\
NEP/SEP & North/South Ecliptic Pole \\
NIR & Near IR \\
NSF & National Science Foundation \\
PAH & Polycyclic Aromatic Hydrocarbon\\
RMS & Root Mean Square \\
SCUBA & Sub-mm Common-User Bolometer Array \\
SDSS & Sloan Digital Sky Survey \\
SED & Spectral Energy Distribution \\
SFR & Star Formation Rate \\
SNAP & SuperNovae Acceleration Probe\\
UDF & Ultra Deep Field \\
WMAP & Wilkinson Microwave Anisotropy Probe \\
ZL & Zodiacal Light \\
\enddata
\label{acronyms}
\end{deluxetable}

\section{Theoretical preliminaries}

The near-IR CIB arises mainly from the stellar component of
galaxies and probes evolution of stellar component of galaxies at
early times. The mid- and far-IR CIB originates from dusty
galaxies reprocessing stellar light and other energetic output. As
will be discussed in Sec. 5, the recent mutually consistent
detections of the near-IR CIB from the COBE/DIRBE and Japan's IRTS
datasets (see Hauser \& Dwek 2001 for review) indicated a
surprisingly high amplitude of both the CIB fluctuations
(Kashlinsky \& Odenwald 2000a, Matsumoto et al 2000,2002) and mean
levels (Dwek \& Arendt 1998, Gorjian \& Wright 2000, Wright \&
Reese 2000, Cambresy et al 2001). In the mid-IR firm upper limits
on the CIB have been found by various indirect methods, but
because foreground emission is so high no direct detection  has
been possible. In the far-IR there are mutually consistent
detections of the CIB from the DIRBE (Schlegel et al 1998, Hauser
et al 1998) and FIRAS (Puget et al 1996, Fixsen et al 1998)
datasets.

In this section we provide a general mathematical and
observational basis for the underlying parameters and physics that
determine the CIB and its structure. Specific cosmological and
galaxy evolution models are mentioned only briefly (Sec. 3.4) and
we attempt to make the discussion as general as possible.

\subsection{Mean level}

In the Friedman-Robertson-Walker Universe with flat geometry and
the Robertson-Walker metric,
$ds^2=c^2dt^2-(1+z)^{-2}[dx^2+x^2(d\theta^2+\sin^2\theta
d\phi^2)]$, the comoving volume occupied by a unit solid angle in
the redshift interval $dz$ is $dV/dz = (1+z)^{-1} d_L^2(z)
cdt/dz$, where $d_L\equiv x(z)/(1+z)$ is the luminosity distance.
Thus the flux density in band $\lambda$ from each galaxy with
absolute bolometric luminosity $L$ at redshift $z$ is
$\frac{L}{4\pi d_L^2(1+z)}f_\lambda(\frac{\lambda}{1+z}; z)$. Here
$f_\lambda d\lambda$ is the fraction of the total light emitted in
the wavelength interval $[\lambda;  \lambda +d\lambda]$ and the
extra factor of $(1+z)$ in the denominator accounts for the fact
that the flux received in band $\lambda$ comes from a redshifted
galaxy. The contribution to the total CIB flux from the redshift
interval $dz$ is given by
\begin{equation}
\frac{dF}{dz} = \frac{R_H}{4\pi} \frac{1}{(1+z)^2}
\frac{d(H_0t)}{dz} \sum_i {\cal L}_i(z) [\lambda f_{\lambda,
i}(\frac{\lambda}{1+z}; z)], \label{dfdz}
\end{equation}
where the sum is taken over all galaxy populations contributing
flux in the observer rest-frame band at $\lambda$, and $f_\lambda$
characterizes the spectral energy distribution (SED) of galaxy
population $i$. Here $R_H=cH_0^{-1}$ and the (present-day) luminosity density is given by:
\begin{equation}
{\cal L}_{\nu}(0) = \sum_i \int \Phi_{0,i}(L_\nu) L_\nu dL_\nu
\label{lumden0}
\end{equation}
where $\Phi_{0,i}$ is the (present-day)luminosity function or the
number density of galaxies of morphological type $i$ in the
$dL_\nu$ interval at frequency band $\nu$.

It is illustrative to study the redshift dependence of the flux
production rate, Eq.\ (\ref{dfdz}). At small redshifts the factor
$(1+z)^{-2}dt/dz$  varies little with $z$, and the flux production
rate, $dF/dz$, is governed by the comoving bolometric luminosity
density ${\cal L}(z)$ and the SED of the galaxy emission
$f_\lambda$. If the luminosity evolution at these redshifts is
small, the flux production rate is governed by the SED shape. If
$f_\lambda(\lambda)$ increases toward shorter wavelengths then
$dF/dz$ increases with $z$. For $\lambda f_\lambda=$ const, and no
luminosity evolution, the rate is roughly constant with small $z$.
At sufficiently high redshifts, modest luminosity evolution in
Eq.\ (\ref{dfdz}) would be offset by the factor $(1+z)^{-2}dt/dz$,
so that the flux production rate would be cut off at sufficiently
large $z$. This factor is responsible for resolving Olbers'
paradox even for a flat SED.

There are three broad parameters that determine the possible modes
of galaxy evolution at higher redshifts: 1) how bright individual
galaxies shine (luminosity evolution), 2) their spectral energy
distribution (SED) which determines how much of the luminosity was
emitted at the rest frame of the galaxy (K-correction), and 3) how
numerous the galaxies were (number density evolution).
Consequently, one can separate the $z$-dependence in ${\cal
L}_\nu(z)$ into terms due to K-correction (${\cal K_\nu}$), pure
luminosity evolution (${\cal E_\nu}$) and pure number density
evolution (${\cal N_\nu}$) (e.g. Yoshii \& Takahara 1988), i.e.
\begin{equation}
{\cal L}_{\nu}(z) = 10^{-0.4[{\cal K_\nu}(z) + {\cal
E_\nu}(z)+{\cal N_\nu}(z)]} {\cal L}_{\nu}(0) \label{lumden}
\end{equation}
Then the total CIB flux emitted by evolving galaxy populations
becomes:
\begin{equation}
F_{\rm total} = \sum_i \frac{{\cal L}_{\nu,i}(0)R_H}{4\pi} \int
\frac{1}{(1+z)^2}\frac{d(H_0t)}{dz} 10^{-0.4({\cal
K}_{\nu,i}+{\cal E}_{\nu,i}+{\cal N}_{\nu,i})} dz \label{fcib_tot}
\end{equation}
where $dt/dz$ is given by eq. \ref{rdot} with the expansion factor
$R\equiv (1+z)^{-1}$.

The details of galaxy spectral energy distributions (SED) will be
discussed later, but a simple analysis can already be made using
eq. (\ref{dfdz}). The SED at rest-frame wavelengths $\lambda \lsim
10$\um\ is dominated by stellar emission, with a peak at visible
wavelengths and a decrease for $\lambda
> 0.7$ \um. Consequently, assuming no-evolution would mean that
most of the predicted J band CIB comes from redshifts $z\sim
0.3-1$, which shifts the visible emission of normal stellar
populations to $\sim 1$ \um . In the M band at 5 \um, most of the
predicted CIB comes from $z>1-2$. Evolution will likely push these
redshift ranges toward earlier times. At 10 \um $\lsim \lambda
\lsim 200$ \um, the emission is dominated by galactic dust and the
situation is reversed, so $f_\lambda$ increases with wavelength
roughly as $\lambda^\alpha$ with $\alpha \sim 1.5$. Hence, the
dusty star-burst galaxies observed by IRAS at low redshifts should
make the dominant contribution to the 10 \um\ CIB. In the far-IR,
the K-correction is strongly negative (Sec. 6.3) and the measured
CIB found can have large contributions from high redshifts.

It is useful to make a simple estimate of the expected CIB flux.
Measurements of the galaxy luminosity function in both optical
(Loveday et al 1992; Blanton et al 2001) and near-IR (Gardner et
al 1997; Kochanek et al 2003,; Cole et al 2003) indicate
approximately that the density of bright galaxies is $\Phi_* \sim
10^{-2} h^3{\rm Mpc}^{-3}$. Each of these galaxies emits in the
rough neighborhood of $L_* \sim 10^{37}$ W. The flux-dimensional
quantity composed of $\Phi_*,L_*$ and the Hubble constant is $F
\sim \frac{1}{4\pi} \Phi_* L_* cH_0^{-1} \simeq 25$ \nwm2sr . This
is a crude estimate, but it already shows that in order to measure
reliably the mean levels of the CIB, foregrounds must be
eliminated to well below $\sim 10$ \nwm2sr levels. Section 4 shows
how difficult a task that is in the infrared bands.

\subsection{CIB anisotropies}

Because of the difficulty of accurately accounting for the
contributions of bright foregrounds such as Galactic stars,
interplanetary and interstellar dust, which must be subtracted
from the observed sky background (Arendt et al, 1998, Kelsall et
al, 1998), Kashlinsky, Mather, Odenwald \& Hauser (1996) have
proposed to measure the structure of the CIB or its fluctuations
spectrum. For a relatively conservative set of assumptions about
clustering of distant galaxies, fluctuations in the brightness of
the CIB have a distinct spectral and spatial signal, and these
signals can be more readily discerned than the actual mean level
of the CIB. The most common source of luminosity in the universe
arises in galaxies, whose clustering properties at the present
times are fairly well known and are consistent with the
$\Lambda$CDM model predictions (Efstathiou, Sutherland \& Maddox
1990, Percival et al 2002, Tegmark et al 2004). The CIB, being
produced by clustered matter, must have fluctuations that reflect
the clustered nature of the underlying sources of luminosity. This
signature will have an angular correlation function (or angular
power spectrum) that distinguishes it from local sources of
background emission such as zodiacal light emission, and
foreground stars in the Milky Way.
 Moreover, distant contributions
of CIB will have a different redshift, and therefore spectral
color, than nearby galaxies and sources of local emission. From
galaxy evolution and cluster evolution models  it is possible to
match the predicted slopes for the power spectrum of the CIB
against the power spectrum of the data.

On the largest angular scales (the dipole component), there would
be an additional source of anisotropy due to our peculiar motion
with respect to the inertial frame of the Universe. This component
may be measurable over a certain range of wavelengths and is
discussed after the more canonical source of CIB anisotropies, the
galaxy clustering.

\subsubsection{CIB anisotropies from galaxy clustering}

Whenever CIB studies encompass relatively small parts of the sky
(angular scales $\theta < 1$ sr) one can use Cartesian formulation
of the Fourier analysis. The fluctuation in the CIB surface
brightness can be defined as $\delta F(\mbox{\boldmath$\theta$})=
F(\mbox{\boldmath$\theta$}) - \langle F \rangle$, where $F =
\lambda I_\lambda$, $\mbox{\boldmath$\theta$}$ is the two
dimensional coordinate on the sky and $\langle F \rangle$ is the
ensemble average. The two-dimensional Fourier transform is $\delta
F(\mbox{\boldmath$\theta$})= (2\pi)^{-2} \int \delta F_q
\exp(-i\mbox{\boldmath$q$}\cdot \mbox{\boldmath$\theta$})
d^2\mbox{\boldmath$q$}$.

If the fluctuation field, $\delta F(\mbox{\boldmath$x$})$, is a
random variable, then it can be described by the moments of its
probability distribution function. The first non-trivial moment is
the  {projected 2-dimensional} correlation function $C(\theta) =
\langle \delta F(\mbox{\boldmath$x$}+\theta) \delta
F(\mbox{\boldmath$x$})\rangle$. The 2-dimensional power spectrum
is $P_2(q) \equiv \langle |\delta F_q|^2\rangle$, where the
average is performed over all phases. The correlation function and
the power spectrum are a pair of 2-dimensional Fourier transforms
and for an isotropically distributed signal are related by
\begin{equation}
C(\theta)= \frac{1}{2\pi} \int_0^\infty P_2(q) J_0(q\theta) q dq,
\label{e1}
\end{equation}
\begin{equation}
P_2(q)= 2\pi \int_0^\infty C(\theta) J_0(q\theta) \theta d\theta,
\label{fourier}
\end{equation}
where $J_n(x)$ is the $n$-th order cylindrical Bessel function. If
the phases are random, then the distribution of the brightness is
Gaussian and the correlation function (or its Fourier transform,
the power spectrum) uniquely describes its statistics. In
measurements with a finite beam, the intrinsic power spectrum is
multiplied by the window function $W$ of the instrument.
Conversely, for the known beam window function, the power spectrum
can be de-convolved by dividing the measured power spectrum by the
beam window function.

Another useful and related quantity is the mean square fluctuation
within a finite beam of angular radius $\vartheta$, or zero-lag
correlation signal, which is related to the power spectrum by
\begin{eqnarray}
C(0) = \langle (\delta F)^2 \rangle_\vartheta = \frac{1}{2\pi}
\int_0^\infty P_2(q)
W_{TH}(q\vartheta) q dq \nonumber \\
\sim \frac{1}{2\pi} q^2P_2(q)|_{q\sim {\pi}/{2} \vartheta} .
\label{c0}
\end{eqnarray}
For  a top-hat beam the window function is
$W_{TH}=[2J_1(x)/x]^2=0.5$ at $x\simeq \pi/2$  where $x=q
\vartheta$,  and hence the values of $q^{-1}$ correspond to
fluctuations on angular scales of diameter $\simeq \pi/q$.

At small angles $< 1$ sr, the CIB power spectrum is related to the
CIB flux production rate, $dF/dz$, and the evolving 3-D power
spectrum of galaxy clustering, $P_3(k)$ via the Limber equation.
In the power spectrum formulation it can be written as (e.g.
Kashlinsky \& Odenwald 2000):
\begin{equation}
P_2(q)=\int \left(\frac{dF}{dz}\right)^2 \frac{1}{c\frac{dt}{dz}
d_A^2(z)} P_3(qd_A^{-1}; z) dz \label{limber_z}
\end{equation}
where $d_A(z)$ is the angular diameter distance and the
integration is over the epoch of the sources contributing to the
CIB. Eq. \ref{limber_z} can be rewritten as:
\begin{equation}
P_2(q) = \frac{1}{c} \int \left( \frac{d
I_{\nu^\prime}}{dt}\right)^2 \frac{P_3(qd_A^{-1};z)}{d_A^2} dt
\label{limber}
\end{equation}
This is equivalent to:
\begin{equation}
\frac{q^2P_2(q)}{2\pi} = \pi t_0 \int \left( \frac{d
I_{\nu^\prime}}{dt}\right)^2 \Delta^2(qd_A^{-1};z) dt
\label{limber_del2}
\end{equation}
where $t_0$ is the time-length of the period over which the CIB is
produced and
\begin{equation}
\Delta^2 (k) = \frac{1}{2\pi^2} \frac{k^2 P_3(k)}{ct_0}
\label{Delta3}
\end{equation}
 is the fluctuation in number of sources within a volume $k^{-2}
 ct_0$. \footnote{The $\Delta(k)$ defined by eq. \ref{Delta3} and used throughout the review
 should not be confused
 with another quantity - $\sqrt{k^3P(k)/2\pi}$ - sometimes encountered in the literature
 under the same symbol.}

To within a factor of order unity, the square of the fractional
fluctuation of the CIB on angular scale $\simeq \pi/q$ is
$\delta_{\rm CIB}^2 = \langle (\delta I_\nu)^2\rangle/I_\nu^2
\simeq I_\nu^{-2} q^2P_2(q)/2\pi$. The meaning of eq.
\ref{limber_del2} becomes obvious if we assume
$dI_{\nu^\prime}/dt$=constant during the lifetime of the emitters
$t_0$. In this case the fractional fluctuation due to clustering
of early galaxies becomes:
\begin{equation}
\delta_{\rm CIB}^2 = \frac{\pi}{t_0} \int \Delta^2(qd_A^{-1};z) dt
\label{spec_cib}
\end{equation}
In other words,
 the fractional fluctuation on angular scale $\pi/q$ in the CIB is given by
 the average value of the r.m.s. fluctuation from spatial clustering over
 a cylinder of length $ct_0$ and diameter $\sim k^{-1}$.

The Cartesian formulation is equivalent to the spherical sky
representation used in the cosmic microwave background (CMB)
studies on small scales. It was used in some CIB analyses (Haiman
\& Knox 2000, Cooray et al 2003).  In that case one expands the
flux into spherical harmonics: $F(\theta, \phi) = \nu I_\nu =
\sum_{l=0}^\infty \sum_{m=-l}^l a_{lm} Y_{lm}(\theta, \phi)$. The
correlation function of the CIB, $C(\theta) = \langle \delta
F({\mbox{\boldmath$x$}})\cdot \delta
F({\mbox{\boldmath$x$}}+\theta)\rangle $, is then $C(\theta) =\sum
\frac{(2l+1)}{4\pi} {\cal C}_l {\cal P}_l(\cos\theta)$ with ${\cal
C}_l = \langle |a_l|^2 \rangle \equiv (2l+1)^{-1} \sum_{m=-l}^l
|a_{lm}|^2$ and ${\cal P}_l$ denoting the Legendre polynomials
(e.g. Peebles 1980). The angular power spectrum, $P_2(q)$, is the
two-dimensional Fourier transform of $C(\theta)$ and for $l\gg1 $
is related to the multipoles via ${\cal C}_l =
P_2(l+\frac{1}{2})$. This follows because at $l \gg 1$ the
Legendre polynomials can be approximated as Bessel functions,
${\cal P}_l(\cos\theta) \simeq J_0((l+1/2)\theta)$. The magnitude
of the CIB fluctuation on scale $\pi/l$ radian for large $l$ is
then $\sim \sqrt{l^2{\cal C}_l/2\pi}$. In the limit of small
angles ($l\gg 1$) the values of ${\cal C}_l$'s are related to the
power spectrum of galaxy clustering and the CIB flux production
rate via:
\begin{equation}
{\cal C}_l = \frac{1}{c} \int \left( \frac{ d I_{\nu'} }{ dt}
\right)^2 \frac{1}{d_A^2(z)} P_3\left(\frac{ l+\frac{1}{2} }{
d_A(z)}; z \right) dt \label{cl}
\end{equation}

In surveys with arcsec angular resolution it is possible to
identify and remove galaxies brighter than some limiting
magnitude, $m_{\rm lim}$. Because on average fainter galaxies are
at higher $z$, by improving sensitivity and angular resolution,
one can isolate contributions to the CIB fluctuations from
progressively earlier epochs (Kashlinsky et al 2002). The
luminosity density, ${\cal L} =\int_0^{L(m_{\rm lim})} \Phi(L)
dL$, is then peaked at some particular redshift - at lower $z$ the
(brighter than apparent magnitude $m_{\rm lim}$) galaxies are
removed and at larger redshifts the luminosity density gets
dominated by the bright end of the luminosity function with a
sharp drop-off in the galaxy number density. The following toy
model is useful in estimating the effect: in the visible to
near-IR bands the present day galaxy luminosity function is of the
Schechter (1976) form $\Phi = \Phi_* L_*^{-1}
(L/L_*)^{-\alpha}\exp(-L/L_*)$. Measurements of the galaxy
luminosity function from B to K bands indicate that  within the
statistical uncertainties $\alpha \simeq 1$ (Loveday et al 1992,
Gardner et al 1997), leading to ${\cal L} = \Phi_* L_*
(1-\exp[-L(m_{\rm lim})/L_*])$. This then defines the redshift
window in eqs. \ref{limber},\ref{lumden} which contributes most to
the power spectrum of the CIB. Contribution from low $z$ galaxies,
for which $L(m_{\rm lim})<L_*$ is ${\cal L} \simeq \Phi_* L(m_{\rm
lim}) \propto d_L^2(z)$. In practice the typical redshift at which
most of the contribution arises can be estimated as $L(m_{\rm
lim}) \sim L_*$. If one can further remove galaxies lying in
narrow bins, $\Delta m$, around progressively fainter apparent
magnitude $m_{\rm lim}$, one can hope to isolate contributions to
the CIB by epoch (Kashlinsky 1992, Kashlinsky et al 2002).

\subsubsection{Shot noise fluctuations from individual galaxies}

In addition to fluctuations from galaxy clustering there would
also be a shot-noise component
 arising from discrete galaxies occasionally entering the beam.
 The relative amplitude of these shot noise fluctuations will be
 $\sim N_{\rm beam}^{-1/2}$ where $N_{\rm beam}$ is the average
 number of galaxies in the beam.
 This component is important in surveys with good angular
 resolution where $N_{\rm beam} \lsim $ a few.

If galaxies are removed down to some magnitude $m$, the shot noise
contribution from the remaining sources to the flux variance,
$C(0)$, in measurements with beam of the area $\omega_{\rm beam}$
steradian is given by:
\begin{equation}
\sigma_{\rm sn}^2 = \frac{1}{\omega_{\rm beam}} \int_m^\infty
F^2(m) \frac{dN_{\rm gal}}{dm} dm \label{shotnoise}
\end{equation}
Here $F(m)\equiv F_0 10^{-0.4m}$ is the flux from galaxy of
magnitude $m$ and $dN_{\rm gal}/dm$ is the number of galaxies per
steradian in the magnitude bin $dm$.

Fourier amplitudes of the shot noise are scale-independent and
equation \ref{c0} implies that the shot noise contribution to the
power spectrum of the CIB would be given by:
\begin{equation}
P_{\rm sn}= \int_m^\infty F^2(m) \frac{dN_{\rm gal}}{dm} dm
\label{power_shotnoise}
\end{equation}
Because galaxy clustering has power spectrum that increases
towards large scales, the shot noise component becomes
progressively more important at smaller angular scales.

\subsubsection{Cosmic variance for CIB anisotropies}

Any measurement of the angular power spectrum will be affected by
the sample or cosmic variance in much the same way as the cosmic
microwave background measurement (Abbot \& Wise 1984). This
results from the fact that in the best of situations we can only
observe $4\pi$ steradian leading to poor sampling of the long
wavelength modes. If the power spectrum is determined from
fraction $f_{\rm sky}$ of the sky by sampling in concentric rings
of width $\Delta q$ in angular wavenumber space, the relative
uncertainty on $P_2(q)$ will be $N_q^{-1/2}$, where $N_q \propto
q\Delta q$ is the number of ring elements in $[q;q+\Delta q]$
interval. Therefore, the relative uncertainty from cosmic variance
in the measured power spectrum on scale $\theta\simeq \pi/\theta$
will be:
\begin{equation}
\frac{\sigma_{P_2}^{\rm CV}}{P_2}|_{\rm cosmic \; variance} \simeq
\frac{1}{2\pi} \; \frac{\theta}{180^\circ}\; \sqrt{\frac{q}{\Delta
q}}\; f_{\rm sky}^{-1/2}
\end{equation}
In order to get reliable and independent measurements at a given
scale, it is useful to have narrow band $\Delta q/q \sim
0.05-0.1$. Therefore, for reliable measurements on scales up to
$\theta$, one has to cover an area a few times larger.

\subsubsection{CIB dipole component}

The dipole anisotropy of the CIB arises from our local motion with
respect to the inertial frame of the Universe, rather than galaxy
clustering. It carries important cosmological information and its
amplitude and wavelength dependence can be predicted in a
model-independent way and may one day be measurable.

If all of the dipole anisotropy of the CMB is produced by peculiar
motion of the Sun and the Local Group with respect to the inertial
frame provided by a distant observer (the last scattering surface
in the case of the CMB or early epochs, high $z$, in the case of
the CIB), the CIB should have dipole anisotropy of the
corresponding amplitude and in the same direction. The amplitude
of the dipole anisotropy can  be characterized by the first term,
${\cal C}_1$ in expanding the sky in spherical harmonics. Since
$I_\nu/\nu^3$ is an optical constant along the ray's trajectory,
the motion of the terrestrial observer at speed $v_{pec}$ with
respect to the observed background will produce dipole fluctuation
of the amplitude $\delta I_\nu/\langle I_\nu\rangle = (3 -
\alpha_\nu) (v_{pec}/c) \cos\theta= {\cal C}_1 \cos\theta $ (e.g.
Peebles \& Wilkinson 1968). Here $\theta$ is the angle between the
line-of-sight and the direction of motion and $\alpha_\nu \equiv
\partial\ln I_\nu/\partial\ln \nu$ is the spectral index of the
radiation.

For CMB measurements in the Rayleigh-Jeans part of the CMB
spectrum the index is $\alpha_{\rm CMB} \simeq 2$. Hence, the CIB
dipole is related to that of the CMB via ${\cal C}_{1,{\rm CIB}} =
(3-\alpha_\nu) \; {\cal C}_{1,{\rm CMB}}/T_{\rm CMB} \; F_{\rm
CIB}$. The CMB dipole is known very accurately to be ${\cal
C}_{1,{\rm CMB}} =3.346 \pm 0.017$ mK (Bennett et al 2003). Hence
the CIB dipole amplitude is expected to be
\begin{equation}
\delta F_{\rm CIB, dipole} \simeq 1.2 \times 10^{-3}
\;(3-\alpha_\nu) \; F_{\rm CIB}
 \label{dipole}
\end{equation}

\begin{figure}[h]
\centering \leavevmode \epsfxsize=0.6 \columnwidth
\epsfbox{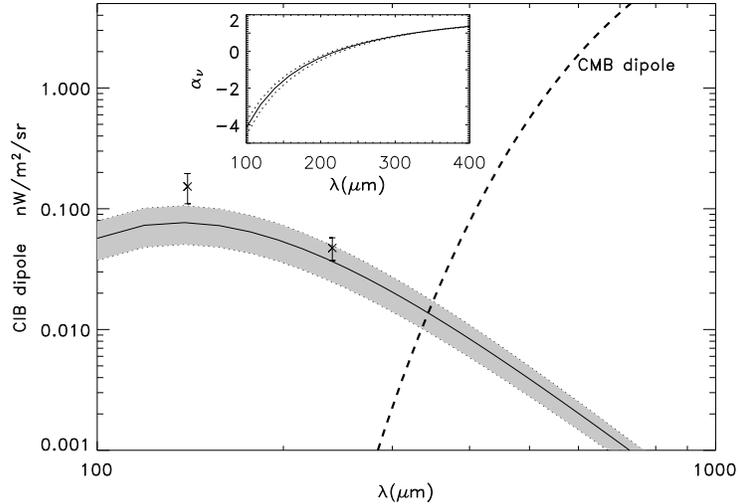} \caption[]{ \scriptsize{ The insert shows the
mean spectral index of the far-IR CIB according to eq.
\ref{firas_cib} (solid line) with its approximate uncertainty
(dotted lines). Main figure shows the expected amplitude of the
CIB dipole anisotropy normalized to the CMB dipole. The mean CIB
dipole according to eq. \ref{firas_cib} is shown with solid line
and the shaded area denotes its approximate uncertainty. Crosses
with errors show the dipole expected for the DIRBE measured levels
of the CIB at 140 and 240 \um (Hauser et al 1998), which are also
plotted with crosses in Fig. \ref{cib_dc}. Thick dashed line shows
the latest measurements of the CMB dipole (Bennett et al 2003).} }
\label{dipole}
\end{figure}

The current measurements of the CIB are discussed at length in
Sec. 5. In the far-IR they show that the energy spectrum of the
CIB has a 'window' where $\alpha_\nu$ is strongly negative (see
Fig. \ref{cib_dc} and eq. \ref{firas_cib}). Fig. \ref{dipole}
shows the spectral index of the CIB derived from the FIRAS based
measurements of the CIB, eq. \ref{firas_cib}, (Puget et al 1996,
Fixsen et al 1998) and the resultant CIB dipole amplitude
normalized to the observed CMB dipole. This component may be
measurable at wavelengths where the CIB has a strongly negative
$\alpha_\nu$ and a sufficiently high flux level. The measurements
show that $\alpha_\nu$ is strongly negative for $\lambda < 200
\mic$ reaching the levels of $\alpha_\nu \sim -2$ to $-4$ there.
Around $\sim 100$ \um\ the CIB spectrum has $\alpha_\nu \simeq -4$
and the CIB dipole should be $\simeq 7$ times more sensitive than
the CMB dipole. The CIB flux at these wavelengths is $\sim 10-20$
\nwm2sr and the CIB dipole at this range of wavelengths should
have a non-negligible amplitude of $\sim 0.1-0.2$ \nwm2sr and may
be detectable in some future measurements. At longer wavelengths
the CIB dipole will be difficult to measure due to steep increase
in the residual uncertainty from the CMB dipole which is also
shown in the Figure.

The CIB dipole anisotropy could be additionally enhanced because
the over-density that provides our peculiar acceleration also
presumably has excess IR luminosity. Indeed, there are persistent
claims of large bulks flows on scales of $\sim 150-200h^{-1}$Mpc,
whose direction roughly coincides with that of the CMB dipole (see
Willick 2000 and references cited therein). Furthermore, the
dipole of the distribution of rich (Abell) clusters does not
converge out to $\sim 200h^{-1}$Mpc, while its direction roughly
coincides with the CMB dipole (Scaramella et al. 1991).

The measurement of the CIB dipole anisotropy should be possible in
the wavelength range 100--300 \um and will be important to provide
additional information on the peculiar motions in the local part
of the Universe and will serve as additional, and perhaps
ultimate, test of the cosmological nature of any CIB detection at
that wavelength.

\subsection{Cosmological paradigms}

CIB levels and structure depend on the history of energy
production in the post-recombination Universe. The energy
production is driven mainly by nucleosynthesis which is related to
the history of the baryonic component of the Universe. Ultimately,
it is the gravity that drives baryon evolution and the latter is
determined by the evolution of the density inhomogeneities, the
nature of dark matter and the cosmological parameters.

With the WMAP \cite{wmap} measurements the structure of the last
scattering surface has been mapped and a firm cosmological model
has now emerged: the Universe is flat, dominated by the vacuum
energy or an exotic quintessence field and is consistent with the
inflationary paradigm and a cold-dark-matter (CDM) model. In this
paper, we adopt the $\Lambda$CDM model (Efstathiou et al 1990)
with cosmological parameters from the WMAP and other observations:
$\Omega_{\rm baryon}=0.044, h=0.71, \Omega_{\rm m}=0.3,
\Omega_\Lambda=0.7, \sigma_8=0.84$. Following the last scattering
at $z\sim 1000$, the Universe entered an era known as the "Dark
Ages", which ended with the star formation which produced
Population III stars. The WMAP polarization results \cite{kogut}
show that the Universe had an optical depth since last scattering
of $\tau \sim 0.2$, indicating an unexpectedly early epoch of the
first star formation ($z_* \sim 20$). From the opposite direction
in $z$, optical and IR telescopes are now making progress into
understanding the luminosity history during the most recent epoch
of the universe ($z < 5$), but the period from recombination to
the redshift of the galaxies in the Hubble Deep Field (HDF)
remains largely an unexplored era. CIB offers an alternative and
powerful tool to probe those epochs.

The evolution of dark halos, destined to convert baryons into
stars, is fixed by the power spectrum of the (dark) matter
inhomogeneities. The latter are believed to have been imprinted
during inflationary era and COBE DMR and WMAP observations confirm
that it started out with the scale invariant spectrum of the
Harrison-Zeldovich slope. Within the CDM framework the later
evolution of the density fluctuations is fairly well understood:
during radiation-dominated era fluctuations inside the horizon
remain frozen, whereas super-horizon modes grow self-similarly.
When the Universe became matter dominated, all modes grew at equal
rate. The epoch of the matter radiation equality thus determines
the overall shape of the power spectrum with the horizon scale at
that time ($\propto \Omega_{\rm matter}  h^2$) being the only
scale imprinted. The initial power spectrum is modified by the
so-called transfer function and, in linear approximation, depends
only on the cosmological parameters. Various approximations exist
for its shape; we chose the approximation from Sugiyama (1995)
with normalization to the COBE and WMAP CMB anisotropies for the
numbers that follow.

The amplitude of matter density fluctuations on a given scale
$\pi/k$ is $\simeq \sqrt{k^3P(k)/2\pi^2}$, where $P(k)$ is the
power spectrum of density perturbations. Fourier modes evolution
in the post-recombination Universe is determined by the
cosmological parameters and the equation of state, $p=w\rho$.
Cosmological constant, or vacuum energy, would lead to $w=-1$;
various quintessence models generally require $-1 < w < -1/3$ and
the ''normal" matter dominated Universe requires $w=0$. The growth
of fluctuations, $\Psi\equiv \delta(z)/\delta(0)$  is governed by
the differential equation:
\begin{equation}
\ddot{\Psi} + 2 \frac{\dot{R}}{R}\dot{\Psi}-4\pi G \bar{\rho}_{\rm
matter}\Psi=0 \label{ddot}
\end{equation}
where $R=(1+z)^{-1}$ is the expansion factor. This is coupled with
the cosmic time -- redshift relation, or Friedman equation, which for the flat Universe becomes:
\begin{equation}
\dot{R}^2 = H_0^2 \; \left( \frac{\Omega_m}{ R} +\frac{1-\Omega_m}{R^{1+3w}}\right)
\label{rdot}
\end{equation}

\begin{figure}[h]
\centering \leavevmode \epsfxsize=0.95 \columnwidth
\epsfbox{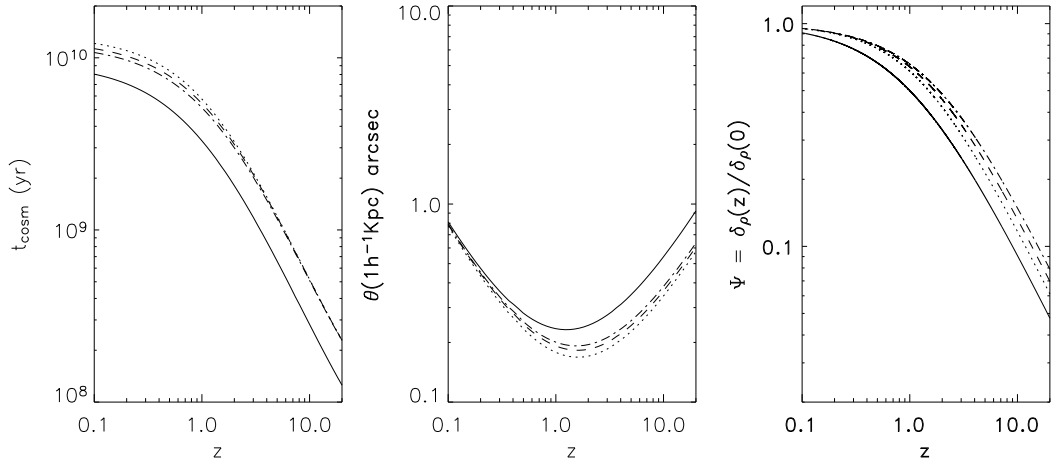} \caption[]{ \scriptsize{ {\bf Left}: cosmic time
as function of redshift. {\bf Middle} angular scale subtended by
the physical scale $1h^{-1}$Kpc vs $z$; and {\bf Right} the growth
factor for linear density fluctuations. Solid lines in all panels
correspond to $\Omega_m = 1$ with $w=0$ and sets of broken lines
to flat Universe with $\Omega_m=0.3$ and equation of state
$P=w\rho c^2$ with $w=-1, -2/3, -1/2$.} } \label{tcosm}
\end{figure}

Fig. \ref{tcosm} shows the various parameters out to $z=20$, the
redshift of the first star formation indicated by the WMAP
polarization measurements at large angular scales, which will be
useful throughout the review: left panel shows the cosmic time
from the Big Bang to the given redshift $z$ (WMAP analysis
suggests the present age is $\simeq$ 14 Gyr), middle panel shows
the angular scale subtended by physical scale 1$h^{-1}$Kpc, and
the right panels shows the growth evolution of density
fluctuations, $\Psi(z)$. The middle panel also shows the necessity
of the CIB studies: e.g. the candidate $z\sim 7$ galaxy (Kneib et
al 2004) is estimated to have the total extent of $\lsim 1$ Kpc
and is likely to be among the largest systems at those epochs.
Even if one resolves such compact systems, one would have to
compromise on simultaneously observing a sufficiently large part
of sky to gather robust statistics about the abundance and
large-scale distribution of (very) high redshift galaxies. Thus
the bulk of stellar material at high redshifts may be inaccessible
to current and even future direct studies unless such work is
complemented by the measurements of the CIB and its structure.

\subsection{From cosmological paradigm to galaxies.}

\begin{figure}[h]
\centering \leavevmode \epsfxsize=0.65 \columnwidth
\epsfbox{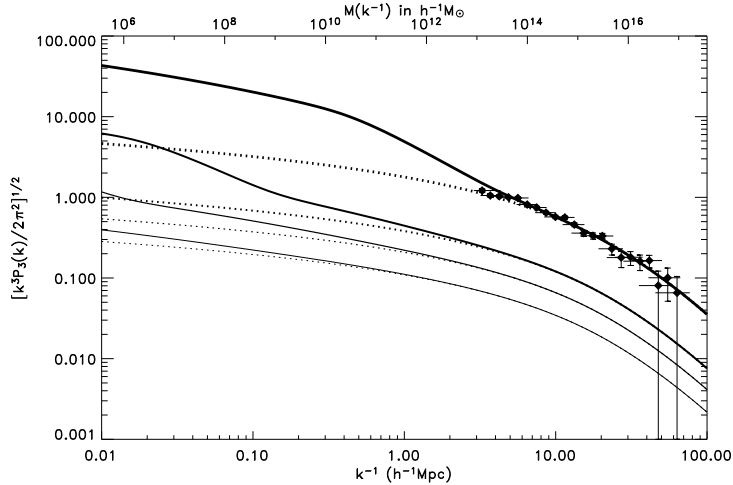} \caption[]{ \scriptsize{ $\Lambda$CDM density
field at $z=20, 10, 5$ and 0 (thin to thickest) using the Peacock
\& Dodds (1996) approximation for non-linear evolution. Dotted
lines corresponds to linear density field and solid lines include
non-linear evolution. Filled diamonds with errors show the
measurements of the present-day power spectrum of galaxy
clustering from SDSS survey (Tegmark et al 2004); assuming no
biasing it would coincide with the power spectrum of mass
fluctuations. The top axis shows the mass in $h^{-1}M_\odot$
contained within a comoving radius of $1/k$ for $\Omega=0.3$. }}
\label{p3}
\end{figure}

Eq. \ref{ddot} applies only in the linear regime, where the
density contrast $\delta_\rho \equiv \delta\rho/\rho <<1$. As time
goes on density fluctuations grow until they become non-linear,
turn around, separate from the general expansion frame and
collapse to form compact objects. For CDM models the typical
density contrast increases toward smaller scales, so the small
scale objects collapse earlier. At present epoch the RMS
fluctuation in the counts of galaxies is close to unity at
$r_8=8h^{-1}$Mpc (Davis \& Peebles 1983) and the power spectrum of
galaxy clustering has been accurately measured on scales up to
$\sim 100 h^{-1}$Mpc from the 2dF (Percival et al 2002) and SDSS
(Tegmark et al 2004) surveys and found in good agreement with the
$\Lambda$CDM model (Efstathiou et al 1990). The power spectrum
should be normalized to reproduce at present ($z=0$) the RMS
fluctuations of $\sigma_8=0.84$ over a sphere of radius $r_8$. On
non-linear scales gravitational effects would modify the shape and
amplitude of the mass power spectrum. We modeled its evolution
with $z$ using the Peacock \& Dodds (1996) approximation. Fig.
\ref{p3} shows the evolution of the mass power spectrum, or the
RMS density contrast $\delta_{\rm RMS} \simeq \sqrt{k^3
P_3(k)/2\pi^2}$, for $\Lambda$CDM model normalized to WMAP
cosmological parameters as function of scale. Non-linear scales
correspond to $\sqrt{k^3 P_3(k)/2\pi^2}\gsim 1$ and in the
Harrison-Zeldovich regime $\sqrt{k^3 P_3(k)/2\pi^2} \propto
k^{2}$. The total mass contained in a given scale is $M\simeq 1.25
\times 10^{12} (r/1h^{-1}{\rm Mpc})^3 \Omega h^{-1}M_\odot$; the
baryonic mass would be smaller by $\Omega_{\rm baryon}/\Omega_{\rm
matter}$. Adopting spherical model for the evolution and collapse
of density fluctuations would give that any mass that, in linear
approximation, reached the density contrast of $\delta_{\rm
col}=1.68$ could collapse. Therefore $\eta = \delta_{\rm
RMS}/\delta_{\rm col}$ gives the number of standard deviations a
given object had to be in order to collapse at the given $z$. CDM
and most inflation inspired models predict that the primordial
density was Gaussian, so the probability that a given mass has
collapsed at redshift $z$ is given by $P_M = {\rm
erfc}(\eta(z)/\sqrt{2})$.

Once the fluctuation on a given scale turns around and collapses,
gaseous processes will be critically important in determining how,
when and which compact objects will form. Hoyle (1953) was the
first to point out the importance of cooling in determining the
galaxy masses, Rees \& Ostriker (1977) have set it in the
cosmological context. Galaxy morphology (roughly speaking,
ellipticals vs disk galaxies) may be related to either dissipative
collapse and fragmentation with angular momentum determining the
efficiency of fragmentation and star formation (Kashlinsky 1982)
or mergers (Toomre \& Toomre 1972).

A further complication is that the distribution of luminous
systems may not trace that of the mass, or the so-called biasing.
This is likely to be important at early times when stars formed in
rare regions with fluctuations of the necessary amplitude. Biasing
amplifies the 2-point correlation function in locations traced by
the objects (Kaiser 1984). Various biasing schemes exist (e.g.
Politzer \& Wise 1986, Jensen \& Szalay 1986, Kashlinsky 1987,
1991, 1998) which may (or may not) at least be good approximations
to real situations.

The era of metal-rich stellar populations composed of Population I
and II stars, which we term 'ordinary', was likely preceded by a
(possibly brief) period of the first zero metallicity stars, the
so-called Population III. Theoretically this is a special case,
much less rooted in observational data, but also very important
for understanding the later formation and evolution of ordinary
galaxy populations. These populations and their imprint on the CIB
are discussed in a separate sec. 7.

In order to interpret the measurements of the CIB, one needs to
understand or model the details related to individual galaxy
formation and evolution. Partridge \& Peebles (1967) were the
first to discuss the levels of the CIB expected from galaxy
formation in the Big Bang model. Tinsley (1976, 1980) pioneered
the studies of evolution of stellar populations, their SEDs and
metallicity. Bond, Carr \& Hogan (1980) have discussed a wide
range of scenarios under which a measurable CIB can be produced,
from contribution by primeval galaxies to those of decaying
elementary particles; they also provided early estimates of the
angular power spectrum of the expected CIB anisotropies. Later
works continued the empirical approach or using a combined N-body
and analytical machinery. The number of papers relevant to this
brief theoretical section by far exceeds the limit of this review
and in what follows we concentrate on only a few representative
works to illustrate the tools used and the results obtained.

$\bullet$ {\it Backward and forward evolution}: Lonsdale (1996)
divided empirical modeling of CIB into the forward and backward
evolution classes. In the backward evolution approach one uses the
measurements of the CIB, galaxy luminosity function, their assumed
or modeled SEDs, observed galaxy counts, etc to translate galaxy
evolution back in time. The forward evolution approach takes the
opposite route, starting with galaxies at some initial epoch and
evolving their properties to the present time. Very often, both
approaches are needed.

Galaxy modeling generally consists of the following stages: 1) One
specifies galaxy morphology and morphological evolution with
cosmic time; 2) One then needs to specify the stellar IMF, which
may depend on cosmic time and vary among the various galaxy types;
3) Star formation rate must be assumed for each morphological type
and $z$; 4) Stellar evolution tracks must be distributed according
to the IMF and SED for each morphological type reconstructed at
each cosmic time; 5) Stellar emissions depend on the cosmic
abundances and chemical evolution of the ISM must be accounted for
and dust formation treatment is critically important at mid- to
far-IR; 6) The various parameters described above may vary among
galaxies of different mass or luminosity and this must be
specified also; 7) For CIB fluctuations one must assume the model
for the power spectrum of galaxy clustering, biasing and  their
time evolution. Additionally, these parameters depend on the total
age of the Universe, the redshift of galaxy formation $z_f$, which
may vary for different galaxies, etc. The models must be
normalized to reproduce as many observational data as possible and
are constrained by the data on galaxy morphology, metallicity $Z$
(including at high $z$),  observations of galaxy counts at various
wavelengths, colors, etc. Despite the many parameters (and
uncertainties) involved in the above construction, one can arrive
at meaningful and fairly accurate limits on galaxy formation and
evolution from the CIB.

Yoshii \& Takahara (1988) used the stellar evolution models from
Arimoto \& Yoshii (1987) in order to compute the near-IR CIB (and
the EBL) as function of the galaxy formation epoch and the
deceleration parameter. The evolving galaxies were divided into
five morphological types and their colors and counts were computed
for the various cosmological and evolutionary parameters. They
found that at visible bands most contribution to the EBL comes
from late types (Sa to Sd), whereas at the NIR bands early types
(E/S0) and late types give comparable contributions. Totani et al
(1997) have extended the Yoshii \& Takahara modeling for various
morphological types to model the observed evolution of the
luminosity density from UV to 1 \um (Lilly et al 1996). They noted
that if the proportion of galaxy mixes remains constant from
$z\sim 1$, then a flat Universe dominated by a cosmological
constant is preferred by the data.

 Franceschini et al (1991) made an early detailed study with
 predictions of future extragalactic surveys from near- to far-IR.
 They modeled chemical evolution of early and late-type galaxies
 normalized to K galaxy counts and allowed for various evolutionary
 modes of AGNs. Mid- to far-IR emission from dust was modeled with data from
 IRAS galaxies allowing for realistic distribution of dust grain sizes and temperatures
 and included emission from PAHs. They presented estimates of
 confusion limits from both Galactic stars and high-$z$ galaxies.

Fall, Charlot \& Pei (1996) presented an original way to relate
the CIB to other observed parameters. They assumed that the dust
at each redshift, having the same spatial distribution as stars,
is traced by the neutral hydrogen. In turn, the neutral hydrogen
column density can be normalized to reproduce the observed
comoving density of HI in damped Ly-$\alpha$ systems out to
$z\simeq 4$. They assumed a Salpeter type IMF for all galaxies and
solved for chemical evolution under various approximations. The
resultant CIB at long wavelengths was in good agreement with the
detection from FIRAS data (Puget et al 1996). In the near-IR the
predicted CIB is produced by stars, and they discuss how warm dust
can produce substantial levels of mid-IR CIB without significantly
affecting the far-IR part of the CIB spectrum, which was argued to
come from the $z\lsim 4$ galaxies responsible for the damped
Ly-$\alpha$ systems. They note that, for a wide range of the IMF,
the agreement of the computed emissivities with estimates from the
Canada-France Redshift Survey or CFRS (Lilly et al 1996) requires
that the initial density parameter of material that went into
stars and dust at $z\lsim 4$ was between $10^{-3} h^{-1}$ and
$8\times 10^{-3}h^{-1}$.

Jimenez \& Kashlinsky (1999) used synthetic models of stellar
populations to analyze the contribution of normal galaxies to
near-IR CIB and its fluctuations. The galaxies were assumed to
form at fixed $z_f$ and were divided into five morphological types
from E to Sd/Irr. Stars in the disks were assumed to form with the
Scalo IMF and in spheroids with the Salpeter IMF. Late type
galaxies were modeled to follow the Schmidt law for star formation
with the timescale which depended on the bulge-to-disk ratio.
Ellipticals were assumed to form at the fixed high $z_f$ in a
single burst of star formation and passively evolve. Because early
type galaxies contribute a significant fraction of the near-IR CIB
one has to be careful in normalizing to the observed properties of
these populations; hence the early-type galaxies were normalized
to the present-day fundamental plane of elliptical galaxies
assumed to mimic the metallicity variations along their luminosity
sequence. They found that, despite their simplicity, the models
gave good fits to the observed near-IR galaxy counts, the CFRS
measurements of the luminosity density evolution (Lilly et al
1996), and gave good matches to the observed metallicities of the
damped Ly-$\alpha$ systems (Pettini et al 1997) and galaxy colors.
This then allowed to refine the range of predictions for the CIB
and its fluctuations in the near-IR.

In the far-IR the bulk of the CIB comes from dust emission.
Beichman \& Helou (1991) constructed a model of dust emission,
which coupled with the observed IRAS galaxy luminosity function
and evolutionary assumptions for the various components of galaxy
emission, allowed (pre-COBE) theoretical estimates of the far-IR
CIB. Dwek et al (1998) normalized the mid- to far-IR emissions
from galaxies to the mean SED of IRAS galaxies and reproduce the
far-IR CIB by assuming a simple evolution of thereby constructed
galaxies. Haiman \& Knox (2000) computed the angular spectrum of
the far-IR CIB fluctuations assuming that the dust temperature is
determined by the UV radiation, i.e. that both the rate of dust
production and its temperature are determined by the (measured)
SFR (Star Formation Rate). They get that for a wide range of
models the far-IR CIB fluctuations should be $\sim 10$\% of the
mean CIB levels on degree and sub-degree scales consistent with
more general arguments of Kashlinsky et al (1996) and Kashlinsky
\& Odenwald (2000). Haiman and Loeb (1997) discussed the imprint
in the CIB produced by smooth dust in IGM spread by remnants of
the first stars. Knox et al (2001) have further expanded the
analysis exploring the possibilities of measuring the FIR CIB
anisotropies with the Planck space mission. Lagache et al (2003)
considered a phenomenological model at far-IR to sub-mm
wavelengths, constructing sample SEDs for starburst and normal
galaxies and assuming only the luminosity function evolution with
$z$. The model gave successful fits to the data on galaxy counts
and redshift distribution, the far-IR CIB and they used it to give
predictions for future observations with Herschel and Planck
missions.

 $\bullet$ {\it Semianalytical
galaxy formation} modeling presents another track for theoretical
insight and revolves around various numerical codes to generate
dark matter evolution for a given hierarchical clustering model
(usually CDM) and trace the merging history for present galaxy
haloes. Prescriptions for galaxy formation inside the formed dark
matter haloes are put in following gas dynamics and radiative
processes and hydrodynamics with the choices constrained by data
on e.g. Tully-Fisher (1977) relation, Lyman-$\alpha$ galaxies,
galaxy colors, SFR history etc (e.g. Kauffman, White \& Guiderdoni
1993, Baugh et al 1998, Sommerville, Primack \& Faber 1999).
Guiderdoni et al (1998) used semi-analytic modeling, normalized to
reproduce the observed far-IR/sub-mm part of the CIB, to study a
plausible range of redshift distributions and faint galaxy counts
at these wavelengths for the various evolutionary assumptions.

\section{Obstacles to measurement: confusion and foregrounds}

\begin{figure}[h]
\centering \leavevmode \epsfxsize=0.9 \columnwidth
\epsfbox{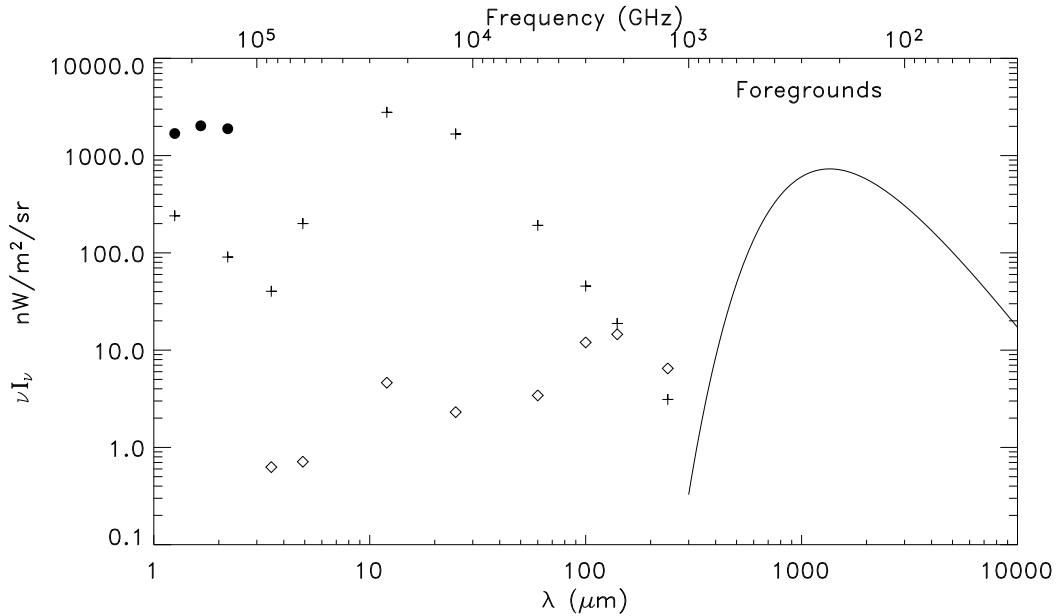} \caption[]{ \scriptsize{ IR emission from the
various foregrounds in CIB measurements: zodiacal light at high
Ecliptic latitudes (crosses, Kelsall et al 1998), Galactic cirrus
at high Galactic latitudes (diamonds, Arendt et al 1998),
atmospheric emission fluctuations at 1$^{\prime\prime}$ after
$\simeq 8$ sec integration (filled circles,
http://pegasus.phast.umass.edu/adams/airglowpage.html), and CMB
(solid line) } } \label{foregrounds}
\end{figure}

CIB measurements are very difficult because while the CIB signal
is relatively weak, the various foregrounds conspire to be fairly
bright as one moves to the IR range of wavelengths. Fig.
\ref{foregrounds} summarizes the various foregrounds and the
remainder of this section discusses them and their structure. The
figure illustrates the difficulty of eliminating the foreground
emission down to levels of $\sim 10$ \nwm2sr in the infrared.
While some of the foregrounds may be small at certain wavelengths,
their sum-total always conspires to be well above the required
level of $\sim 10$ \nwm2sr .

Foreground emission from the Galaxy and the solar system is the
main problem in unveiling the expected CIB.  At wavelengths less
than 10 \um, the dominant foreground after removing the zodiacal
light model is emission from stars in our Galaxy. At sub-mm
wavelengths cosmic microwave background (CMB) emission dominates
everything else. In ground-based measurements atmospheric emission
is important.

Intensity fluctuations of the Galactic foregrounds are perhaps the
most difficult to distinguish from those of the CIB. Stellar
emission may exhibit structure from binaries, clusters and
associations, and from large scale tidal streams ripped from past
and present dwarf galaxy satellites of the Milky Way. At long IR
wavelengths, stellar emission is minimized by virtue of being far
out on the Rayleigh--Jeans tail of the stellar spectrum (apart
from certain rare classes of dusty stars). At near-IR wavelengths
stellar emission is important, but with sufficient sensitivity and
angular resolution most Galactic stellar emission, and related
structure, can be resolved and removed.

\subsection{Atmospheric emission}

For ground-based observations, the largest contribution to the sky
background comes from the atmosphere itself, which at K$_s$ band
amounts to 295 MJy/sr.  At wavelengths longer than $\sim 2.5$ \um\
the spectrum of the emission is characteristic of a black-body
with a typical atmospheric temperature near 250 K. The emission at
1--2.5 \um\ range is dominated by  many intrinsically narrow OH
lines with some contribution from molecular hydrogen (at 1.27 \um\
) and other species. The atmospheric seeing is typically $\sim
1^{\prime\prime}$. On sub-arcminute angular scales fluctuations in
the atmospheric emission have white noise spectrum both in the
spatial and time domains and hence scale $\propto \theta^{-1}
t^{-1/2}$. On larger angular scales atmospheric gradients become
important (http://pegasus.phast.umass.edu/adams/airglowpage.html).
Solid circles in Fig. \ref{foregrounds} show typical atmospheric
fluctuations at 1$^{\prime\prime}$ during nighttime observations
by 2MASS survey after 7.8 sec of integration.

\subsection{Galactic stars}

For low resolution experiments (e.g. DIRBE) Galactic stars are a
major contributor to foreground emission at wavelengths $\lsim 3$
\um. In narrow beam observations beam observations they can be
excised out to fairly faint magnitude. In larger beam
measurements, one can use their statistical properties for removal
of their cumulative emission. For purposes of measuring the CIB,
regions of the sky within $20^\circ-30^\circ$ of the Galactic
plane can be ignored.

In the near-IR Galactic stars, have two useful properties: 1) at
the Galactic poles star counts have a simple scaling with
magnitude given by:
\begin{equation}
\frac{dN}{dm} \propto 10^{Bm} \label{dnstars} \label{dngalstars}
\end{equation}
 and 2) outside the Galactic plane ($|b_{\rm Gal}|\geq 20^\circ$) and
 away from the Galactic center ($90^\circ < \ell_{\rm Gal} <
 270^\circ$)
 the Galaxy can be approximated as plane-parallel. In a
plane-parallel Galaxy in which the radial structure can be
neglected,  the  differential counts in the direction $x={\rm
cosec}|b|$ can be related to those at the Galactic pole by
(Kashlinsky \& Odenwald 2000)
\begin{equation}
\frac{dN}{dm}|_x = x^3 \frac{ dN(m-5\log_{10} x)}{dm}|_{\rm Pole}.
\label{planegal}
\end{equation}

\begin{figure}[h]
\centering \leavevmode \epsfxsize=0.8 \columnwidth
\epsfbox{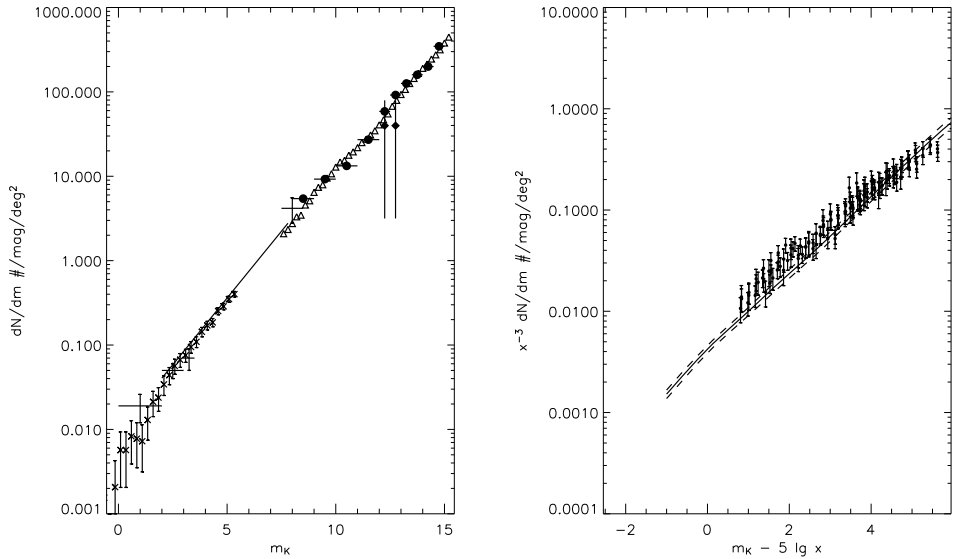} \vspace{0.5cm} \caption[]{ \scriptsize{ {\bf
Left} -- Number counts measurements of Galactic stars in K band at
the Galactic Poles. DIRBE based counts (crosses) are from
Kashlinsky \& Odenwald (2000) (at fainter magnitudes the DIRBE
counts are affected by the confusion noise of the large DIRBE beam
- see Fig. 7 in Kashlinsky \& Odenwald 2000). Filled triangles are
differential counts  from Elias (1978) NGP measurements. Filled
circles are differential NGP counts from 2MASS  (Jarrett 1998).
Open triangles are cumulative 2MASS counts from Beichman (1998)
multiplied by $0.3\ln 10$ to convert to differential counts for
$dN/dm \propto 10^{0.3m}$. Filled diamonds with error bars are
South Galactic Pole counts from Fig.  1 of Minezaki et al. (1998).
Solid line shows the
$B=0.3$ fit.  \\
{\bf Right} -- DIRBE star counts at 2.2 \um plotted in coordinates
where a plane-parallel Galaxy would collapse on a single line;
$x=\cosec|b|$. Data are from the $\simeq 20\circ\times 20^\circ$
DIRBE patches with $|b| \geq 20 \deg$ and $90\deg < l < 270\deg$,
with Poisson errors shown.  For the DIRBE beam the confusion noise
affects counts at  $K >5.5$. Lines show the model Galaxy: solid is
mean $x^{-3}dN/dm$ and dashes are the $\pm$1-sigma spread.

} }
\label{gstars}
\end{figure}

 The right panel of Fig. \ref{gstars} illustrates the degree of accuracy of
 the plane-parallel approximation for the Galaxy (eq. \ref{planegal})
 using DIRBE data (Kashlinsky \& Odenwald 2000).
Using these approximations, once the counts at the Galactic pole
are measured, one can evaluate the expected number counts  in any
direction $b$ and then compute the flux and fluctuations in the
flux they produce via:
\begin{equation}
F_{\rm stars}(<m) \propto \int_m^\infty 10^{-0.4 m}
\frac{dN}{dm}|_x dm \; \; \; ; \; \; \; \sigma^2(x)  \propto
\int_m^\infty 10^{-0.8 m} \frac{dN}{dm}|_x dm. \label{e22}
\label{fgalstars}
\end{equation}
For $B<0.4$ both are dominated by the brightest stars remaining in
the field.

The left panel in Fig. \ref{gstars} shows compilation of stars
counts at 2.2 \um (K band) in the North Galactic Pole (NGP) region
from various measurements. The data show that $B\simeq 0.3$; the
value of $B=0.6$ would correspond to homogeneous distribution of
sources or stars coming from well within the scale height of the
Galactic disk. The NGP star counts were observed directly by Elias
(1978). We show his data at  $K= 1$, 2.5, 3.25 and 8 with
$N^{1/2}$ error bars and our binning of his data. Further NGP data
were obtained by the 2MASS survey in K$_s$ band, almost identical
to the DIRBE Band 2, and were kindly provided to us by Tom Jarrett
(1998, private communication) who created it using the 2MASS point
source catalog, now available to the public. The cumulative counts
from these measurements were shown in Fig. 1 of Beichman (1997)
out to $K_s
>15$, who found that they follow $dN/dm \propto 10^{0.3 m}$  (cf.\
his Table 4). Actual 2MASS star counts from a region of 5 square
degrees centered on the NGP are plotted in Fig. \ref{gstars}. The
agreement between the DIRBE counts, the Elias (1978) and Jarrett
(1998, private communication) data, and the $B=0.3$ extrapolation
is excellent over 15 magnitudes, or six decades in flux. South
Galactic Pole counts from Minezaki et al.(1998) are also shown. At
$m_K < 1.5$, the counts tend to the slope of $B=0.6$ coming from
stars much closer than the scale height; if $B$ were less than
0.6, the integrated star brightness would diverge at the bright
end.

The star counts agree with model predictions. Both Beichman (1996)
in his Fig.\ 1 and Minezaki et al. (1998) in their Fig.\ 1 show
that the counts are fitted well by extensions of either the
Bahcall \& Soneira (1980) or  Wainscoat et al.\  (1992) models. An
eyeball fit to their data gives $B=0.3-0.32$ at $K=11$.  The
Wainscoat et al.\ model at $K=11$ shown in Fig.\ 1 of Minezaki et
al.\ (1998) gives $\log dN/dm \simeq 1.35$, whereas continuation
of the solid line in the left panel of Fig.\ \ref{gstars} to $K =
11$ gives $\log dN/dm = 1.3$ if $B=0.3$ and 1.4 if $B=0.33$. The
agreement between the two slopes and normalizations is thus very
good. Even the large-beam DIRBE instrument sees far beyond the
scale height of the bright K band stars.

\subsection{Zodiacal emission}

Zodiacal emission from interplanetary dust (IPD) is the brightest
foreground at most IR wavelengths over most of the sky. There are
some structures in this emission associated with particular
asteroid families, comets and comet trails, and an earth--resonant
ring, but these structures tend to be confined to low ecliptic
latitudes or otherwise localized (Reach et al 1995). The main IPD
cloud is inclined at $\sim 2^\circ$ with respect to the Ecliptic
and is generally modeled with a smooth density distribution that
also contains additional features of a circumsolar ring, a density
enhancement in the Earth's wake and the dust bands at several AU
(Kelsall et al 1998). Combining DIRBE and FIRAS observations,
Fixsen \& Dwek (2002) find a sharp break in the dust distribution
at radius of $\sim 30$\um and that the zodiacal energy spectrum
beyond $\sim 150$ \um can be fitted with a single black-body with
$\lambda^{-2}$ emissivity and temperature of 240K.
Observationally, intensity fluctuations of the main IPD cloud have
been limited to $<0.2$\%
 at 25 $\mu$m (\'Abrah\'am, Leinert, \& Lemke
1997). Because the Earth is moving with respect to (orbiting
within) the IPD cloud, the zodiacal light varies over time.
Likewise, any zodiacal light fluctuations will not remain fixed in
celestial coordinates. Therefore repeated observations of a field
on timescales of weeks to months should be able to distinguish and
reject any zodiacal light fluctuations from the invariant Galactic
and CIB fluctuations.

\subsection{Galactic cirrus}

IR emission from the ISM (cirrus) is intrinsically diffuse and
cannot be resolved. Cirrus emission is known to extend to
wavelengths as short as 3 $\mu$m. Statistically, the structure of
the cirrus emission can be modeled with power--law distributions,
$P_2(q) \propto q^n$ (Gautier et al 1992), and has the power index
$n\simeq -2,-2.5$ (Wright 2000, Kashlinsky \& Odenwald 2000,
Ingalls et al 2004). This power index lead to almost
scale-independent fluctuations in cirrus emission which are
typically $\lsim 10^{-2}$ of the mean flux level. Using the mean
cirrus spectrum, measurements made in the far-IR can be scaled to
3.5 $\mu$m, providing estimates for the fluctuation contribution
from cirrus. The extrapolation to shorter wavelengths is highly
uncertain, because cirrus (diffuse ISM) emission has not been
detected at these wavelengths, and the effects of extinction may
become more significant than those of emission.

\subsection{Cosmic microwave background}

Ironically, the cosmic microwave background (CMB) has to be put in
the category of ``foregrounds" when searching for the CIB. CMB is
a relic from the Big Bang and has a strict black-body energy
spectrum correspond to $T_{\rm CMB} = 2.725 \pm 0.001$K with very
small, if any, deviations from the black-body spectrum
\cite{cmb_firas,firas_cmb,fixsenanmather}. CMB is also very
homogeneous with the largest fractional spatial deviation being
$\sim 10^{-3}$ on dipole scales; the other angular moments are
some two orders of magnitude smaller (Bennett et al 2003). Its
brightness, shown with solid line in Fig. \ref{foregrounds},
overwhelms all other emissions at wavelengths longer than $\sim
500$ \um and is even a non-negligible source of noise in many
everyday electric appliances. Because its energy spectrum is known
so accurately and its spatial structure is so small, the CMB can
be subtracted down to levels $\ll$ 1 \nwm2sr .

\section{Current CIB measurements}

This section discusses the current observational status of both
the mean CIB levels and its structure (fluctuations).

Observationally, the CIB is difficult to distinguish from the
generally brighter foregrounds contributed by the local matter
within the Solar system, and the stars and ISM of the Galaxy. A
number of investigations have attempted to extract the isotropic
component (mean level) of the CIB from ground- and satellite-based
data. These analyses of the {\it COBE} data have revealed the CIB
at far-IR wavelengths $\lambda > 100$\um (Hauser et al 1998), and
probably at near-IR wavelengths from 1 - 3 \um\  (Dwek \& Arendt
1998, Gorjian et al 2001, Cambresy et al 2001) with additional
support from analysis of data from {\it IRTS} (Matsumoto et al
2000, 2003). However, it would probably be fair to say that none
of the reported detections of the isotropic CIB are very robust
(especially in the NIR), because all are dominated by the
systematic uncertainties associated with the modeling and removal
of the strong foreground emission of the zodiacal light and
Galactic stars and ISM. Furthermore, the near-IR colors of the
mean CIB do not differ greatly from those of the foregrounds, thus
limiting the use of spectral information in distinguishing the
true CIB from residual Galactic or solar system emission.

In order to avoid the difficulty of exactly accounting for the
contributions of these bright foregrounds in direct CIB
measurements, (and the difficulty in detecting {\it all} the
contributing sources individually), Kashlinsky, Mather, Odenwald
and Hauser (1996) have proposed measuring the structure or
anisotropy of the CIB via its angular power spectrum. They noted
that for a relatively conservative set of assumptions about
clustering of distant galaxies, fluctuations in the brightness of
the CIB have a distinct spectral and spatial signal, and these
signals can be perhaps more readily discerned than the actual mean
level of the CIB.

\begin{figure}[h]
\centering \leavevmode \epsfxsize=0.9 \columnwidth
\epsfbox{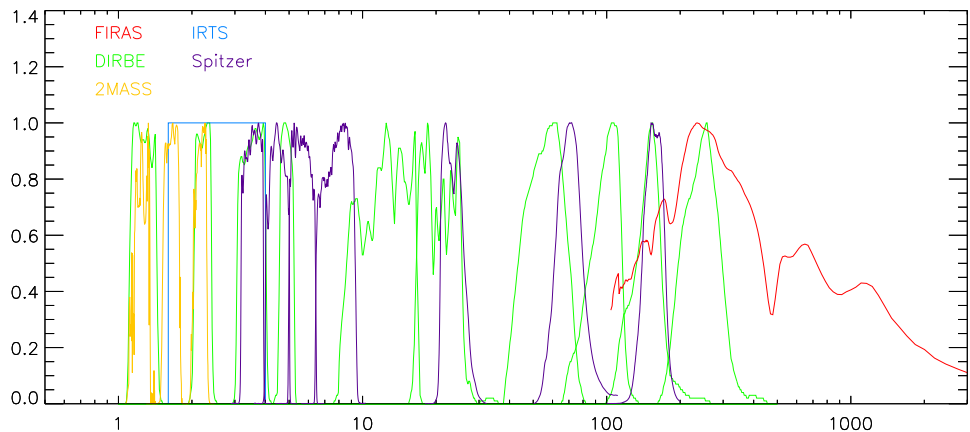} \caption[]{ \scriptsize{Filter responses for the
various instruments discussed in the paper. The curves were
normalized to a unity peak. DIRBE, FIRAS, 2MASS, IRTS, Spitzer
(IRAC and MIPS) filters correspond to the colors shown in the
upper left corner. The multiple ISO filters filters are not shown
for clarity.} } \label{filters}
\end{figure}

Figure \ref{filters} shows the filters and the IR wavelengths
covered by the measurements from the instruments described below.

\subsection{COBE DIRBE}

The primary mission of the DIRBE instrument on board COBE
satellite was to find, or set very strict limits on, the CIB
contribution from near- to far-IR. The DIRBE was described by
Boggess \etal (1992) and is a 10-band, photometer system covering
the wavelength range from 1.25 to 240 $\mu$m with an angular
resolution of 0.7\deg. The 1.25, 2.2 and 3.5 \um DIRBE bands are
similar, although not identical, to the ground based J, K and L
bands. It was designed to achieve stray light rejection of less
than 1 \nwm2sr and a low absolute brightness calibration
uncertainty. Accurate beam profile and beam response maps were
obtained in flight by observing multiple transits of bright point
sources. All spectral channels simultaneously observed the same
field-of-view on the sky. Its cryogenic stage lasted from November
1989 to September 1990 collecting 41 weeks of data.

Absolute photometry was obtained through frequent (32 Hz) chopping
between the sky and an internal zero flux  surface maintained at
temperatures below 2 K. System response was monitored every 20
minutes throughout the 10 month mission  by observing an internal
thermal reference source. The DIRBE was designed to achieve a
zero-point to the photometric scale with an uncertainty below $\nu
I(\nu) = 10^{-9} \;\rm W \rm m^{-2} \rm {sr}^{-1}$ at all
wavelengths; tests conducted prior to launch and in flight
indicate that this goal was indeed reached.

Absolute calibration of the DIRBE photometric scale, and long-term
photometric consistency, was established by monitoring a network
of stable celestial sources. The primary calibrator for the $J$,
$K$ and $L$-bands was the star Sirius. Although DIRBE is a
broad-band instrument, the measured in-band intensities are
reported as spectral intensities at the nominal effective
wavelengths of each band. Sources that have an SED significantly
different from a flat $\nu I(\nu)$ would require color
corrections. In the $J$, $K$ and $L$-bands, these color
corrections are of the order of a few percent or less over the
temperature range from 1500 to 20,000 K. Long-term stability of
the instrumental photometric scale at each wavelength was better
than 1\%. The stability of the band-to-band intensity ratios, or
colors, was 1.4\%. The absolute photometric uncertainty in the
DIRBE $J$, $K$ and $L$-bands was $\approx 4$\%.

The calibrated DIRBE observations were binned into 20 arcmin
pixels of approximately equal area for each of the forty one weeks
and are available from
(http://lambda/product/cobe/dirbe\_overview.cfm). Each week
covered about half the sky and four months of data gave one full
sky coverage.

In a real tour d'force the background and foreground results of
the combined 41 week maps of DIRBE sky observations have been
analyzed by the DIRBE team (Hauser et al 1998, Arendt et al 1998,
Kelsall et al 1998, Dwek et al 1998). From these maps, the {\it
models} of contributions from the interplanetary dust (IPD) cloud
and the Galaxy, both its interstellar dust and stellar components,
were subtracted for each direction.

The DIRBE IPD model is described in Kelsall et al (1998). It
represents the zodiacal sky brightness as the integral along the
line of sight of the dust emissivity times the 3-D dust density
distribution function. The emissivity arises from both thermal
emission and scattering; the former dominates at $\lambda \gsim 5
\mic$ and the latter at shorter wavelengths. The thermal emission
at each location assumes a single dust temperature for all IPD
cloud components and is assumed to drop as a power law of the
distance to the Sun. The density model included contributions from
a smooth cloud, three pairs of asteroidal dust bands and a
circumsolar dust ring. Parameters of the analytical functions that
enter the model were determined by matching the observed temporal
brightness variations in various directions on the sky. Once the
optimal set of model parameters was determined, the IPD model was
integrated along the line of sight at the mean time of each week
of observations. The calculated IPD brightness map was then
subtracted from each week of the DIRBE weekly maps. Because of the
brightness of the zodiacal emission, the residual sky maps are
dominated by the uncertainties in the IPD model at between 12 and
60 \um .

The Galactic emission was modeled by Arendt et al (1998) as made
up of emission of bright and faint Galactic stars and discrete
sources and the interstellar medium, or cirrus cloud emission. It
was removed from the combined the maps after the zodiacal
component removal. Bright(er) sources above a wavelength dependent
threshold were removed from the maps at all bands. The
contribution to the integrated light from faint sources was
removed using a statistical source count Faint Source Model (FSM -
Arendt et al 1998) which is based upon Wainscoat et al (1992).
Prior to subtracting the FSM, bright stars with flux greater than
15 Jy (corresponding to K$\simeq 4$) have been subtracted
directly. Stellar contribution is negligible at mid- and far-IR
DIRBE bands. The model for the ISM emission was taken to be a
product of the standard spatial template times a single spectral
factor at each wavelength. At 100 \um the spatial template was a
map of HI emission constructed from three different 21 cm HI
surveys (see Arendt et al 1998 and references cited therein); at
other wavelengths the 100 \um data were used.

Hauser et al (1998) presented the limits on the CIB after removal
of the zodiacal light model, the FSM and the Galactic cirrus
contributions from the combined DIRBE maps. In the residual maps,
they selected for the analysis ''high-quality" regions that are
located high Galactic and ecliptic latitudes, are free of
foreground removal artifacts and cover at least 2\% of the sky.
The smallest and best region contains some 8,140 pixels in both
northern and southern hemispheres allowing to test for isotropy of
the residual signal. Their criteria for a CIB detection were that
the residuals had to be in excess of 3-$\sigma$, $\sigma$ being
{\it both} systematic and statistical uncertainty, and be
isotropic in the high-quality region. They reported firm
detections at 140 and 240 \um\ and upper limits at shorter
wavelengths. The 100 \um\ residual they find at the level of
$21.9\pm 6.1$ \nwm2sr also has high statistical significance, but
is not isotropic. The dominant errors in the DIRBE CIB
measurements are the systematic errors associated with the
foreground subtraction. Dwek et al (1998) show in detail that the
isotropic signal detected at 140 and 240 \um\ is unlikely to arise
from some Solar system and Galactic sources. Adopting the DIRBE
detections at 140 and 240 \um, Dwek et al (1998) claim a {\it
lower} limit of 5\nwm2sr at 100 \um\ assuming that the CIB
emission arises from the coldest possible dust fitted to the 140,
240 \um detections.

\begin{figure}[h]
\centering \leavevmode \epsfxsize=0.7 \columnwidth
\epsfbox{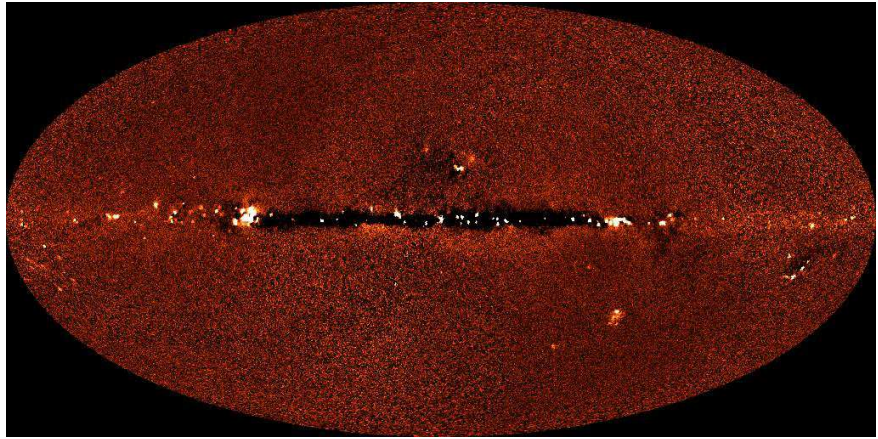} \caption[]{ \scriptsize{DIRBE map of the CIB at
240 \um shown in Galactic coordinates with the South Galactic Pole
pointing toward the bottom of the figure.} } \label{dirbe240}
\end{figure}

Figure \ref{dirbe240} shows the map of the 240 \um residual
emission from Hauser et al (1998); it is plotted here in Galactic
coordinates (cf. Arendt et al 1998). The produced map corresponds
to the original 240 \um map from which the zodiacal light was
removed along with a two-component ISM model, corresponding to a
weighted subtraction of the DIRBE 140 and 100 \um maps. The bright
sources near the Galactic plane are clearly visible as are the
Small and Large Magellanic clouds.  The remaining emission is
isotropic and was identified with the CIB. The 140 \um map also
passed the isotropy limits, although is significantly more
anisotropic outside the high-quality region used for identifying
the CIB.

Chronologically, the DIRBE discovery of the far-IR CIB was
preceded by a few months by an independent analysis of Schlegel et
al (1998), who have combined the well calibrated DIRBE data that
has relatively poor resolution ($\sim 0.5^\circ$) with the IRAS
data, which have much better angular resolution ($\sim 5^\prime$)
but poor absolute calibration. The combined maps at 100 \um\
provides a very accurate to-date template of the diffuse Galactic
emission and dust distribution. To construct the map, they have
removed 1) zodiacal emission from the DIRBE maps, 2) striping
artifacts from the IRAS data, 3) confirmed point sources, and then
combined the maps preserving the DIRBE calibration and IRAS
resolution. The ratio of the 100 \um\ to 240 \um\ emission gave
the dust temperature leading to estimate of column density of the
radiating dust.  The 25 \um\ channel data was used to model the
zodiacal light emission from the interplanetary dust (IPD) cloud.
Fitting black-body functions from 12 to 60 \um\ they find
variations in the IPD temperature of less than 10 \%. As the
zodiacal light contributes ca. 300 \nwm2sr at 100 \um\ the
temperature variations of the cloud lead to absolute errors of
$\sim 30$ \nwm2sr if the 25 \um\ template is used linearly in
subtracting from the longer wavelength maps. After accounting for
the (non-linear) corrections for the IPD temperature variations in
subtracting the zodiacal light emission, the CIB was determined as
the zero offset in cross-correlation between the residual
''de-zodied" emission with the Galactic HI column density from the
Leiden-Dwingeloo 21 cm survey (Hartmann \& Burton 1997). The
analysis has been applied to the 100, 140 and 240 \um\ maps
restricted to lie outside the ecliptic plane leading to
null-detection at 100 \um\ and robust CIB values at the two longer
bands for $|\beta| \gsim 20^\circ$.

Finkbeiner, Schlegel \& Davis (1998) extended the analysis to the
DIRBE data at 60 and 100 \um. They applied two different methods
to remove the zodiacal light component: The first of these methods
used the North-South asymmetry with a 1 year period of the
zodiacal component and concentrated on the data within $5^\circ$
of the ecliptic poles, where effects dependent on solar elongation
cancel out. The method further assumed that the zodiacal light
North-South asymmetry is the same in all bands. Cirrus was taken
from the 100 \um\ map produced in the Schlegel et al (1998)
analysis. They decomposed each of the North and South datasets
into time-dependent (zodi) and time-independent (cirrus, CIB etc)
components and obtain robust solutions for the latter. In the
second method they remove zodiacal component by using its
dependence on the ecliptic latitude in each of the DIRBE's 41
weeks of data with solar elongation 90$^\circ$. A statistic was
then constructed that ratios the flux in the various diagonal
directions to that toward the ecliptic poles and whose advantage
is that it has a known dependence on the ecliptic latitude
($\propto {\rm csc}|\beta|$) with a negligible annual variation.
The CIB flux was determined as a term at ${\rm csc}|\beta|=0$.
Both methods gave consistent and robust, if high compared to other
measurements, levels of the CIB at 60 and 100 \um.

Because of the difficulty in measuring the isotropic component of
the CIB, Kashlinsky, Mather, Odenwald \& Hauser (1996) have
proposed to probe the CIB structure instead. For a reasonable set
of assumptions about the underlying distribution of
emitters/galaxies, the fluctuations in the CIB can provide
important information about the rate of production of the CIB and
luminous activity in the early Universe. The method leads to
particularly strong constraints at mid- to far-IR where the
foregrounds are very smooth (Kashlinsky, Mather \& Odenwald 1996).
The analysis, however, could not have been applied to the DIRBE's
longest channels (140 and 240 \um), which required the use of
bolometers instead of the photoconductors, resulting in much
larger noise than at shorter wavelengths.

Kashlinsky \& Odenwald (2000) have applied the method to the final
DIRBE data release leading to detection of the near-IR CIB
anisotropies. They started with the final DIRBE sky maps from
which the zodiacal model from Kelsall et al (1998) was subtracted.
The sky was divided into (384) patches of 32 pixels ($\simeq
10^\circ$) on the side which were individually clipped of Galactic
stars and other point sources by various procedures all of which
led to the same final results. The RMS flux fluctuation, $\delta
F_{\rm RMS} \equiv \sqrt{C(0)}$, or fluctuation on the DIRBE beam
scale ($\simeq 0.6^\circ$) was computed in each of the clipped
patches. The residual fluctuation is the sum of contributions from
the remaining Galactic stars, CIB, instrument noise, etc. When the
data in the near-IR 1.25 to 5 \um were selected from high Galactic
latitude ($|b_{\rm Gal}| \geq 20^\circ$) and away from the
Galactic center ($90^\circ \leq \ell_{\rm Gal} \leq 270^\circ$),
they showed a very tight correlation between cosec($|b_{\rm
Gal}|$) and $C(0)$. That NIR fluctuation was explained in terms of
the {\it observed} Galactic star counts in the plane parallel
Galaxy, such as given by eqs. \ref{dngalstars},\ref{planegal},
shifted by a constant (isotropic) component. The latter presumably
arises from CIB fluctuations as well as instrumental noise.
Extrapolation of the correlation to cosec($|b_{\rm Gal}|)=0$, or
zero contribution from Galactic stars, gave a robust value of the
near-IR CIB anisotropies, which was demonstrated to be independent
of the clipping methods, sky maps processing and other effects.
The instrument noise was evaluated in Kashlinsky et al (1996),
Kashlinsky \& Odenwald (2000) and Hauser et al (1998) and was
shown to be significantly lower than the detected near-IR signal
at wavelengths from 1.25 to 5 \um. At mid- and far-IR wavelengths
the results of the analysis led to slightly tighter upper limits
than in the previous analysis (Kashlinsky, Mather \& Odenwald
1996) which used an intermediate zodiacal light model from the
DIRBE team.

Dwek \& Arendt (1998) have presented another original method to
isolate the isotropic component of the near-IR CIB by using prior
information of the flux in one of the bands. They noted that the
observed galaxy contribution to the K band CIB pretty much
saturates at $\simeq$9 \nwm2sr . They started with the DIRBE maps
at 1.25, 2.2, 3.5 and 4.9 \um from which they subtracted the
zodiacal light model from Kelsall et al (1998). From DIRBE near-IR
maps the regions were selected with small errors in the
subtraction of the foreground emission (Hauser et al 1998, Arendt
et al 1998). Small cirrus emission at 3.5 and 4.9 \um was removed
by scaling the average 100 \um intensity in DIRBE maps as in
Arendt et al (1998). The 2.2 \um DIRBE map minus the assumed CIB
at that band was taken as a template of the Galactic stellar
emission. That template was then cross-correlated with the maps at
the other wavelengths to give a positive and statistically
significant residuals (zero intercepts) summarized in Table
\ref{dirbetable}. This technique was pushed farther by Arendt \&
Dwek (2003) attaining a more isotropic removal of Galactic
foregrounds, but at the expense of increased uncertainty in the
absolute zero point.

In order to reduce the uncertainty in the Galactic stars
contribution to the NIR DIRBE maps, Gorjian et al (2000) have
imaged a high Galactic latitude region of the sky of $2^\circ$ on
the side in K and L bands with arcsec resolution to K$\simeq 9$
and $L\simeq 8$ magnitudes reducing the stellar contribution to
confusion noise a factor of $\sim 16$ below that of DIRBE. They
subtracted a modified zodiacal light model proposed by Wright
(1997) which comes in two flavors: a ''weak" zodi model which
requires that the DIRBE high Galactic latitude sky at 25 \um be
isotropic in addition to its non-ZL components remaining constant
in time, and a ''strong" zodi model which requires that the former
component in addition is negligibly small. This reduced the
residual 25 \um intensity by a factor of 7 compared to the DIRBE
model (Kelsall et al 1998). They then selected 17 DIRBE pixels
where overlap with star position is sufficient for robust
statistics in subtracting the Galactic stellar component. The CIB
was evaluated after subtracting the zodiacal components, the
measured flux from the Galactic stars and the model flux from
fainter and unobserved stars. Wright \& Reese (2000) used the
above ZL model subtraction and the Arendt et al (1998) method for
removing the ISM contribution to the residual emission. They found
a constant flux offset in the flux histograms at 2.2 and 3.5 \um
compared to the Wainscoat et al (1992) model after testing the
latter with 2MASS observations in several selected fields. The
flux offset is identified as the CIB and is consistent with the
Gorjian et al (2000) results. Wright (2000) used the 2MASS catalog
stars in order to subtract Galactic star contribution to the DIRBE
maps flux in selected regions near the Galactic poles. After the
subtraction he finds a statistically significant flux excess at
2.2 \um that is in agreement with the Gorjian et al (2000) and
Wright \& Reese (2000) results.

Cambresy et al (2001) used the 1,400 deg$^2$ of the 2MASS data
from standard exposures (7.8 sec) in J, H, K$_s$. They degraded
the 2MASS maps (2$^{\prime\prime}$) to $5^\prime$ resolution.
After subtracting the zodiacal light from DIRBE weekly maps using
the zodiacal subtracted mission average maps, the residual
intensity maps were averaged together. The subset of the 2MASS
that was used to account for Galactic contribution to the DIRBE
intensities came from high Galactic latitude ($|b_{\rm Gal}| \geq
40^\circ$) and high ecliptic latitude ($|\beta_{\rm ecl}| \geq
40^\circ$) regions of low cirrus emission for which a complete
2MASS coverage existed in each DIRBE pixel. After removing bright
stars, which contaminate more than one DIRBE pixel, the final area
for analysis was 1,040 deg$^2$. The intensities in the DIRBE sky
were cross correlated with the observed fluxes from 2MASS stars.
The zero-intercept of this correlation minus the (small)
contribution from faint stars, which was computed from Galaxy
modeling, is the CIB flux.

\begin{deluxetable}{c c c c c c c c c c c }
\tabletypesize{\scriptsize} \rotate
\tablecaption{SUMMARY OF DIRBE MEASUREMENTS (\nwm2sr )
\label{dirbetable} } \startdata
 Ref & 1.25 \um & 2.2 \um & 3.5 \um & 4.9 \um & 12 \um & 25 \um & 60 \um & 100 \um & 140 \um & 240 \um \\
\hline \cutinhead{\hspace{-2cm}$\nu I_\nu$}
 $^1$ & & & & & & & & & $32\pm 13$ & $17 \pm 4$ \\

 $^2$ & $< 75$ & $<39$& $<23$ & $<41$ & $< 468$ & $<504$ & $<75$ & $<34$ & $25.0\pm 6.9$ & $13.6 \pm 2.5$ \\

$^3$ & 26.9+2.3$F_{2.2 \mu{\rm m}}$& -- & 9.9+0.3$F_{2.2 \mu{\rm m}}$ & 23.3+0.1$F_{2.2 \mu{\rm m}}$ & & & & & \\

 & $\pm$20.9&  & $\pm$2.9& $\pm$6.4 & & & & & \\

$^{4,\$}$ & & & & & & & $28.1$ & $24.6$ & & \\

 & & & & & & & $\pm 1.8 \pm 7$ & $\pm 2.5 \pm 8$ & & \\

$^{5}$ & & $22.4 \pm 6$ & $11.0 \pm 0.3$ & & & & & & \\

$^{6}$ & & $23.1 \pm 5.9$ & $12.4 \pm 3.2$ & & & & & & \\

$^{7}$ & & $20.2 \pm 6.3$ & & & & & & & \\

$^8$ & $54.0\pm16.8$ & $27.8\pm 6.7$ & & & & & & & \\

\hline \cutinhead{\hspace{-2cm}$\sqrt{C(0)}$ }
 $^{9,10,*}$ & $15.5^{+3.7}_{-7.0}$ & $5.9^{+1.6}_{-3.7}$ & $2.4^{+0.5}_{-0.9}$ & $2.0^{+0.25}_{-0.5}$ & $\lsim 1.$ & $\lsim 0.5$ & $\lsim 0.7$ & $\lsim 1$ & -- & -- \\
\enddata
\tablecomments{: Refs: $^1$ Schlegel, Finkbeiner \& Davis (1998),
$^2$ Hauser et al (1998), $^3$ Dwek \& Arendt (1998) --
$F_{2.2\mu{\rm m}}$ is the flux at 2.2 \um in \nwm2sr , $^4$
Finkbeiner, Davis \& Schlegel (2000) -- $^\$ $ random and
systematic errors shown, $^{5,6,7}$ Gorjian, Wright \& Chary
(2000), Wright \& Reese (2000), Wright (2000), $^8$ Cambresy et al
(2001), $^{9,10}$ Kashlinsky, Mather \& Odenwald (1996),
Kashlinsky \& Odenwald (2000) -- $^*$ 92\% confidence level
uncertainties are shown.}
\end{deluxetable}

Table \ref{dirbetable} summarizes the results of the CIB
measurements based on the DIRBE data.

\subsection{COBE FIRAS}

Far-Infrared Absolute Spectrometer (FIRAS) on board COBE was
designed for measuring the energy spectrum of the CMB and the
far-IR CIB. It is a four-port Michelson interferometer. At each
port (Left and Right) a dichroic filter split the beam into Low
(30-660 GHz) and High (600-2880 GHz) frequency beams producing
spectra with a resolution of 4.2 GHz or 16.9 GHz. The FIRAS data
have angular resolution of 7 degrees, with pixel size of 2.6
degrees. FIRAS is described in Boggess et al (1992) and Mather et
al (1993). The FIRAS Pass 4 data consist of spectra between 104
\um\ and 5000 \um\ in each of 6,063 out of the 6,144 pixels of the
sky (81 pixels that have no data are omitted). They were
calibrated using the method described in Fixsen et al. (1994,
1996).

The Pass 4 FIRAS data have approximately half of the noise of the
previous data releases (Pass 3). This is partly due to combining
all FIRAS frequency bins and scan modes and partly due to improved
understanding of the systematic errors. Particularly important was
the reduction in the systematic errors since these have a
complicated spectrum in the spatial domain. Since the fluctuations
of the far-IR CIB may themselves look like noise with some a
priori uncertain spatial power spectrum, finding a noise floor
leaves one uncertain whether it is the fluctuations in the far-IR
CIB or systematic effects in the instrument. The Pass 4 FIRAS data
have many of the known systematics removed, which enables a direct
and substantial lowering of the noise spectrum.

Puget et al (1996) made a first cut at detecting the far-IR CIB
from the Pass 3 FIRAS data. They used the particular spatial and
spectral distribution of the different far-IR components of the
sky emission in subtracting them from the Pass 3 FIRAS maps.
First, DIRBE maps were degraded to the FIRAS resolution in order
to map and subtract the zodiacal emission. In the DIRBE 25 \um
channel the emission is dominated by the zodiacal light and they
noted the average brightness ratio $B_\nu(100\mu{\rm
m})/B_\nu(25\mu{\rm m})=0.167$ and used the FIRAS observations of
the spectral dependence of the zodiacal light, $B_{\nu,{\rm
zodi}}\propto \nu^3$ (Reach et al 1995), to model the zodiacal
light emission along each line-of-sight as $0.167 (\lambda/100
\mu{\rm m})^{-3}B_\nu(100\mu{\rm m})$. The CMB along with its
dipole anisotropy was then subtracted from the raw FIRAS spectrum
at each pixel. The Leiden-Dwingeloo HI survey (Hartmann \& Burton
1997) degraded to the FIRAS resolution was used to subtract the
Galactic interstellar dust emission which at far-IR correlates
well with the neutral hydrogen emission (Boulanger et al 1996). At
high Galactic latitude ($|b_{\rm Gal}|
> 20^\circ$) and low HI column density lines-of-sight the residual
was positive, approximately isotropic, and homogeneous with
amplitude of $\simeq 3.4 (\lambda/400 \mu{\rm m})^{-3}$ \nwm2sr in
the 400-1000 \um range.

Fixsen et al (1998) used Pass 4 FIRAS data with significantly
lower noise and systematic errors. From the FIRAS data they
subtracted the DIRBE zodiacal model (Kelsall et al 1998)
extrapolated to FIRAS bands and the low-frequency FIRAS CMB model
with dipole anisotropy. In order to separate the Galactic emission
they used three independent methods: 1) They noted that after
subtracting CMB and zodiacal model, the residual emission has
approximately identical spectral energy distribution across the
sky. This spectral template was to model emission of the Galaxy
for further subtraction. Using all the FIRAS channels and assuming
a non-negative prior for the CIB, they find a statistically
significant residual. 2) HI (21 cm) and CII (158\um) Galactic
lines were used together with a quadratic fit to the Galactic HI
map degraded to FIRAS resolution from the AT\&T  survey by Stark
et al (1992) to subtract Galactic IR emission. After the
subtraction, the darkest 1/3 of the sky, where the HI data was
sufficiently accurate, showed a statistically significant CIB
residual at longer wavelengths. 3) Finally, they constructed
Galactic templates from the DIRBE 140 \um and 240 \um maps with
zodiacal emission subtracted together with the levels of the CIB
detected by DIRBE at these bands. The template was degraded to
FIRAS resolution. The CIB was obtained by extrapolating the the
correlations with the DIRBE templates to the DIRBE measurements at
140 \um and 240 \um . All three methods gave consistent results.

An attempt to study far-IR CIB fluctuations has been undertaken by
Burigana \& Popa (1998) on the FIRAS Pass 3 data between 200 and
1,000 \um. From the data they subtracted CMB monopole and dipole
components, a two-temperature dust model and a zodiacal light
model using the DIRBE 25 \um\ channel and the $\nu^{-3}$ frequency
scaling. The analysis resulted in upper limits on the CIB
fluctuations on scales greater than the FIRAS beam ($\simeq
7^\circ$).

\subsection{IRTS}

Japan's InfraRed Telescope in Space (IRTS) was launched in March
1995. The Near InfraRed Spectrometer (NIRS) on board IRTS was
designed to obtain spectrum of the diffuse background emissions
(Noda et al 1994). It covers the wavelengths from 1.4 \um to 4 \um
in 24 independent bands with spectral resolution of 0.13 \um .
IRTS observations covered ca. 7\% of the sky in 30 days (Murakami
et al 1996). The resolution of IRTS is $\simeq 8^\prime$,
significantly better than DIRBE and allows better removal of faint
stars.

The results from the IRTS CIB analysis have been presented in
Matsumoto (2000, 2003). A narrow strip at high Galactic latitude
was chosen for the CIB analysis in order to reduce the
contribution from Galactic stars. Contribution of Galactic stars
not resolved by the IRTS beam was estimated from the point source
sky model of the Galaxy by Cohen (1997). They used the DIRBE
physical model of the interplanetary dust (Kelsall et al 1998) to
remove the zodiacal component corresponding to the NIRS bands and
the observed sky directions. There was a good correlation between
the star-subtracted data and the modeled zodiacal component. The
isotropic component was determined as the intercept of that
correlation at zero modeled zodiacal brightness.  The brightness
of the isotropic emission found in this analysis was about 20 \%
of the sky brightness with an in-band energy of $\sim 30$ \nwm2sr
and was estimated to be a factor of $\simeq 2.5$ greater than the
integrated light of faint stars.

In order to determine the spatial spectrum of the fluctuations,
Matsumoto et al (2000, 2003) have added the NIRS short wavelength
bands (with smaller noise) from 1.43 \um to 2.14 \um . They then
calculated the correlation function of the emission, $C(\theta)$,
which was converted into the power spectrum by Fourier
transformation. The RMS fluctuations of the isotropic component
was, after subtracting read-out noise and stellar fluctuations,
consistent with the Kashlinsky \& Odenwald (2000) results from the
DIRBE sky analysis. The color of fluctuations was essentially the
same as that of the isotropic component, consistent with both
having the same origin.

\subsection{ISO}

Infrared Space Observatory (ISO) was operated by the European
Space Agency between November 1995 and April 1998. Its infrared
camera (ISOCAM) had a $32\times 32$ pixel array and two IR
channels with multiple filters: short wavelength channel operated
between 2.5 and 5.2 \um and the long wavelength operated between 4
and 17 \um. The other instrument relevant to CIB studies was
ISOPHOT which covered the wavelength range from 2.5 to 200 \um but
had much smaller arrays. The ISO telescope had no shutter making
absolute calibration difficult for studies of the isotropic
component of the CIB.

There were several important findings concerning the mid- to
far-IR CIB from the ISO surveys.

IR number counts indicate that the IR-luminous galaxies evolved
more rapidly than their optical counterparts and make a
substantial contribution to the star formation at higher $z$
(Elbaz et al 1999) which in turn makes them the dominant
contributors to the CIB at mid-IR bands. Elbaz et al (2002)
computed the contribution from distant galaxies observed by deep
ISOCAM extragalactic surveys at the central wavelengths of 6.75
and 15 \um to the CIB. The 6.75 \um sample was strongly
contaminated by Galactic stars, whereas at 15\um the stars are
readily separated from galaxies using optical-MIR color-color
plots. They estimate that in this way they accounted for the CIB
from dust in the starburst galaxies out to $z\sim 1.5$ and ISO
measurements at 15 and 170 \um reveal that the bulk of CIB at
these wavelengths comes from galaxies at $z\lsim 1.2$ (Dole et al
2001, Elbaz et al 2002). They find that galaxies above their
completeness level of 50 $\mu$Jy produce at least $2.4\pm 0.5$
\nwm2sr contribution to the total CIB at 15 \um. (For comparison,
the IRAS galaxies at 60 \um with completeness limit of 0.5 mJy
contribute only $\sim0.15$ \nwm2sr to the CIB at 60 \um.)

Lagache \& Puget (2000) have analyzed a sub-field from the deep
extragalactic FIRBACK survey with the ISOPHOT instrument at 170
\um. The survey covered 4 deg$^2$ and they selected for analysis a
subfield of low cirrus emission covering 0.25 deg$^2$. 24
extragalactic sources were detected and subtracted up to the
confusion limit of $3\sigma = 67$ mJy. The pixel size of the maps
was $\theta_p=89.4^{\prime\prime}$ implying the Nyquist frequency
of $k_{\rm Nyquist} = (2 \theta_p)^{-1}$ = 0.3 arcmin$^{-1}$. On
scales corresponding to spatial frequencies below the Nyquist
limit $k < 0.3 $ arcmin$^{-1}$ the results lead to upper limits on
the CIB fluctuations at 170 \um that correspond to about 0.5
\nwm2sr at 0.5 arcmin.

\subsection{2MASS}

The 2-Micron All Sky Survey (2MASS) uses two 1.3-m Cassegrain
telescopes, one at Mt. Hopkins in the Northern Hemisphere, and one
at CTIO in the Southern Hemisphere. Each telescope is equipped
with a three-channel camera, capable of observing the sky
simultaneously at J (1.25 \um ), H (1.65 \um ), and K$_s$ (2.17
\um ) at a scale of $2\dasec$ per pixel. As the telescopes are
scanned in declination, individual 1.3-second sky frames are
imaged on an overlapping grid by stepping 1/6 the array. The
frames are combined six at a time, to form the standard 2MASS
Atlas images of size 512 $\times$ 1024 pixels with re-sampled
1\dasec  pixels, and an effective integration time of 7.8 seconds
per pixel. Hereafter, we will use the term `image' to refer to the
calibrated 2MASS Atlas images which have been co-added to an
effective integration time of 7.8 seconds. The 2MASS photometric
stability is $<$0.02 mag in all bands. A detailed description of
the calibration process can be found in the 2MASS Explanatory
Supplement (Cutri et al 2003).

The standard 2MASS images are too shallow to be useful for CIB
studies. But a limited number of standard stars had to be observed
repeatedly each night, and for several months at a time, to
establish the photometric zero-points  for the data. The 2MASS
standard stars were drawn from near infrared standard star
catalogs (e.g. Persson et al. 1998, Casali and Hawarden 1992).
Each calibration observation consisted of six independent scans of
a calibration field. Each scan is a mosaic of 48 images, and the
scans were made in alternating north-south directions, each
displaced 5 \dasec  in RA from the previous one to minimize
systematic pixel effects. Repeated observations of calibration
fields during a night at a variety of elevation angles were also
used to develop long-term atmospheric extinction statistics. By
collecting the calibration scan data, which spanned nearly 6
months of repeated observation, effective integration times
exceeding one hour could be achieved for a small number of sky
locations.

Kashlinsky et al (2002) and Odenwald et al (2003) used the long
exposure 2MASS data for CIB analysis and reported the first
detection of small angular scale fluctuations in the near-IR CIB.
They have selected data from the 2MASS Second  Incremental Data
Release between 1998 March 19  and 1999 February 20, and used the
2MASS, on-line catalog of completed calibration fields in order to
identify fields that had the  largest number of repeated
measurements. One of these, 90565N, with a total exposure $\gsim$1
hour was selected for further study. The field has Galactic and
Ecliptic coordinates (243$^\circ , 27^\circ$) and ($21^\circ,
35^\circ$) respectively, around the star P565-C in the
constellation Hercules, 5$^\circ$ north of the bright star
$\lambda$ Ophiuchus. The data analyzed by them consisted of 2080
calibrated 8.6$^\prime\times$15$^\prime$ frames covering an
$8.6^\prime \times 1^\circ$ swath oriented north-south, obtained
during observations at CTIO between Apr and Aug 1998. The images
from individual exposures were co-added to produce a $8.6^\prime
\times 1^\circ$ field from which resolved sources (both stars and
distant galaxies) and other artifacts were removed by various
methods to ensure consistency and robustness of the final results.
After assembling the final image with a total exposure $>$3,700
secs the data were calibrated and the field was divided into seven
square patches $512^{\prime\prime}$ on the side. In each patch,
individual sources (stars and galaxies) were removed by an
iterative procedure where each pixel with surface brightness
exceeding 3 standard deviations for the patch was excised along
with 8 neighboring pixels. The 2MASS beam of $\sim
2^{\prime\prime}$ subtends comoving scale $\gsim$30$h^{-1}$Kpc at
$z$=1 corresponding to a large galaxy so the sources near the the
clipping threshold are unresolved. Because of the variation in
background level and associated noise, the patches were clipped to
different point source absolute flux levels. In K$_s$ band these
varied from 18.5 to 19 Vega magnitudes. The clipping left $>$90\%
of the pixels in the patches, which provided a good basis for a
Fourier analysis of the diffuse light. For the final images, the
fluctuation spectrum of the residual diffuse emission was
computed; its slope in the co-added images is consistent with that
of galaxy clustering. They estimated the contributions to the
power spectrum from atmospheric fluctuations, remaining Galactic
stars and cirrus emission, zodiacal light, instrument noise and
extinction. These components had different slopes and negligible
amplitudes compared to the detected signal, suggesting that the
diffuse component fluctuations in the final image are dominated by
the CIB. Because in this type of analysis fairly faint galaxies
($K\lsim 19$) have been removed, the resultant CIB provides a
probe of still fainter sources which are presumably located at
high redshifts, and allows to start isolating the contributions to
the CIB from various cosmic epochs.

\begin{figure}[h]
\centering \leavevmode \epsfxsize=0.9 \columnwidth
\epsfbox{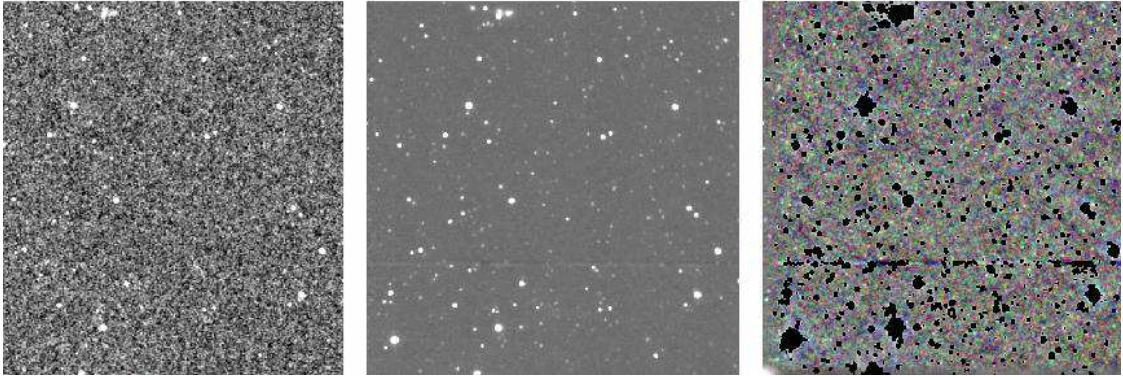} \caption[]{ \scriptsize{ {\bf Patch 3} of
$512^{\prime\prime}\times 512^{\prime\prime}$ in size from the
90565 field used in the deep 2MASS data analysis of Kashlinsky,
Odenwald et al (2002) and Odenwald, Kashlinsky et al (2003)}. {\bf
(Left)} Individual 2MASS images for 7.8-second integration in
K$_s$ band. {\bf (Middle)} Corresponding deep-integration co-added
K$_s$ band image for an effective integration time of 3,900
seconds. {\bf (Right)} 3 color image of the patch after co-adding,
clipping and de-striping in Fourier space. Red, green and blue
correspond to J, H, K$_s$ bands. The structure in the image is
identified with near-IR CIB from galaxies fainter than $K_s \simeq
19$. } \label{patch3}
\end{figure}

Fig. \ref{patch3} shows one of the patches in K$_s$ band after 7.8
sec integration (left panel - standard 2MASS product) and after
the integration used in the analysis (middle panel). The right
panel shows the final 3-color image (J, H and K$_s$) which was
identified with the map of the near-IR CIB from galaxies fainter
than $K_s \simeq +19$. An important step forward in this
measurement is the removal of progressively fainter and more
distant galaxies. This allows to begin to identify how much of the
CIB comes from earlier times. Observations and theory both suggest
that $K\gsim +19$ galaxies are typically located at $z\gsim 1$
(Cowie et al 1996, Kashlinsky et al 2002).

\subsection{Results}

\begin{figure}[h]
\centering \leavevmode \epsfxsize=0.7 \columnwidth
\epsfbox{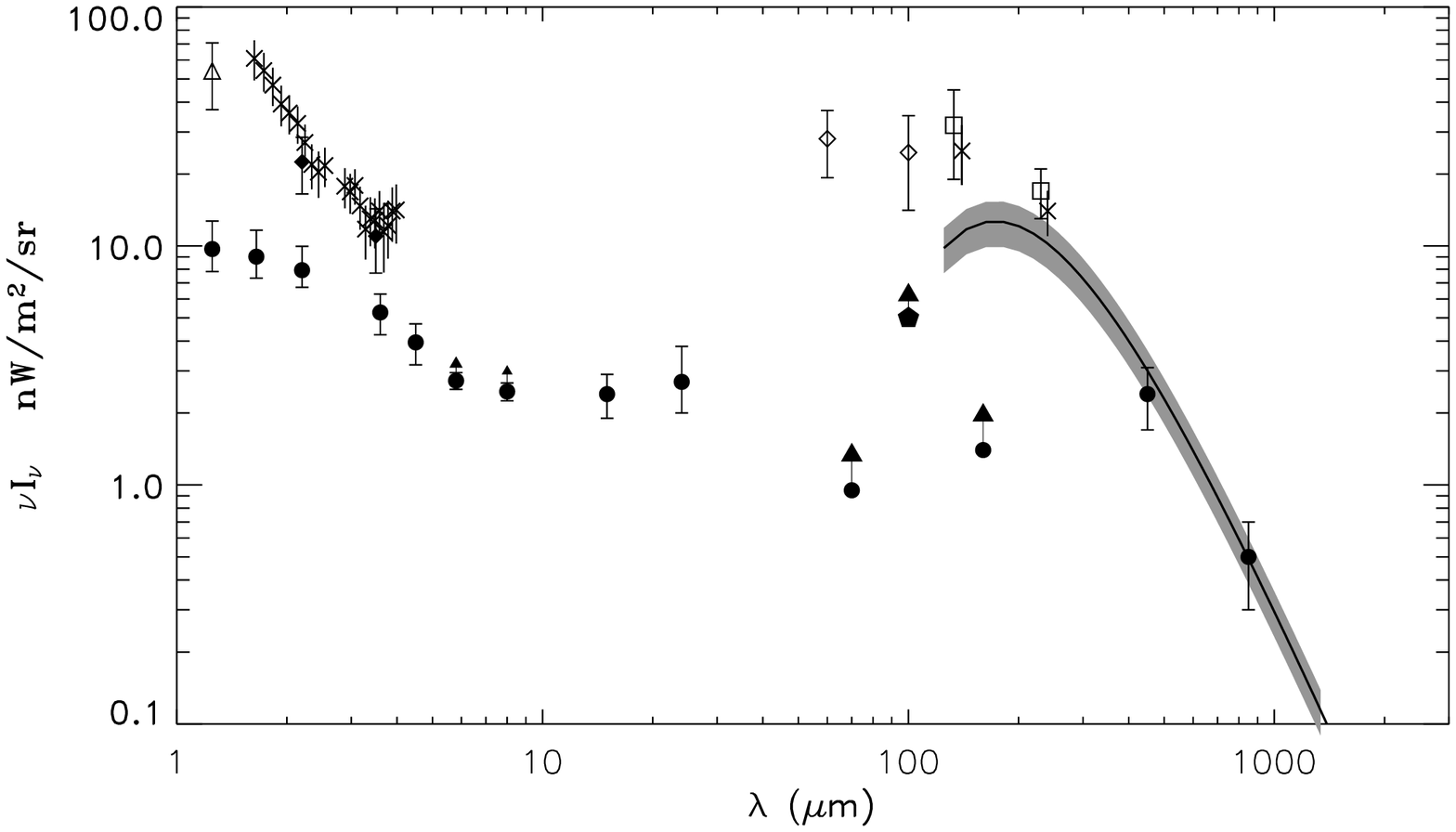} \vspace{0.5cm} \caption[]{ \scriptsize{ {\bf
NIR}: Mean levels of the near--IR CIB from the IRTS (crosses) from
\cite{irts} and open symbols from COBE/DIRBE data. Open triangle
shows the CIB results of the DIRBE J-band analysis (1.65 \um) from
\cite{cambresy} and filled diamonds at 2.2 \um\ from
\cite{gorjian}, at 3.5 \um\ from \cite{dwek}. {\bf FIR}: Thick
solid line with shade showing the uncertainty shows the results
from COBE FIRAS data analysis from Puget et al (1996) and Fixsen
et al (1998). DIRBE data detection at 140 and 240 \um\ is shown
with crosses from Hauser et al (1998) and open squares from
Schlegel et al (1998). Filled pentagon shows the lower limit on
the CIB from the DIRBE data at 100 \um\ from Dwek et al (1998).
Open diamonds correspond to results of DIRBE analysis by
Finkbeiner et al (2000). Different results at the same wavelength
are slightly shifted for clearer display. {\bf Cumulative fluxes
from {\it observed} deep galaxy counts} are shown with the filled
circles from Table \ref{galcounts}. The cases where the flux from
progressively fainter galaxies does not saturate or the saturation
is not as  clear are marked with an upward arrow. } }
\label{cib_dc}
\end{figure}

The results on the mean CIB levels from the measurements described
above are shown in Fig. \ref{cib_dc}. Only detections are shown;
upper limits, in the absence of detections, are discussed below
for each range of wavelengths. Hence the gap in the figure at the
MIR wavelengths where zodiacal foreground is brightest. Total
fluxes from the observed galaxy population are also shown for
comparison; they are discussed at length in Sec. 6.

\begin{figure}[h]
\centering \leavevmode \epsfxsize=0.5 \columnwidth
\epsfbox{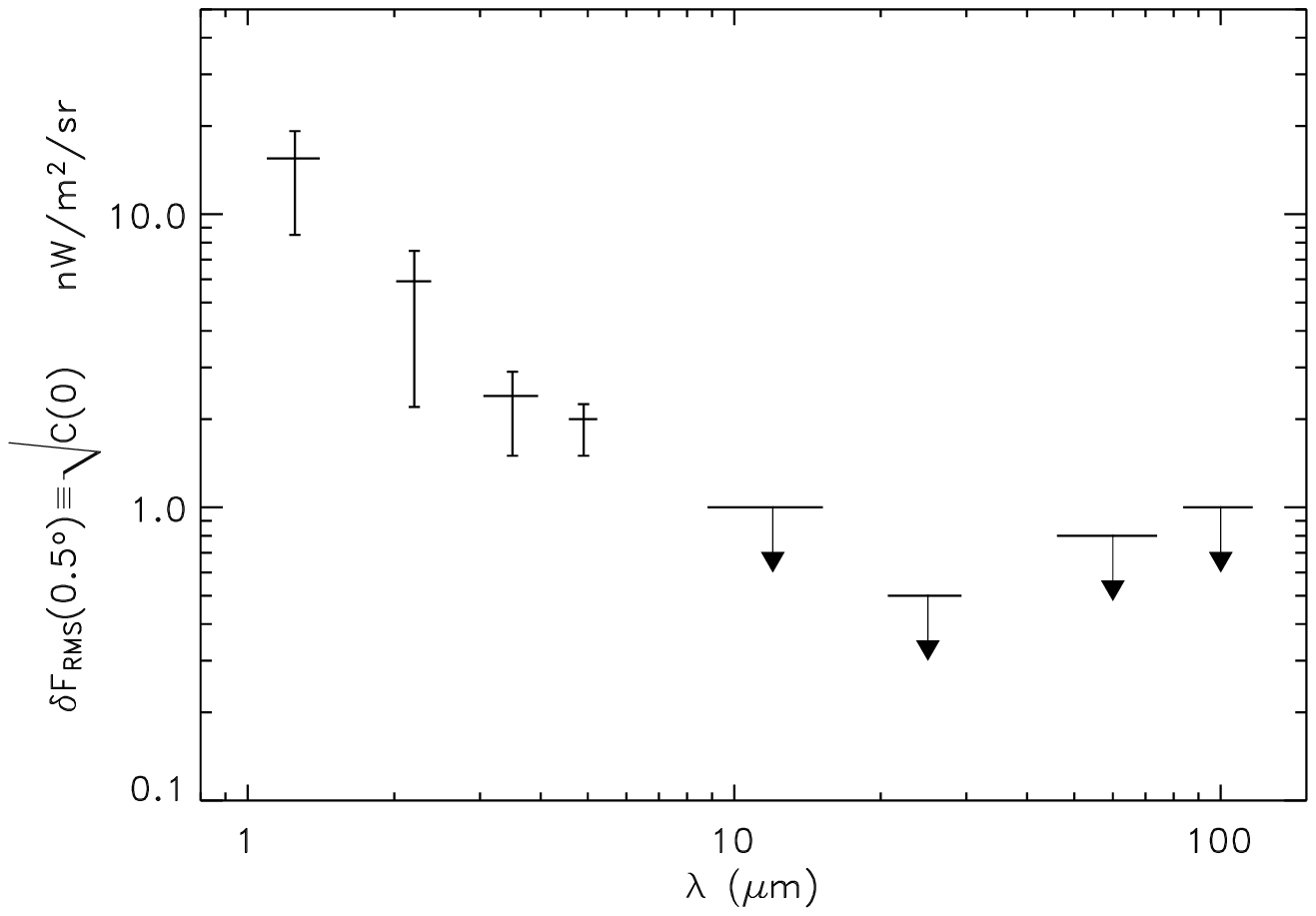} \caption[]{ \scriptsize{Detections of and upper
limits on the RMS CIB fluctuation  at 0.5$^\circ$ from the DIRBE
data analysis by Kashlinsky \& Odenwald (2000).} } \label{paper3}
\end{figure}

Displaying measurements of CIB fluctuations is more complicated
because of the extra spatial dependence and some of their
measurements will be discussed for each wavelength range
separately. Figure \ref{paper3} shows the amplitude of CIB
anisotropies at $\sim 0.5^\circ$ from the DIRBE data analysis of
Kashlinsky \& Odenwald (2000) and Fig. \ref{komsc} shows the
results for CIB fluctuations from the deep 2MASS data analysis
described in Sec. 5.5.

Below we itemize the status of CIB measurements in three IR
wavelength ranges: NIR, MIR and FIR.

\subsubsection{Near-IR}

In the near-IR there are detections of the mean isotropic part of
the CIB based on analysis of DIRBE and IRTS data (Dwek \& Arendt
2000; Matsumoto et al 2000, 2003, Gorjian et al 2000, Wright \&
Reese 2000, Cambresy et al 2001). The measurements agree with each
other, although the methods of analysis and foreground removal
differ substantially. They also agree with the measured amplitude
of CIB fluctuations. The results seem to indicate fluxes
significantly in excess of those from observed galaxy populations.
Total fluxes from galaxies are shown with filled circles in Fig.
\ref{cib_dc}; in near-IR they saturate and are a factor is $\sim
2-3$ below the detected CIB.

Detections of CIB fluctuations come from three independent
experiments and are consistent with each other. At 0.5$^\circ$
Kashlinsky \& Odenwald (2000) find a statistically significant CIB
anisotropy in the DIRBE first 4 channels. Matsumoto et al (2000,
2003) measured a spectrum of CIB anisotropies on degree scales
from IRTS data band-averaged to $\sim $ 2 \um; their results are
consistent with Kashlinsky \& Odenwald (2000). Because the beam
was large in both instruments, no galaxy removal was possible in
the data and the CIB anisotropies arise from {\it all} galaxies,
i.e. from $z=0$ to the earliest times. Based on the amplitude of
the present-day galaxy clustering, the NIR detections exceed the
expectations from galaxies evolving with no or little evolution by
a factor of $\sim 2-3$ and are consistent with the measured (high)
mean CIB levels in the NIR.

\begin{figure}[h]
\centering \leavevmode \epsfxsize=0.7 \columnwidth
\epsfbox{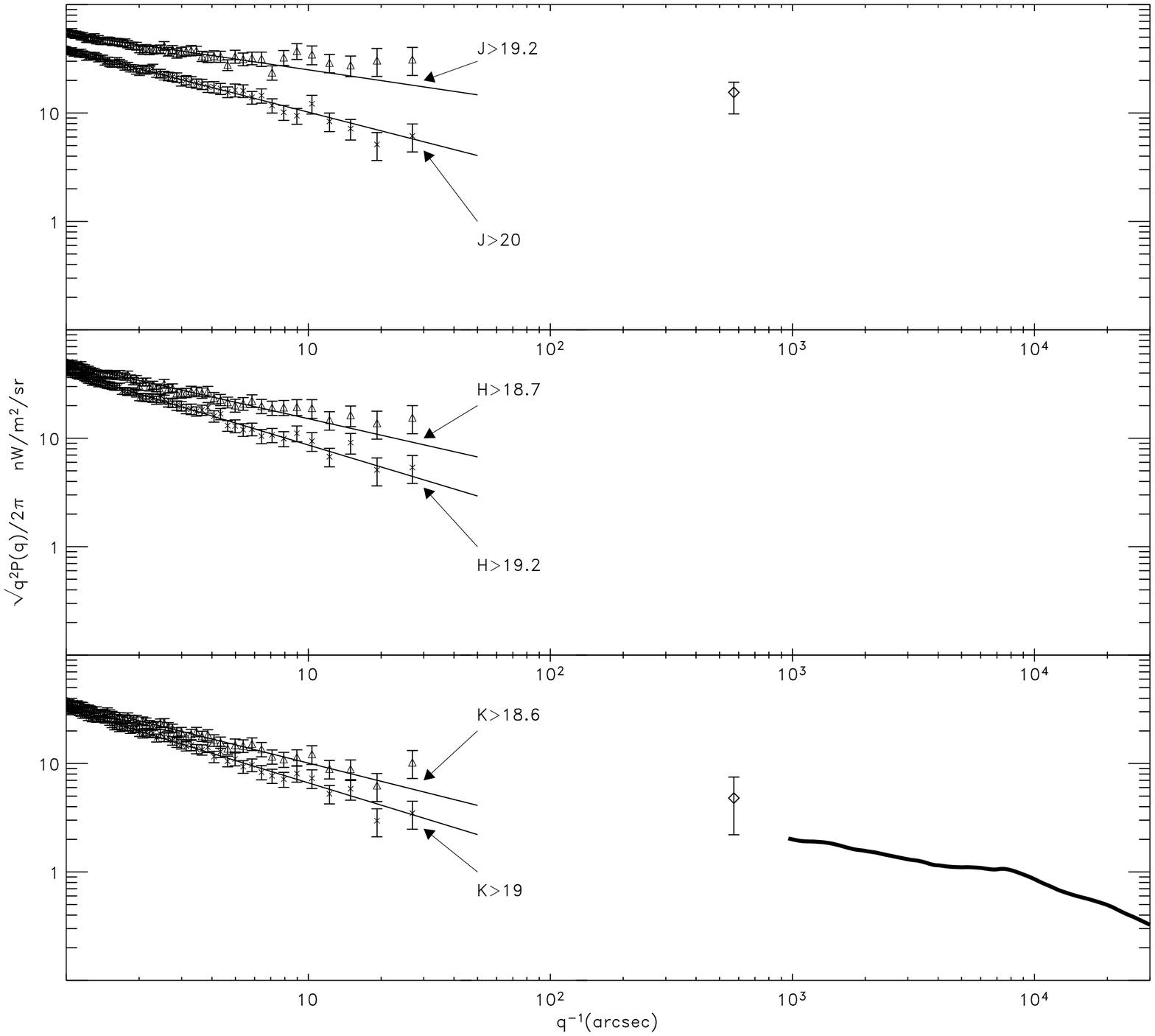} \caption[]{ \scriptsize{ The angular spectrum of
the CIB fluctuations from the 2MASS analysis by Kashlinsky,
Odenwald et al (2002) and Odenwald, Kashlinsky et al (2003) at J,
H and K bands. Triangles correspond to the results from the patch
from which galaxies out to $K_S \lsim 18.6$ have been removed and
crosses to another $512^{\prime\prime}\times 512^{\prime\prime}$
patches with galaxies removed out to $K_S \lsim 19$. Open diamonds
with 92\% errors show the fluctuation on larger angular scale from
Kashlinsky \& Odenwald (2000) analysis of COBE DIRBE data. Thick
solid line shows the CIB result from IRTS analysis (Matsumoto et
al 2000, 2003). Both the DIRBE and IRTS analyses included all
galaxies and are consistent with each other and the 2MASS data
when account is made of the contribution from the removed
(brighter) galaxies (Odenwald, Kashlinsky et al 2003). } }
\label{komsc}
\end{figure}

Figure \ref{komsc} summarizes the 2MASS-based results and compares
them with the other measurements. Analysis of deep 2MASS data
(Kashlinsky et al 2002, Odenwald et al 2003) has allowed to remove
galaxies out to $K\simeq (18.5-19)$ and measure the spectrum of
CIB anisotropies in J, H, and K 2MASS bands on sub-arcminute
angular scales. These galaxies are typically located at
cosmological times when the Universe was less than $\sim$ half its
present age and when extrapolated to the present-day the 2MASS
based results are consistent with the DIRBE- and IRTS-based
measurements. This indicates that the near-IR CIB excess, if real,
must originate at still earlier times.

All NIR CIB results are mutually consistent, exceed the simple
extrapolations from the observed galaxy populations with little
evolution and, if true, probably imply significant energy-release
activity at early cosmic times.

\subsubsection{Mid-IR}

Because the zodiacal light is so bright at these wavelengths there
is no direct detection of the CIB between $\sim 10$ and $\sim 50$
\um. There are low upper limits though. Some come from the
fluctuations analysis of the DIRBE data by Kashlinsky, Mather \&
Odenwald and use the fact that zodiacal light, while very bright,
is also very smooth. Whereas light emitted by galaxies must be
clustered with amplitude of about $\sim 10$\% in flux fluctuations
on the DIRBE beam scale. This leads to upper limits of about $\sim
10$ \nwm2sr at these wavelengths.

Other upper limits come from scattering of $\gamma$-ray and CIB
photons producing electron positron pairs if $E_{\gamma}
E_{\gamma_{\rm CIB}} \geq 2 (m_ec^2)^2$. For $\gamma$-rays the
cross-section is peaked in the IR wavelength range, $\lambda_{\rm
peak} \simeq 2.5 \frac{E_\gamma}{1 {\rm TeV}}$ \um (Stecker  et al
1992). If a drop in the spectrum of an extragalactic $\gamma$-ray
source in TeV range can be observed this should lead to limits on
the CIB in the corresponding wavelength range (Stecker \& de Jager
1993). Dwek \& Slavin (1994) applied this analysis to the
$\gamma$-ray spectrum of Mrk 421. The best {\it upper} limits come
from analysis of the 1997 $\gamma$-ray outburst of the blazar Mrk
501 and are $\simeq 5$ \nwm2sr from 5 to 15 \um (Stanev \&
Franceschini 1998, Renault et al 2001).

The ISO and Spitzer deep counts provide important {\it lower}
limits on the CIB at these wavelengths which are very close to the
upper limits from $\gamma$-ray and fluctuations analyses. It would
be thus very surprising if the MIR CIB is not inside this 2.5
\nwm2sr $\lsim \nu I_\nu \lsim 5$ \nwm2sr range.

\subsubsection{Far-IR}

Far-IR CIB measurements come from the COBE FIRAS (Puget et al
1996, Fixsen et al 1998) and COBE DIRBE (Schlegel et al 1998,
Hauser et al 1998, Finkbeiner et al 2000) data analysis. They are
broadly consistent although the COBE DIRBE detections at 140 and
240 \um give larger fluxes (at almost 2-$\sigma$ level) and the
Finkbeiner et al (2000) analysis gives an even bigger discrepancy
at 60 \um if the COBE FIRAS results are simply extrapolated to
that wavelength. Hauser et al (1998) discuss these differences and
point that the differences are probably due to different
instrumental calibrations.

Fixsen et al (1998) suggest a simple fit to the CIB at $\lambda
\gsim 100$ \um:
\begin{equation}
I_\nu = A \left(\frac{\nu}{\nu_0}\right)^\beta B_\nu (T)
\label{firas_cib}
\end{equation}
where $A=(1.3\pm 0.4)\times 10^{-5}, \beta=0.64 \pm 0.12,
T=18.5\pm 1.2$ K, $\nu_0=100 {\rm cm}^{-1}$ = 3000 GHz and $B_\nu$
is the Planck function. In eq. \ref{firas_cib} the uncertainties
are highly correlated with correlations $+0.98$ for $A$ and
$\beta$, $-0.99$ for $A,T$ and $-0.95$ for $\beta,T$ (Fixsen et al
1998).

The bulk (likely all) of the CIB at these wavelengths comes from
dust in high-$z$ and present-day galaxies and the SCUBA galaxy
counts at 850 \um (Smail et al 1997) compare favorably with eq.
\ref{firas_cib}.

\subsubsection{Bolometric CIB flux}

\begin{deluxetable}{c | c | c | c }
\tabletypesize{\scriptsize}
\tablecaption{Bolometric/integrated CIB flux (\nwm2sr )
\label{fluxbol} } \startdata NIR: 1\um $\lsim \lambda \lsim$ 4 \um
& MIR: 10\um $\lsim \lambda \lsim$ 50 \um & FIR: 120
\um $\lsim \lambda \lsim$ 1,000 \um &  Total: 1\um $\lsim \lambda \lsim$ 1 mm\\
\hline
37 & $\lsim$ 8 & 13 & 50 $\lsim F_{\rm total} \lsim$ 58 \\
\enddata
\end{deluxetable}

Table \ref{fluxbol} shows the integrated IR fluxes over the three
ranges of wavelengths where direct CIB measurements were possible
to meaningfully integrate (hence, e.g. the gap between 4 and 10
\um). In the NIR out of 37 \nwm2sr approximately 13 \nwm2sr is
contributed by the Cambresy et al (2001) measurement at the
DIRBE's J band; the remaining come from the IRTS data and other
measurements out to $\simeq 4$ \um. In order to evaluate the
contribution in the MIR region we assumed an upper limit between
$\sim 10$ and 50 \um of $\nu I_\nu \lsim$ 5\nwm2sr from the
$\gamma-$ray absorption. (An upper limit from fluctuations
analysis would be a little higher). In the far-IR we integrated
the approximation eq. \ref{firas_cib}; including the other DIRBE
detections will increase this number.

The table shows that CIB carries very substantial fluxes
testifying to significant energy release throughout the evolution
of the Universe. For comparison the total flux from the CMB is
only $\sim$ one order of magnitude higher. Table \ref{galcounts}
and discussion in Section 6 show that normal stellar populations
observed in deep galaxy counts surveys contribute about as much
CIB flux in the near-IR as the total far-IR CIB in Table
\ref{fluxbol}, meaning that approximately half the energy produced
by stars has been absorbed and re-emitted in the far-IR by dust in
(high-$z$) galaxies. The observed NIR galaxies  in J,H,K bands
contribute only $\simeq 5$\nwm2sr .  The total fluxes from the
deepest counts in Spitzer data are also shown although at 5.8 and
8 \um does not saturate as clearly as at 3.6 and 4.8 \um. If one
integrates the flux from
 galaxies in J, H, K bands and the Spitzer 3.6 \um one gets 8.6 \nwm2sr .
 Extending the range of integration to the other Spitzer IRAC channel at 4.5 \um
 where the total flux from galaxies saturates (see Sec. 6) gives 9.7 \nwm2sr
 and integrating over galaxies from 2MASS and all Spitzer IRAC channels gives total
 flux of $\simeq 11.4$ \nwm2sr between 1 and 10 \um. The observed
 galaxy populations contribute only a small fraction of the
 observed flux.

 This means a substantial excess of the total CIB measurements
 over that observed to come from total galaxies. It then appears
 likely that either 1) the total contributions from ordinary galaxies are
 significantly underestimated, or 2) the bulk of the CIB comes
 from still earlier epochs than probed by the systems accessible
 to current telescopic studies, e.g. from the Population III era, or 3)
 all the the NIR measurements contain systematic errors mistaken for
 CIB.
 The first of these possibilities is discussed in Sec. 6 and
 shown not very likely, the second is discussed in Sec. 7 and the
 third possibility should always be borne in mind when interpreting the very difficult
 measurements of the CIB.

 Historically the excess CIB flux in the near-IR has been nearly
 simultaneously and independently detected
 for the isotropic part of the CIB by Dwek \& Arendt (1998) and
 for fluctuations by Kashlinsky \& Odenwald (2000).
 \footnote{The paper by Kashlinsky \& Odenwald (2000) has been
 received by the Ap.J on April 30, 1998, but due to an an editorial
 mishandling the manuscript was not sent to an
independent referee for review until over a year later, in May
1999.}

\section{'Ordinary' contributors to CIB}

In this review we use the term ``ordinary" galaxy populations to
refer to galaxies made of the observed normal {\it metal-enriched}
stellar populations containing Population I and II stars and the
dust produced by them. Observations from the Hubble Deep (and
Ultra Deep) Field suggest that metal enrichment has happened
fairly early on in the history of the Universe. Similarly, data on
quasar emission and intrinsic absorption lines indicate that solar
metallicities in the surrounding medium have already been reached
at $z\gsim 4$ (Hammann \& Ferland 1999). This is also suggested by
the existence of the oldest Population II stars with non-zero
metal abundances. Star formation processes in the metal enriched
medium (metallicity $Z\gsim 10^{-3}-10^{-2} Z_\odot$) should lead
to formation of populations with stellar properties not radically
different from those observed today. These populations must have
been preceded by completely metal-free (first) stars, the
so-called Population III that formed at very high redshifts,
$z\gsim 10$. Population III stars likely form a completely
different class of stellar objects, probably have a very different
range of masses and also form in a different environment. Their
case is special, necessarily more speculative, and their possible
contribution to the CIB is discussed in the following section.

Before we proceed a simple ``back-of-the-envelope" evaluation is
in order for the amount of the near-IR CIB produced by 'ordinary'
stellar populations. If such stars had Salpeter-type IMF most of
the light would be produced by the high mass end of the stellar
mass spectrum, i.e. by stars that produced the metal abundances
($Z\sim Z_\odot$) observed today. If galaxies whose stellar
contents contribute $\Omega_* \simeq 3 \times 10^{-3}h^{-1}$ to
the critical density ($\rho_{\rm cr}$), then during their main
sequence the stars would have produced energy density of $\epsilon
\Omega_* \rho_{\rm cr}c^2$, where $\epsilon $ is the radiative
efficiency. For hydrogen burning $\epsilon =0.007$ and this
efficiency is reached for massive stars, $M\gsim 10 M_\odot$. For
present-day stellar populations with a Salpeter IMF and a lower
cutoff of 0.1$M\odot$, the effective efficiency is only $\epsilon
\simeq 0.001$ (Franceschini et al 2001) \footnote{Solar mass stars
burn a core of about $\sim 10-15$ \% of their hyrdogen mass during
main-sequence stage. During their red giant stage the hydrogen
burning core is $\simeq 0.5M_\odot$ for $M\lsim 2.5 M_\odot$
(Sweigart, Greggio \& Renzini 1990). For the IMFs where the bulk
of stellar mass is locked in the low end of the stellar
mass-spectrum this results in effective $\epsilon$ considerably
smaller than the canonical value of 0.007 (Sweigart, private
communication)}. A fraction $f(\lambda \geq \frac{1}{1+z}\mu{\rm
m})$ of the emitted energy will contribute to the near-IR CIB at
wavelengths longer than 1\um, assuming it was not absorbed or
re-processed. This would lead to the CIB flux of:
\begin{equation}
\nu I_\nu \sim \frac{c}{4\pi} \epsilon Z \Omega_{*} \rho_{\rm
cr}c^2 \langle\frac{f(\lambda \geq \frac{1\mu{\rm
m}}{1+z})}{1+z}\rangle = 8 h
\;\frac{Z}{10^{-2}}\; \frac{\epsilon}{0.007}\;\frac{\Omega_*}{3\times
10^{-3}h^{-1}}\;\langle\frac{f(\lambda \geq \frac{1\mu{\rm
m}}{1+z})}{1+z}\rangle \frac{\rm nW}{\rm m^2 sr} \label{cib_pop2}
\end{equation}
The stars responsible for producing metals are predominantly hot
stars with surface temperature $T_*\gsim (1-2)\times 10^4$ K. In
this case $f(\lambda \geq \frac{1}{1+z}\mu{\rm m})$ the CIB flux
comes from the Rayleigh-Jeans part of the spectrum, where $f
\propto T_*^{-3} (1+z)^{3}$ and the redshift-dependent term on the
right-hand-side of eq. \ref{cib_pop2} increases toward high $z$
because a larger part of the emitted spectrum is shifted into the
IR today. With $T_* \simeq 10^4$ K and $z\sim 2-3$ one gets
contribution to the CIB around a few \nwm2sr , but if there were
very hot (massive) stars at very early times, their contribution
may well exceed that of the 'normal' stellar populations.

Silk (1996) makes a similar argument to eq. \ref{cib_pop2}
differentiating between the metals (iron) produced in early and
late type galaxies and their respective mass-to-light ratios, and
Hauser \& Dwek (2001) normalize instead to helium production; both
get similar numbers. Helium is produced by less massive (and
longer living) stars, and they could produce an amount of CIB
comparable to (or less) than eq. \ref{cib_pop2}.

Comparison with the near-IR levels of CIB implies that, if the
measurements are true, it is difficult to explain these high
levels by left-over emissions from galaxy populations with IMF
similar to that of today. The levels of the near-IR CIB can be
increased in one (or all) of the following ways over eq.
\ref{cib_pop2}: 1) If there exists a population of stars that can
emit significant amounts of light without producing metals (as
happens e.g. in stars more massive than $\simeq 240 M_\odot$,
Heger 2003), one can have extensive period of energy production
that would not violate the observed metallicities. 2) If emission
comes from populations that contribute a larger fraction of
$\Omega_{\rm baryon} \simeq 0.044$ than the stellar populations of
today which have $\Omega_* \ll \Omega_{\rm baryon}$.
Interestingly, most of the baryons today are in dark form
(Fukugita, Hogan \& Peebles 1998), the dark baryons possibly being
in remnants of the Population III era. 3) And if there existed at
high $z$ a population of very hot stars, the bulk of their
emission will today be shifted into the NIR bands. We discuss such
possibility in the context of Population III stars contribution in
the following section.

The remainder of this section discusses the various aspects
necessary for estimating the contributions from ordinary
(metal-rich) stellar populations in  galaxies to the CIB from NIR
to FIR.

\subsection{IMF and star formation history}

SFR indicators include UV and emission-line luminosities
indicative of light from short-lived massive stars. Conversion to
SFR depends on the assumed IMF, but once the IMF assumption is
made, the relative SFR is less sensitive to IMF. The conversion of
the UV luminosities into the SFR must also correct for the
absorption of  UV photons by dust which is generally associated
with the same young stars.

In the first calculation of the stellar IMF, Salpeter (1953)
approximated the Solar neighborhood measurements with IMF having a
power law of slope $-1.35$ and it still is commonly used in
modeling synthetic stellar populations. Scalo (1986), from
measurements of the Solar neighborhood, gives an IMF with a mass
fraction peak at $\simeq 0.5-1 M_\odot$. While very useful and
necessary in constructing synthetic stellar population models, it
is not clear whether the assumption of the {\it universal} IMF is
valid in the real Universe. E.g. it cannot reproduce H$\alpha$
luminosities (Kennicutt et al 1994) or match mean galaxy colors
(Madau et al 1998) because it has too few stars of $M\gsim 10
M_\odot$.

In galaxies with on-going star formation, the composite UV
spectrum is approximately flat. If this is combined with the
Salpeter IMF between 0.1 and 100 $M_\odot$ in stellar mass, one
can relate the rest frame UV luminosity to SFR (Madau et al 1998)
for galaxies with continuous star formation on timescales longer
than $\gsim 0.1$ Gyr:
\begin{equation}
{\rm SFR}(M_\odot/yr) = 1.4\times 10^{-28} L_\nu ({\rm
erg/sec/Hz})
\label{sfrvsuv}
\end{equation}
Because in starburst galaxies a significant fraction of
luminosity, which is dominated by young stars, is absorbed and
re-emitted in the infrared by dust, the IR luminosity can also be
used as tracer of SFR. Assuming that the bolometric luminosity is
dominated by young stars, Kennicutt (1998) obtained in the
optically thick limit (when IR luminosity measures the total
bolometric luminosity) for continuous star bursts lasting at least
10 Myr, solar metal abundance and a Salpeter IMF:
\begin{equation}
{\rm SFR}(M_\odot/yr) = 1.71 \times 10^{-10} L_{\rm IR}(8-1000
\mu{\rm m}) (L_{\odot, {\rm bol}})
\label{sfrvsfir}
\end{equation}
so luminous IR galaxies with $L_{\rm IR}> 10^{11}L_\odot$ form
more than 20 $M_\odot$/yr in stars. Note that these calibrations
depend on the IMF in star forming galaxies, since the integrated
UV spectrum is dominated by massive stars and is assumed to lead
to dust (re)emission in the mid- to far-IR.

Lilly et al (1996) used galaxies from the Canada-France Redshift
Survey (CFRS) to construct the comoving luminosity,  ${\cal
L}_\nu(z)$, of the Universe out to $z\simeq 1$ at rest-frame
wavelengths of 0.28, 0.44 and 1 \um. The sample included 730
I-band selected galaxies, with luminosities extrapolated to the
above rest wavelengths using interpolations from the $BVIK$
photometry with SEDs interpolated from synthetic spectra designed
to match the observed colors. The present-day I-band corresponds
to the blue-band for rest frame emission at $z\sim 1$.  They found
observational evidence for an increase of ${\cal L}_\nu(z)$ out to
$z\simeq 1$ in the galaxy rest-frame UV to near-IR (1 \um ) bands
and argued that the data requires SFR $\propto (\Delta t)^{-2.5}$
where $\Delta t$ is the time elapsed since the initial star
formation.

With the Lyman-dropout technique (Steidel et al 1996), galaxies
are now found to progressively higher redshifts (Giavalisco 2002).
Flux amplification by gravitational lensing by known clusters of
galaxies can lead to galaxy detections at even higher redshifts as
amplifications of order $\sim 10-100$ can be achieved for distant
galaxies lying on caustics of the cluster potential.
 Kneib et al (2004) measured optical and IR photometry of
 one such candidate galaxy at $z\sim 7$ and find that, while only
 $\lsim 1$ Kpc in extent, it undergoes star formation at the rate of
 $\simeq 2.6M_\odot$/yr. A claim was made recently of a galaxy at $z\simeq 10$
among the objects lensed by a known
 cluster A1835 (Pello et al 2004, but see Bremer et al 2004).
 Bouwens et al (2004) have applied the $z_{850}$ dropout search
to the Hubble Ultra Deep Field (UDF) data and detected four (and
possibly five) potential $z\sim 7-8$ objects that are seen in J
and H bands, but have no detection shortward of 8,500 \AA (one of
the candidates has a weak signal at 8,500 \AA). The objects are
strongly clustered covering a total area of 1 arcmin$^2$ and also
appear very blue in their rest frame. Using VLA observations of
the molecular gas in the underlying galaxy of the current
record-holder $z=6.4$ quasar, Walter et al (2004) detect $\sim
5\times 10^9 M_\odot$ in molecular gas extended to $\sim 2.5$ Kpc.
The amount of gas is comparable to what is found in the nearby
ULIRGs and its brightness temperature is about that of the nearby
starburst centers. Clearly stars must have formed early on in the
Universe's history and at vigorous rates.

Madau et al (1996) have analyzed the Hubble Deep Field (HDF)
observations reaching AB magnitudes $\gsim 28$ in four bands from
3000 \AA\ to 8140 \AA. They modeled the HI opacity and used
stellar population synthesis models to separate low- and high-$z$
galaxies using the Lyman continuum drop-out method (Steidel et al
1996). From the images they constructed a sample of star-forming
galaxies from which the comoving UV luminosity density was
estimated out to $z\sim 4$. Assuming that the UV luminosities give
a complete view of star formation, the SFR density can be
evaluated from the conversion factors for the Salpeter IMF (Madau
et al 1998). The derived SFR, known as the Madau curve, was found
to rise rapidly toward early times possibly peaking at $z \sim
2$. Estimating the comoving UV luminosity density from the high
redshift galaxies found in the UDF data (Bouwens et al 2004)
suggests a factor of $\gsim 3.5$ drop in SFR from $z=3.8$ to
$z\sim 7.5$ consistent with what is found at $z\sim 6$ (Dickinson
et al 2004, Stanway et al 2004).

 Guiderdoni et al (1997) argued that most galaxies are
hidden by dust which leads to substantial underestimating of the
high redshift SFR deduced from optical surveys. They constructed
models of several galaxy populations to separately match the
optical surveys' data and the far-IR CIB deduced from the
COBE/FIRAS data. The galaxy population responsible for the FIR CIB
has a high SFR at early times, information on which will be
missing from SFR deduced from optical bands. Indeed, if most of
energy is (re)emitted at IR wavelengths, the SFR may be
underestimated when deduced from surveys in the UV bands.

Rowan-Robinson et al (1997) analyzed ISOCAM observations of the
HDF at 6.7 and 15 \um (Serjeant et al 1997), where thirteen HDF
galaxies have been detected by ISO in the 12.5 hour exposure. In
eleven of these, there was a substantial mid-IR emission excess,
which they interpreted as dust emission from a strong star-burst.
The SEDs were modeled from 0.3 to 15 \um using the available
photometry from visible to NIR bands and the Bruzual-Charlot
stellar evolution models with a Rayleigh-Jeans fall-off beyond 2.5
\um, and the conversion to SFR was done using starburst modeling
with emission from dust and PAHs. If these star-forming galaxies
are typical of the fainter HDF galaxies, then Rowan-Robinson et al
argued that the true SFR remains flat at $z\gsim 1.5$ instead of
falling off.

\subsection{Normal stellar populations}

Normal stellar populations are defined as those with Population I
and II metallicities and the (locally) observed IMF, usually taken
to be either of Salpeter or Scalo type.

Constructing realistic SEDs of galaxies is important for proper
comparison of contributions of galaxies of various types, from
various redshifts and at various wavelengths to the observed CIB.
In the near-IR such SEDs can be approximated with the spectrum and
its evolution derived from synthetic stellar population models.
This approach requires input of a universal IMF, assumptions about
the rate of star formation, self-consistent treatment of stellar
evolution, metal enrichment etc. The early studies have been
pioneered by Tinsley (1976) with Bruzual (1983) developing the
state-of-the art models that can be compared with detailed
observations. The current versions of the synthetic stellar models
e.g. by Bruzual-Charlot (1993) or PEGASE (Fioc \& Rocca-Volmerange
1997) codes are available on the world-wide-web. The modeling has
to be done carefully and many priors are required, since the SEDs
in turn depend on many parameters: IMF, metallicities, galaxy
ages, dependence of each of them on other, galaxy luminosity, SFR,
etc. Still, such modeling gives a fair representation of reality
and can certainly provide approximate answers.

\begin{figure}[h]
\centering \leavevmode \epsfxsize=1. \columnwidth
\epsfbox{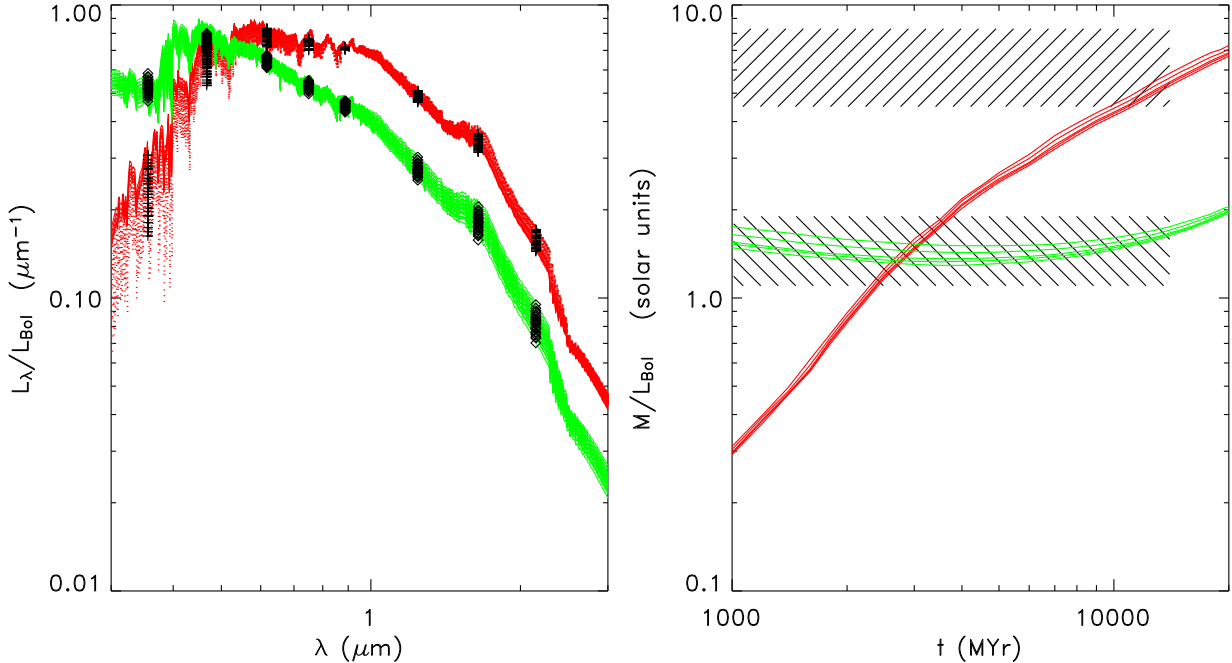} \caption[]{ \scriptsize{ Runs of synthetic
galaxy SED spectra using PEGASE. Green lines correspond to late
types, red lines to early types stellar populations. {\bf Left}:
black symbols correspond to the SEDs integrated over the SDSS and
2MASS filters. The sets of lines are shown for five metallicites
of $Z=0, 0.001,0.002,0.005,0.01$ and five ages between 10 and 14
Gyr. {\bf Right}: Mass-to-(bolometric) light ratios in solar units
for early and late stellar populations are plotted vs their age
for $Z=0, 0.001,0.002,0.005,0.01$. Cross-shaded areas correspond
to typical mass-to-light ratios observed locally at $z=0$ in the
early (top cross-shaded area) and late populations with
uncertainties from Fukugita, Hogan \& Peebles (1999); the observed
variations in the mass-to-light ratios are larger. } }
\label{sed_gal}
\end{figure}

For the purpose of calculations in this section we used the PEGASE
\footnote{http://www2.iap.fr/users/fioc/PEGASE.html} code to
construct SEDs of Early and Late type galaxy stellar populations
as follows: Early type galaxy stellar populations were assumed to
be all described by Salpeter IMF and form at some early time in a
single short burst of star formation lasting 0.5 Gyr. Late type
stellar populations in galaxies were assumed to form via an
on-going star formation with Scalo IMF starting at some early
epoch. The PEGASE models were run for a grid of galaxy ages and
metallicities. The left panel in Fig. \ref{sed_gal} shows the SED
for the two types of populations for ages between 10 and 14 Gyr
and metallicities
 $Z=0, 10^{-3}, 2\times 10^{-3}, 5\times 10^{-3}, 10^{-2}$. The
symbols show the value of the relative flux when averaged over
SDSS and 2MASS filters to compare with data discussed below. (Plus
signs correspond to early types and diamonds to late types.) The
NIR luminosities are dominated by red giants in the old stellar
populations and are directly related to the total stellar mass.

The right panel in Fig. \ref{sed_gal} shows how the mass-to-light
ratio varies with time for such populations. Basically the
differences in the spectra reflect the ratio of young ($\lsim$ 1
Gyr) to old stars (Kennicutt 1998).

Shaded regions show the approximate range of the mass-to-light
ratios of the two populations (e.g. Fukugita, Hogan \& Peebles
1998) plotted for $t_{\rm cosm}\leq 14$Gyrs, the age implied by -
inter alia - WMAP measurements. They agree reasonably well with
the plots especially if early populations are older than a few
Gyrs, although there may not be one 'magical' value to plot for
each of the galaxies (Faber \& Gallagher 1979, Roberts \& Haynes
1994).

In order to compare with CIB measurements, the synthetic SEDs must
be averaged over the entire ensemble of galaxy populations. This
includes averaging over all morphological types and, within each
type, averaging over the luminosity function. The first is
important because early-type galaxies contribute substantially to
the near-IR EBL (Jimenez \& Kashlinsky 1999). The second step is
also important because, at least for early-type galaxies, the
mass-to-light depends on galaxy mass or luminosity as suggested by
the fundamental plane measurements for elliptical galaxies (Faber
et al 1989, Djorgovski \& Davis 1987). In the visible, the
fundamental plane implies $M/L_B \propto L_B^{\kappa_B}$ with
$\kappa_B \simeq 0.25$. The slope varies systematically with
wavelength from 0.35 \um through the K band (Pahre et al 1998).
Also, measurements of cluster galaxies out to $z\sim 1$, or $\sim
\frac{1}{2}$ look-back time, indicate that the fundamental plane
preserves its logarithmic slope while the amplitude changes (Treu
et al 2002, van Dokkum et al 2001, 2004, van der Wel et al 2004).
The origin of this wavelength trend may lie in various quantities,
like systematic variations in the IMF along the luminosity
sequence (Pahre et al 1998), or with metallicity (Jimenez \&
Kashlinsky 1999).

\subsection{Dust emission from galaxies: mid-IR to sub-mm}

The near--IR CIB data provide information on the brightness
and structure of largely unextincted starlight. The mid-- and far--IR observations are
necessary to complete the picture, by providing the corresponding information for starlight
that has been absorbed by dust and thermally re-radiated. In many environments,
even at early epochs, this re-radiated emission is a substantial or even dominant fraction
of the total luminosity associated with star formation.

There are excellent reviews on the subject of IRAS galaxies
(Soifer, Neugebauer \& Houck 1987b), ultra luminous infrared
galaxies (ULIRGs, Sanders \& Mirabel 1996) and sub-mm SCUBA
galaxies (Blain et al 2002) and the reader is referred to them for
more details. This subsection summarizes the established
properties of these sources as they relate to the CIB.

The range of galaxy luminosities in the mid- to far-IR greatly
exceeds that in the UV, optical and NIR. E.g the mid- to far-IR
luminosities of IRAS galaxies vary from below a few times
$10^{39}$ erg/sec to over $5\times 10^{46}$ erg/sec. Many of the
more luminous IRAS galaxies are starbursts. The bulk of the IRAS
detected galaxies are late-type spirals, with only a handful
belonging to the early (E or S0) type. The shape and slope
luminosity function of IRAS galaxies differs from that of the
visible band galaxies and is best approximated as two different
power-laws at the high and low ends of the galaxy flux
distribution (Soifer \& Neugebauer 1991). At the bright end of the
bolometric luminosity function, the bright mid- to FIR galaxies
become the dominant galaxy population in the present-day Universe.
The ratio of the FIR to visible galaxy luminosity correlates with
60 to 100 \um color temperature for most galaxies (Soifer et al
1987a,b). Soifer et al (1987), using the Draine \& Lee (1984) dust
modeling, give a useful approximation for the amount of dust
needed to produce a given FIR luminosity:

\begin{equation}
L_{\rm FIR} = 10^4 \; \frac{M_{\rm dust}}{M_\odot} \; \left(\frac{T_{\rm dust}}{40 {\rm K}}\right)^5\;  L_\odot
\label{lumdust}
\end{equation}
This shows that even a tiny amount of dust can make the galaxy
very bright in the mid- to far-IR. Zubko, Dwek \& Arendt (2004)
have developed sophisticated ISM dust models by fitting
simultaneously the far-UV to NIR extinctions, the diffuse IR
emission and constraining the dust properties from the observed
elemental abundances. Assuming that the dust mass-absorption
coefficient $\kappa(\lambda) \propto \lambda^{-n_{\rm dust}}$, the
specific flux at wavelength $\lambda$ received from a galaxy
containing $M_{\rm dust}$ a distance $d$ away is given by (Dwek
2004):
\begin{equation}
I_\nu (\lambda) = 8 \times 10^3 \frac{M_{\rm dust}}{M_\odot} \; \left(\frac{d}{1\; {\rm Mpc}}\right)^{-2} \; \frac{\kappa(\lambda_0)}{{\rm cm^2 g^{-1}}} \; \left( \frac{\lambda_0}{\lambda} \right)^{n_{\rm dust}} \left( \frac{\lambda}{1 \mu{\rm m}} \right)^{-3} \left[\exp\left(\frac{14387.7}{(\lambda/1\mu{\rm m}) T_{\rm dust}}\right)-1 \right]^{-1} \;\; {\rm Jy}
\end{equation}
where $\lambda_0$ is some reference wavelength and $n_{\rm dust}\sim $ 1 to 2.

Most IR emitting galaxies can be classified as normal spirals,
starbursts or AGNs. Starburst galaxies are especially bright at
mid- to far-IR wavelengths and approximately half the galaxies
brighter than $L_{\rm FIR} \simeq 3\times 10^{10}L_\odot$ have
star-burst optical spectra (Elston et al 1985). For the ULIRGs
their measured star formation rate cannot be maintained for longer
than (at most) a Gyr, as at that rate all their ISM will be
consumed. The most luminous IRAS galaxies show features of both
being a starburst and an AGN. In the less luminous IRAS galaxies
their MIR/FIR emission seems to be unrelated to their current SFR.
For most IRAS galaxies that are not AGNs, the observed color-color
correlations can be accounted by a two-component dust model, where
warmer dust is associated with regions of active star formation
and the cold component comes from cirrus in the galactic disk
(Helou 1986).

\begin{figure}[h]
\centering \leavevmode \epsfxsize=0.8 \columnwidth
\epsfbox{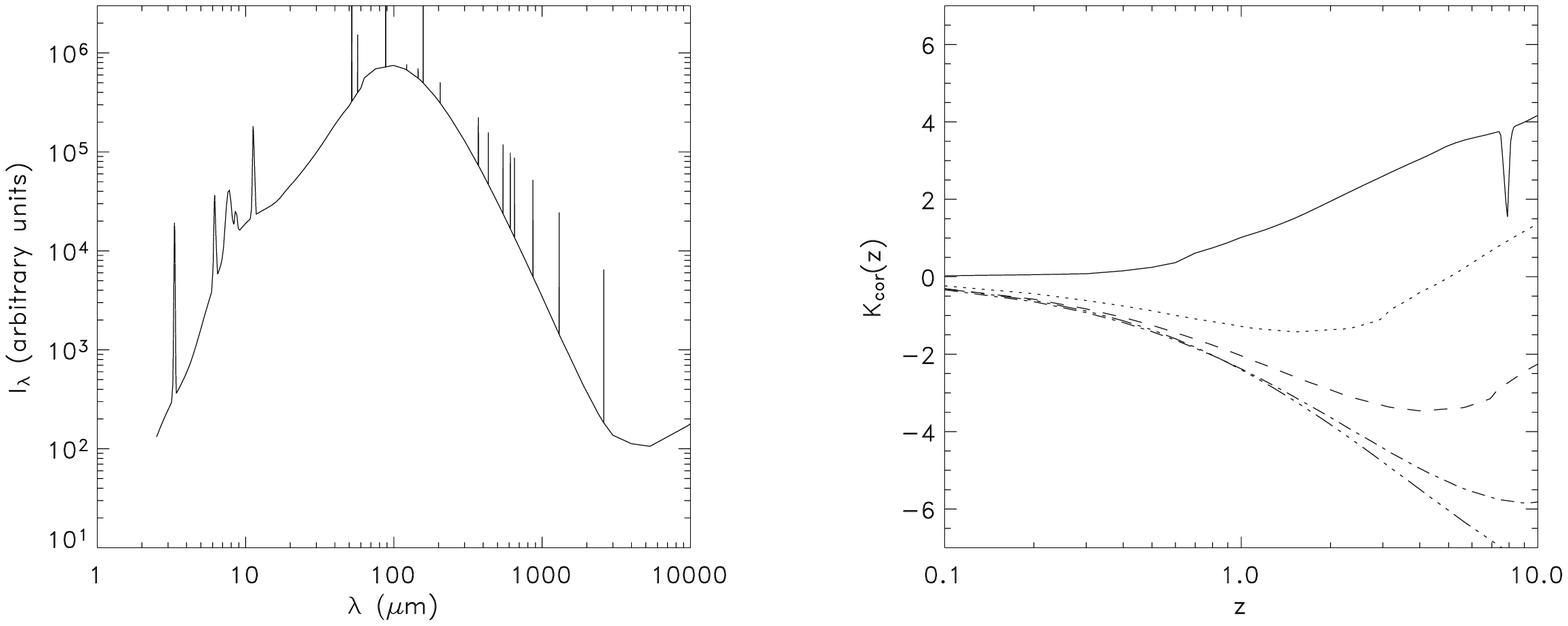} \caption[]{ \scriptsize{ {\bf Left}: Typical
synthetic sub-mm SED of dusty galaxies including
lines from PAHs and ions, atoms and molecules as observed by FIRAS and ISO.\\
{\bf Right}: K-correction vs $z$ at various wavelengths: 100 \um
(solid line), 250 \um (dots), 500 \um (dashes), 1000 \um (dashed
dotted line) and 2000 \um (dash-dot-dot)} } \label{sed_submm}
\end{figure}

One can reconstruct the SED of a typical galaxy in the MIR to FIR
range of wavelengths. Fig. \ref{sed_submm} shows such synthetic
template: it was constructed accounting for continuum dust
emission assuming $T_{\rm dust}=$ 35 K and the free-free emission
at long wavelengths. FIR emission lines were scaled proportional
to the flux and adopted from: the Galactic IR lines observed by
FIRAS (Wright et al 1991), the PAH broad line features at $\lambda
\lsim 12$ \um were taken from the Dwek et al (1997) Galaxy model
of the DIRBE data, and the lines from the ISO Long Wavelength
Spectrometer observations of nearby bright IR galaxies by Fischer
et al (1999). This SED is in good agreement with the IRAS galaxies
average used by Dwek et al (1998), the model of Guiderdoni et al
(1998) and with the template of sub-mm emission from Blain et al
(2002), where comparison is given with the data from galaxy
observations.

The right panel in Fig. \ref{sed_submm} plots the K-correction due
to this SED (defined as  $K_{\rm cor} \equiv -2.5\lg
[I_{\nu(1+z)}(10 {\rm pc})/I_\nu(10 {\rm pc})]$) for 100 \um
(solid line), 250 \um (dots), 500 \um (dashes), 1000 \um (dashed
dotted line) and 2000 \um (dash-dot-dot). One can see the strongly
negative K-correction at $\lambda \gsim 500$ \um and high $z$:
galaxy flux should decrease much slower with increasing redshift
and it can actually increase with $z$. This further implies that
at sub-mm wavelengths the contribution to the CIB should be more
heavily weighted toward high redshift sources.

Optical follow up of the IRAS catalog shows that on average about
30 \% of the bolometric luminosity is emitted in the far-IR via
thermal radiation from dust heated by stars (Soifer \& Neugebauer
1991) and this number rises to up to 95\% for ULIRGs (Sanders \&
Mirabel 1996). Guiderdoni et al (1997) argue that most galaxies
are hidden by dust. The mid- and far-IR number counts indicate
that the IR-luminous galaxies evolved more rapidly than their
optical counterparts and make a substantial contribution to the
star formation at higher $z$ (Elbaz et al 1999). Locally, typical
galaxies radiate a little more in the UV to optical bands than in
mid-IR to far-IR (Soifer \& Neugebauer 1991). But, while luminous
IR galaxies contribute a negligible amount to the local rate of
star formation, they are major contributors at high $z$.

Spinoglio et al (1995) assembled a sample of $\sim 900$ IRAS
galaxies out to the 12 \um limit of $\sim0.2$ Jy and find that the
12 \um luminosity is $\propto L_{\rm IR, total}$ and suggest that
their selection by the $L_{12 \mu{\rm m}}$ is approximately
equivalent to that by the total (bolometric) luminosity . They
find that the 60 and 25 \um luminosities rise more steeply than
linear with the bolometric $L_{\rm IR, total}$ and the opposite is
true for optical bands implying that more luminous disk galaxies
have more dust shrouded stars. Malkan and Stecker (1998) used
these empirical correlations together with assumptions of
luminosity evolution to reconstruct the CIB across from near- to
far-IR.

Several mid-IR deep surveys have been conducted with ISOCAM,
reaching sensitivity levels of $\sim 50 \mu$Jy at 15 \um (Elbaz et
al 1999). Gravitational lensing by known massive clusters was
exploited by Metcalfe et al (2003) to go deeper reaching the flux
limits of 5 and 18 $\mu$Jy at 7 and 15 \um with ISOCAM (Metcalfe
et al 2003). Elbaz et al (2002) demonstrate that MIR luminosities
at 6.75, 12 and 15 \um are strongly correlated with the total IR
luminosity (8 to 1000 \um). They infer the redshift distribution
of galaxies from the spectroscopically complete galaxy sample in
the Hubble Deep Field (North) and find that the correlations hold
out to $z\sim 1$. Chary \& Elbaz (2001) show that a wide variety
of evolutionary models normalized to these correlations and the
observed luminosity functions are consistent with the current CIB
measurements in the mid- to far-IR. At present about 15-20\% of
the bolometric luminosity is emitted in mid- to far-IR, but it is
likely that this fraction increases toward early times. Serjeant
et al (2001) find good agreement for 90\um galaxies with the local
luminosity function from other surveys and evidence for pure
luminosity evolution at the rate of $ {\cal E} \propto (0.98 \pm
0.34) \lg(1+z)$.

The FIRBACK ISO survey (Puget et al 1999) was conducted with the
ISOPHOT instrument  at 175 \um  and reached the confusion levels
of 45 mJy. It provided a catalog of almost 200 galaxies (at
3-sigma level) down to fluxes $\simeq 180$ mJy (Dole et al 2001).
The survey covered four square degrees in three high Galactic
latitude regions. The final catalog of 106 sources (4-sigma)
between 180 mJy and 2.4 Jy was $> 85$ \% complete. The observed
galaxy counts at 170 \um require strong evolution at flux levels
fainter than 500 mJy. Dole et al (2001) estimate that galaxies out
to the flux limit of 135 mJy account for about 5\% of the CIB.
Analysis of the first Spitzer MIPS observations is consistent with
the FIRBACK results (Dole et al 2004).

Significant progress in understanding the sources contributing to
the far-IR CIB has been made with the discovery and subsequent
studies of the sub-mm galaxies with the SCUBA instrument at the
James Clerk Maxwell Telescope.
\footnote{http://www.jach.hawaii.edu/JACpublic/JCMT/Continuum\_observing/SCUBA/index.html}
SCUBA is an array of 37 bolometers at 850 \um\ and 91 at 450 \um
and can observe at both wavelengths simultaneously with angular
resolution of about $15^{\prime\prime}$; the shorter wavelength is
more sensitive to atmospheric conditions. Several hundred SCUBA
sources have by now been detected (Blain et al 2002) and they are
mostly high-$z$ galaxies. Redshifts of the galaxies have to be
determined from optical observations. Because resolution of the
SCUBA instrument is relatively low and the galaxies are at high
$z$, and therefore faint in optical bands, redshift determination
is difficult. Hughes et al (1998) identified optical starburst
counterparts for the sample of five SCUBA sources brighter than 2
mJy, four of which have $z$ between 4 and 5. Scott et al (2000)
detected a sample of ten FIRBACK galaxies which are bright at 170
\um also at the SCUBA wavelengths with fluxes $\gsim 10$ mJy.
Sawicki \& Webb (2004) detect another ten 850 \um SCUBA sources
brighter than $\sim 10$mJy in the general area of the Spitzer
First Look Survey. Chapman et al (2003) have obtained
spectroscopic redshifts of ten representative sub-mm galaxies,
using the VLA to identify the source positions and obtaining the
redshifts with the optical spectroscopy from the Keck telescope.
They derived a median redshift of 2.4 for the sample and the space
densities of $(3.3 \pm 2.3)\times 10^{-6}, (6.5 \pm 2.5)\times
10^{-6}, (2.4 \pm 1.2) \times 10^{-6}$ Mpc$^{-3}$ at redshift bins
$z=[0.5-1.2],[1.8-2.8],[2.8-4]$ respectively. They note that the
median redshift range coincides with the peak quasar activity
suggesting a close relationship between the two. The space density
of the Chapman et al SCUBA sample increases strongly toward higher
$z$ suggesting that they make an important component of the SFR at
$z>2$.

Confusion is a major problem for identifying deep galaxies in
counts at sub-mm wavelengths. It can be reduced with gravitational
lensing by massive clusters which amplifies the flux of galaxies
and at the same time increases their apparent separation on the
sky (Blain 1997). This has been successfully applied to the fields
of known massive clusters to reach flux limits of below 1 mJy at
850 \um (Smail et al 1997, Blain et al 1999, Cowie et al 2002).
The cumulative flux from the galaxy counts out to $\simeq 0.5-1$
mJy adds up to $0.5\pm 0.2$ \nwm2sr accounting for all or most of
the CIB at that wavelength. Serjeant et al (2004) account for most
of the CIB at 450 \um coming from the SCUBA 8 mJy survey maps of
Scott et al (2002). These were combined with a Spitzer
identification and integrating the sub-mm fluxes they find that
the galaxies contribute most of the measured CIB at 450 \um, but
only a small fraction of the CIB at 850 \um.

The surface density of SCUBA galaxies is quite high, $\sim 300$
deg$^{-2}$ and, in principle, one can begin to study their
clustering properties. Blain et al (2004) obtained accurate
positions for a sample of distant SCUBA galaxies. The sample was
followed up with Keck spectroscopy to determine the redshifts of
73 galaxies, constructing a substantially complete ($\sim 70$ \%)
redshift distribution. They find strong clustering for the sample
and, assuming the two-point correlation function
$\xi=(r/r_0)^{-\gamma}$ with $\gamma=1.8$, measure that the SCUBA
galaxies at $z\sim 2-3$ have a correlation length of $r_0 \simeq
(6.9 \pm 2.1)h^{-1}$Mpc. For comparison the present day blue
galaxies have $r_0=5.5 h^{-1}$Mpc (Maddox et al 1990), and the
Lyman break galaxies, that predominantly lie at higher $z$, have
clustering length of $r_0\sim 4 h^{-1}$Mpc (Porciani \& Giavilsco
2002). This should be reflected in larger CIB fluctuations at the
sub-mm wavelengths.

\subsection{Contribution from quasars/AGNs}

A different, from stars, source of energy release is accretion
onto massive and supermassive black holes. The latter manifests
itself in the emission from AGNs and quasars, the bulk of whose
luminosity is produced by accretion onto black holes, not
nucleosynthesis. Severgnini (2000) have compared an X-ray sample
of the [2--10] KeV and 850 \um SCUBA galaxy sources which resolve
most of the backgrounds in the two bands. They find the ratio of
the 850\um to [2--10] KeV fluxes for the sources much smaller than
the value observed for the two backgrounds. The [2-10] KeV
galaxies brighter than $10^{-15}$ erg/cm$^2$ make up $\gsim$ 75 \%
of the X-ray background in this band, but contribute $\lsim 7$ \%
to the sub-mm background. Barger et al (2004) have constructed an
optical and NIR catalog of quasasrs and AGNs from the Chandra Deep
Field out to $z\sim3$. The catalogue currently consists of several
hundred galaxies and extends to $m_{\rm AB}\sim 20.5$ in K band,
but in the future could pave a way to direct measurements of the
AGN contribution to the NIR CIB levels.

The current thinking is that the contribution of the active
galactic nuclei and quasars to the CIB  is most likely small. Its
precise value depends on bolometric correction, the details of the
luminosity function and its evolution at early cosmic times. Madau
\& Pozzetti (2000) discuss the contribution from these sources to
the total EBL. They estimate the total mean mass density of the
quasar remnants today to be $\rho_{\rm BH} \simeq (3\pm2) \times
10^6h M_\odot$Mpc$^{-3}$ which should have contributed $\sim
\frac{c}{4\pi} \epsilon\rho_{\rm BH}c^2\langle (1+z)^{-1}\rangle$
by accretion at redshift $z$ to the total bolometric EBL flux.
Assuming then that these sources radiated with an average
efficiency of $\epsilon\simeq 6$\% corresponding to standard disk
accretion for a Schwarzschild black hole, they argue that, unless
dust-obscured accretion on to supermassive black holes results in
a much larger efficiency, QSOs peaking at $z\sim 2$, as suggested
by observations, are expected to produce no more than 10-20\% of
the total EBL and a still smaller fraction of that will contribute
to the total CIB.

Malkan \& Stecker (1998) estimated the levels of the CIB between 2
and 300 \um using empirically constructed SED spectra and the
observed correlations between the near-IR and mid-IR galaxy
luminosities for various galaxy types including Type 1 and 2
Seyfert galaxies. They also concluded that the contribution from
Seyferts to the CIB is less than 10\%. Lagache et al (2003)
discuss the AGN contribution to the CIB from $\simeq 10$ \um to
sub-mm wavelengths and, assuming that the black holes masses which
power the AGNs are similar to those measured in the HDF, conlcude
that the AGN contribution to the CIB is relatively small compared
to that of stars.

Elbaz et al (2002) have summed the observed galaxy counts at 15
\um from the ISOCAM measurements and obtain the total flux at $2.4
\pm 0.5$ \nwm2sr or just below the upper CIB limits ($\sim 4-5$
\nwm2sr ) discussed above. Of these, they estimate that less than
20 \% comes from AGNs at 15 \um and less than 10 \% at 140\um.
Matute et al (2004) estimate the luminosity function of AGNs and
its evolution from sources in the ELAIS field at 15 \um and
conclude that their contribution to the CIB at that wavelengths is
small ($\sim 2-3$ \%).

\subsection{Present-day luminosity density}

Present day luminosity density is an important normalizing factor
in determining the CIB fluxes via eqs. \ref{dfdz},\ref{lumden}. It
is determined by the galaxy luminosity functions and morphological
types. From UV to the near-IR it is dominated by stellar emission
and at mid- to far-IR by emission from dust. For the Schechter
(1976) luminosity function, $\Phi = \Phi_* (L/L_*)^{-\alpha}
\exp(-L/L_*)$ the luminosity density, eq. \ref{lumden0}, becomes
${\cal L}_{\nu}(0) = L_* \Phi_* \Gamma(\alpha+2)$. Note that the
parameters for the fit are not independent and this parametric
form, while very convenient, may not give an accurate measure of
the overall luminosity density and its uncertainties.

Because the near-IR CIB is produced by stars and the latter emit
most of its energy in the visible bands, any inter-comparison
between the near-IR CIB and the present-day galaxy populations
must be done over the entire visible to NIR range of wavelengths.
Thus in this section we discuss the measurements of the
present-day luminosity density from the recent galaxy surveys from
UV to NIR and their consistency across the wavelength range.

Early measurements of the luminosity function of galaxies
(Broadhurst, Ellis \& Shanks 1988, Marzke et al 1994, Loveday et
al 1992, Gardner et al 1997) have recently been complemented by
much more extensive (and expensive!) surveys in the B band (2dF),
visible bands (SDSS) and the near-IR (2MASS) using the latest
multi-band and detector technologies.

In the visible, the new measurements from the SDSS data come in
five filters from 0.354 \um to 0.913 \um. Blanton et al (2001)
have compiled a substantial catalog of over 10,000 galaxies  to
the depth of $\sim 21-22$ AB magnitudes from SDSS commissioning
observations over 140 deg$^2$. They computed Petrosian magnitudes
over 3$^{\prime \prime}$ apertures arguing that Petrosian
magnitudes best reflect the total flux over all the galaxy types.
Their first findings gave substantially higher total luminosity
densities than earlier measurements. However, a later re-analysis
of the SDSS data with a larger catalog (close to 150,000 galaxies)
and differently treated evolutionary corrections lead to lower
values of ${\cal L}_{\nu}(0)$ in agreement with measurements from
the 2dF (Colless et al 2001) and Millenium Galaxy Catalog (Liske
et al 2003) surveys in $b$ filters using Kron and isophotal
magnitudes.

Kochanek et al (2001) selected over 4,000 galaxies with the median
redshift of 0.02 from the 2MASS catalog and measured the
luminosity function using isophotal magnitudes $K\leq 11.25$. They
also subdivided the sample into early and late type galaxies
(using the RC3 catalog) providing measurements of the luminosity
function for the two morphologies. Cole et al (2001) have combined
the 2MASS and 2dF observations with over 17,000 galaxies with
measured redshifts over $\gsim$ 600 deg$^2$. They computed the
luminosity function using (mostly) the Kron magnitudes. The
analyzes of both Kochanek et al (2001) and Cole et al (2001) agree
within their errors. Huang et al (2003) analyzed a smaller catalog
of 1,056 bright ($K<15$) galaxies with median redshift of 0.14
over $\simeq 8$ deg$^2$ and obtain a significantly higher
present-day luminosity density, but claim that the luminosity
function they measure evolves significantly with time and, hence,
its estimation may be sensitive to proper evolutionary
corrections.

The mid- to far-IR measurements of the luminosity function come
from IRAS galaxies and ISO surveys. IRAS galaxies are detected
most efficiently at 60 \um in the mid- to far-IR data and this
wavelength presents the best band for galaxy identification.
Soifer \& Neugebauer (1991) used the 60 \um IRAS Bright Galaxy
Sample to derive complete flux-limited samples of galaxies at
other IRAS wavelengths. From these samples they derive the
luminosity functions at of galaxies in the local Universe at 12,
25, 60 and 100 \um. ISO observations of the ELAIS (European Large
Area ISO Survey) field covered an area of 12 deg$^2$ at 15 and 90
\um (Oliver et al (2000) and resulted in the largest ISO catalog
(La Franca et al 2004, Rowan-Robinson et al 2004) enabling the
measurements of galaxy luminosity function at these wavelengths by
Serjeant et al (2001, 2004) and Pozzi et al (2004).

\begin{figure}[h]
\centering \leavevmode \epsfxsize=0.8 \columnwidth
\epsfbox{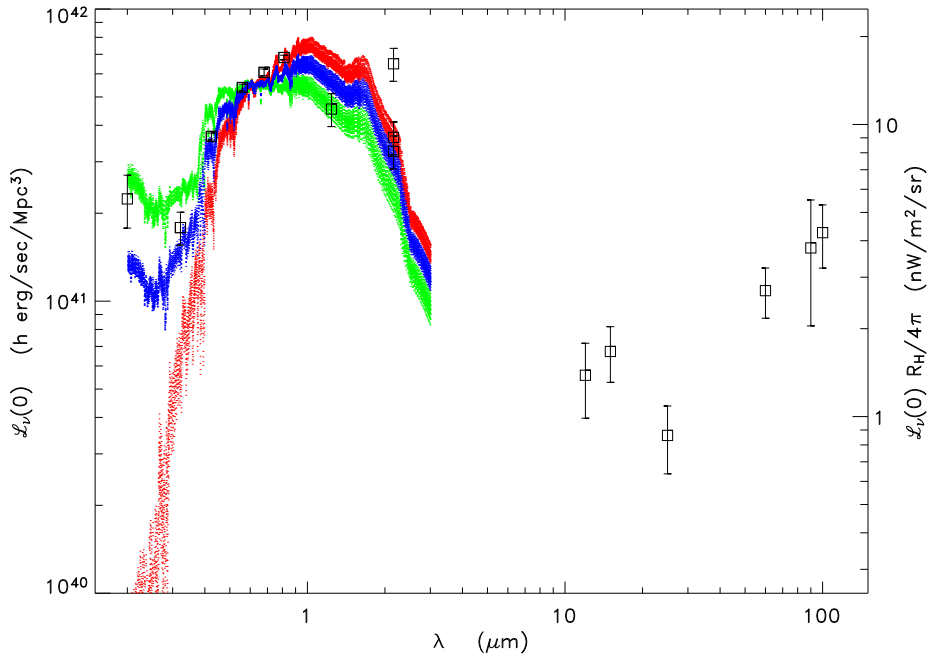} \caption[]{ \scriptsize{Open squares show
measurements of the total present-day luminosity density at the
various wavelengths. The UV measurements at 0.28 \um is from
Sullivan et al (2000). The next five squares in the direction of
increasing wavelength correspond to the SDSS measurements from
Blanton et al (2001). The near-IR points in K band are taken from
Cole et al (2001), Kochanek et al (2001) and the highest point
from Huang et al (2003). The measurement in J band at 1.2 \um is
from Cole et al (2001). The five mid- to far-IR points are the
IRAS measurements of Soifer \& Neugebauer (1991) and the 90 \um
measurements from the ELAIS field ISO observations (Serjeant et al
2001, 2004). Red sets of lines correspond to the contribution from
the early type stellar populations alone with metallicities and
ages shown in Fig. \ref{sed_gal} as described in the main text,
green lines show the same same for late type populations and blue
lines show the simple average of the two contributions.} }
\label{figlumden}
\end{figure}

Fig. \ref{figlumden} summarizes the current measurements of the
present-day luminosity density from the various surveys. Do they
indicate that a substantial flux is missing from the near-IR
measurements when normalized to the measurements in the visible
bands? Such possibility has been suggested by Wright (2001) using
very simplified modeling of galactic SED. Below we discuss this
possibility using more realistic galaxy modeling and including a
simple treatment of galaxy morphology. With this modeling we show
that the data on the luminosity density at various wavelengths are
consistent with each other across the entire range from UV to
near-IR.

Generating realistic SEDs from stellar evolution models is very
important in proper interpretation of the numbers on ${\cal
L}_\nu(0)$ as Fig. \ref{sed_gal} shows. It is interesting to
evaluate the total bolometric flux contained in the luminosity
density measurements. Stellar populations contribute directly to
emission from UV to near-IR with dust contributions dominating the
emissions at longer wavelengths. Integrating the data on the
luminosity density from Fig. \ref{figlumden} from UV to 2.2 \um
and separately over the IRAS wavelengths we get:
\begin{eqnarray}
{\cal L}_{\rm bol} (0.2-2\mu{\rm m}) = (9.8 \pm 1.2)\times 10^{41} h \;
{\rm erg/sec/Mpc}^3 \nonumber \\
{\cal L}_{\rm bol} (12-100\mu{\rm m}) = (1.5 \pm 0.3)\times
10^{41} h \; {\rm erg/sec/Mpc}^3 \label{lumbol_data}
\end{eqnarray}
here we integrated over the near-IR data from Cole et al (2001).
Omitting the UV contribution which mostly comes from very young
stars would give ${\cal L}_{\rm bol} (0.32-2\mu{\rm m}) = (8.5 \pm
0.7)\times 10^{41} h \; {\rm erg/sec/Mpc}^3$. Taking only UV and
visible data gives ${\cal L}_{\rm bol} (0.2-0.8\mu{\rm m}) \simeq
(6.4 \pm 0.3) \times 10^{41} h \; {\rm erg/sec/Mpc}^3$.

In order to do a proper comparison between the data on ${\cal
L}_\nu(0)$ at visible and NIR bands we divide stellar populations
into two types: early (E) and late (L) with the modeling discussed
in Sec. 6.2. If each type contains $\Omega_{\rm E}$, or
$\Omega_{\rm L}$, of the critical density and has stellar
populations with mass-to-light ratio of $M/L \equiv \gamma
M_\odot/L_\odot$, these populations will produce $\Omega_{\rm E/L}
(M_\odot/L_\odot)^{-1}\gamma_{\rm E/L}^{-1} \frac{3H_0^2}{8 \pi
G}$ leading to:
\begin{equation}
\frac{\Omega_{\rm E}}{\gamma_{\rm E}}+\frac{\Omega_{\rm
L}}{\gamma_{\rm L}} = \frac{{\cal L}_{\rm bol}}{10^{45} \; h^2
\frac{\rm erg}{\rm sec\;Mpc^3}} \label{earlylatebol}
\end{equation}
If we adopt the average values of $\Omega_{\rm E}=2\times
10^{-3}h^{-1}$ and $\Omega_{\rm L}=0.6\times 10^{-3}h^{-1}$
suggested by Fukugita, Hogan \& Peebles (1997) with $\gamma_{\rm
E}=5$ and $\gamma_{\rm E}=1.3$ from the right panel of Fig.
\ref{sed_gal}, we get ${\cal L}_{\rm bol, stars} = 9.1\times
10^{41}h\;{\rm erg/sec/Mpc}^3$ in good agreement with eq.
\ref{lumbol_data}. (The old SDSS measurements from Blanton et al
(2001) would require higher values of $\Omega_{\rm E/L}$, but the
latter would still be within the total baryon density parameter
$\Omega_{\rm baryon}$). Thus the measurements of the bolometric
luminosity density from UV to near-IR suggest that no significant
amounts of galaxy fluxes is missing between one wavelength
measurement to the next.

Are then the measurements of the luminosity density across the UV
to near-IR spectrum consistent with each other and with that
expected from realistic stellar populations? In order to see the
consistency between the near-IR and visible bands data on ${\cal
L}_\nu(0)$ we use the SEDs plotted in Fig. \ref{sed_gal}. We
further assume that all the parameters (IMF, SED, $Z$, etc) are
independent of $L$. The total luminosity density then becomes:
\begin{equation}
f_{\lambda,{\rm E}}\frac{\Omega_{\rm E}}{\gamma_{\rm
E}}+f_{\lambda,{\rm L}} \frac{\Omega_{\rm L}}{\gamma_{\rm L}} =
\frac{{\cal L}_\lambda}{10^{45} h^2\;\frac{\rm erg}{{\rm sec \;
Mpc}^3}} \label{omegavslumden}
\end{equation}
Red lines show the least squares fit to eq. \ref{omegavslumden}
for each $(Z,t)$ if only galaxies with the Early type stellar
populations were present. This leads to $\Omega_{\rm E} h^2 \simeq
(4.5-7)\times 10^{-3}$ for the range of $(Z,t)$. Green lines show
the least squares fit to eq. \ref{omegavslumden} for each $(Z,t)$
if only galaxies with the Late type stellar populations were
present. This leads to $\Omega_{\rm L} h^2 \sim (1.9-2.5)\times
10^{-3}$. These numbers are in broad agreement with the above
values on these parameters. It is clear that the luminosity
density at visible and UV wavelengths reflects mainly emission
from late type populations, whereas in the near-IR both early type
populations can dominate. In principle, equation
\ref{omegavslumden} can be solved for both $\Omega_{\rm E}$ and
$\Omega_{\rm L}$, but the answer will be too reflective of the
assumptions made in constructing galactic SEDs to be useful.
Instead we prefer fitting the data in Fig. \ref{figlumden} with
{\it a priori} values of $\Omega_{\rm E}$ and $\Omega_{\rm L}$.
The blue lines show an example of simple mean of the green and red
lines for each $(Z,t)$.

In conclusion, we note again that the answers one gets in this way
are not unique as the realistic SEDs depends on: 1) galaxy
metallicity, $Z$, and its dependence on other galaxy properties,
2) the details of the IMF, particularly in the poorly measured
stellar mass range and its possible dependence on other galaxy
properties, 3) galaxy morphology mixes, 4) galaxy luminosity
function for each galaxy type, 5) galaxy ages for each
morphological type, luminosity/mass etc. If a statistically
significant discrepancy between the measurements of ${\cal
L}_\nu(0)$ is found between the various bands and surveys, it is
probably indicative of the variations in any or all of the above
points, but we have shown that with realistic (i.e. based on
synthetic stellar evolution models) modeling of galactic SEDs the
measurements can already be fit reasonably well.

\subsection{Deep galaxy counts}

The total flux from galaxies measured in deep count surveys gives
another measure of the CIB, or more precisely the contribution to
the CIB from the known sources. This helps to identify the
possible CIB excess and in what cosmological populations and at
what times it might arise. The situation from the current set of
measurements from NIR to sub-mm bands is discussed below.

\subsubsection{Near-IR}

In the near-IR the galaxy counts have been measured now to fairly
faint limits in J,H,K bands with observations coming from ground
and Hubble Space Telescope observations. The integrated flux of
the counts saturates at the levels shown in Table \ref{galcounts}
(Gardner 1996, Kashlinsky \& Odenwald 2000, Cambresy et al 2000,
Madau \& Pozzetti 2000, Fazio et al 2004). Madau \& Pozzetti
(2000) show that the contribution from the HDF galaxies to total
EBL from visible to near-IR  saturates around AB magnitude $m_{\rm
AB} \sim 20$ in all bands and discuss that this is unlikely to
result from underestimating the abundance of distant sources due
to reddening (as their Lyman break shifts into progressively
longer wavelength bands) or absorption.

\begin{figure}[h]
\centering \leavevmode \epsfxsize=0.85 \columnwidth
\epsfbox{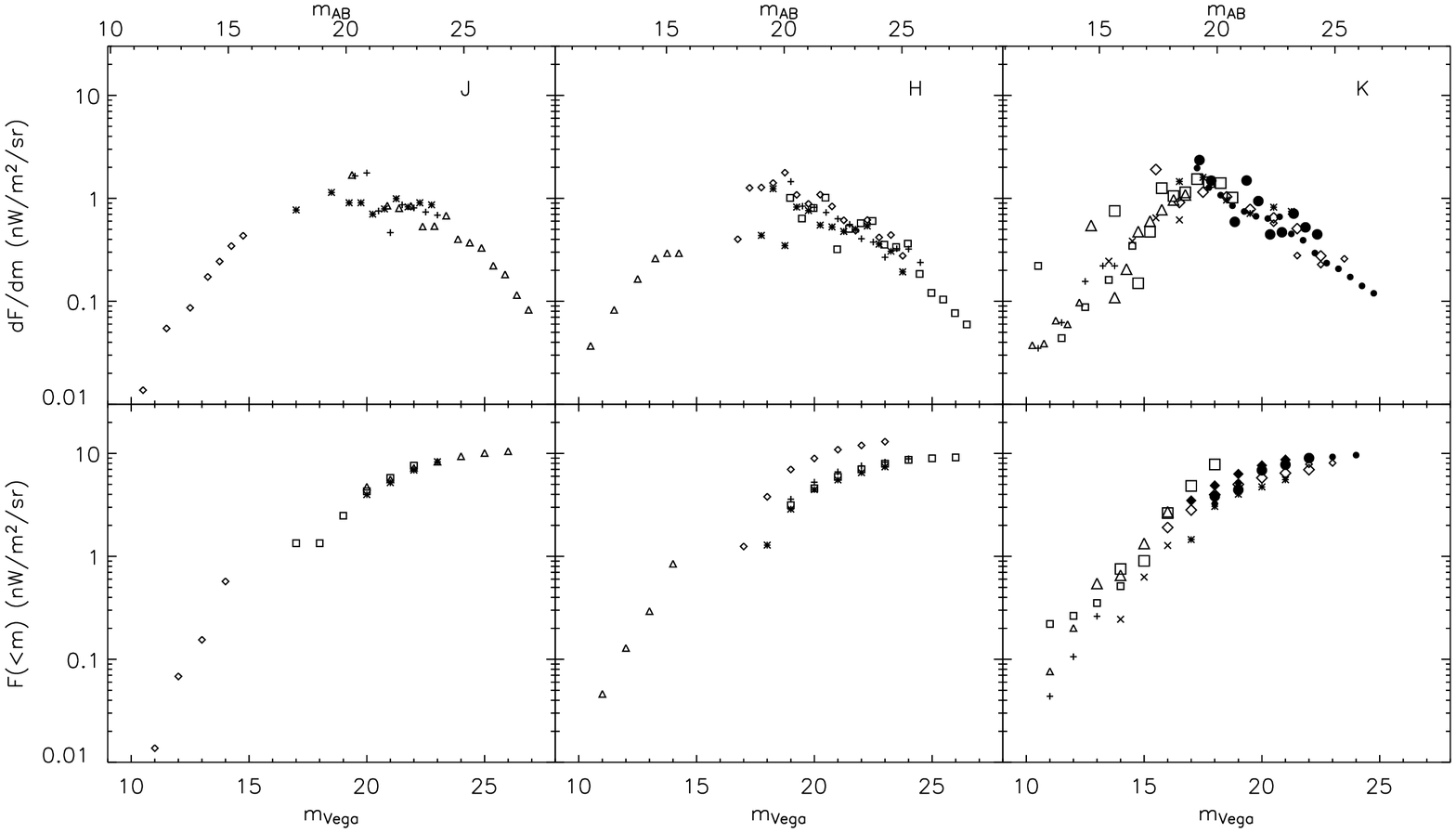} \vspace{0.5cm} \caption[]{ \scriptsize{
Cumulative flux in \nwm2sr contributed by galaxies from a narrow
$dm$ magnitude bin is shown vs the Vega
magnitude from galaxies from deep galaxy surveys in J, H, K. The data are as follows: \\
{\bf J band:} Crosses from Bershady et al (1998), asterisks from Saracco et al (1999),
diamonds from Chester et al (1998) and triangles from Pozetti et al (1996, 1998)
and Madau \& Pozzetti (2000).\\
{\bf H band}: Crosses are from Yan et al (1998), asterisks and diamonds
from Teplitz et al (1998), triangles from Chester et al (1998), and squares
from Pozetti et al (1996, 1998) and Madau \& Pozzetti (2000).\\
{\bf K band}: crosses are from Chester et al (1998), asterisks
from McLeod et al (1995), small open diamonds from Djorgovski et
al (1995), small triangles from Mobasher et al (1986), small and
large open squares, large triangles and large open diamonds from
Gardner et al (1993), $\times$-signs from Glazebrook et al (1994),
filled diamonds from Soifer et al (1994), filled large circles
from Pozetti et al (1996, 1998) and Madau \& Pozzetti (2000), and
filled small circles from Maihara et al (2001). } }
\label{jhk_gal}
\end{figure}

Fig. \ref{jhk_gal} shows the contributions to the total CIB from
galaxies at J,H,K bands. The data come from the compilations
summarized in Madau \& Pozzetti (2000) and Pozzetti et al (1996,
1998) and the Subaru Deep Field data (Maihara et al 2001); see
caption for the figure. The upper panels show $dF/dm$ vs magnitude
and the lower panels show the cumulative flux from galaxies
brighter than magnitude $m$. The fluxes saturate at $m_{AB} \sim
20-23$. The asymptotic values of the fluxes from ordinary galaxies
compared to the observed CIB levels are shown in Fig.
\ref{cib_dc}.

ISO galaxy counts at 6.75 \um reach $\sim 40 \mu$Jy in the HDF
(Oliver et al 1997) and Lockman hole regions (Taniguchi et al
1997) and Sato et al (2003) reach sources as faint as $\sim
10\mu$J in another region. At these wavelengths the counts have
now been complemented by deeper measurements with the Spitzer IRAC
instrument.

\begin{figure}[h]
\centering \leavevmode \epsfxsize=1. \columnwidth
\epsfbox{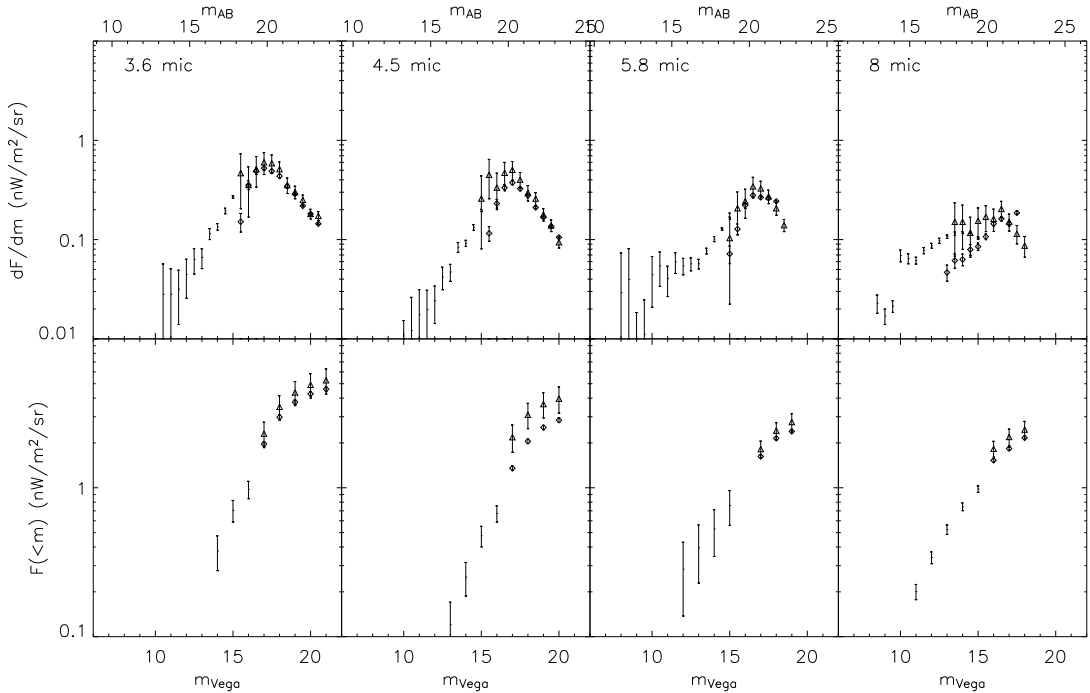} \vspace{0.5cm} \caption[]{
\scriptsize{Cumulative (bottom set of panels) and differential
(top) flux distribution in \nwm2sr vs the Vega magnitude for
galaxies in three different surveys from the Spitzer IRAC galaxy
surveys by Fazio et al (2004): dots with errors correspond to
their Bootes field, diamonds to their EGS data and triangles to
the QSO1700 field observations. AB magnitudes are shown on the
upper horizontal axis.} } \label{spitzer_irac}
\end{figure}

Fig. \ref{spitzer_irac} shows the contributions to the CIB flux
from faint galaxies at 3.6, 4.5, 5.8 and 8 \um from three recent
Spitzer IRAC surveys reported by Fazio et al (2004). The surveys
covered 3 independent fields referred to as QSO field (the deepest
with $\sim$ 9.2 hours total exposure), the EGS (Extended Groth
Strip) field at a higher Galactic latitude and the widest and the
shallowest Bootes region survey. They separated stars from
galaxies and corrected for incompleteness in the faintest counts
(mostly galaxies). At 3.6 and 4.5 \um the total flux seems to
saturate at the levels of shown in Table \ref{galcounts}. At 5.8
and 8 \um the saturation of the cumulative flux from observed
galaxies is not as clear, although a case can be made that the
ceiling of the total flux contribution to the CIB has been reached
as well. At any rate we interpret the total fluxes at 5.8 and 8\um
as lower limits on the contribution from ordinary galaxies to the
CIB and plot them with arrows in Fig. \ref{cib_dc}. The numbers
from the Spitzer IRAC surveys shown in Table \ref{galcounts} are
direct summation of the fluxes from upper panels in Fig.
\ref{spitzer_irac}. When integrating the counts weighted according
to uncertainties, Fazio et al (2004) find total fluxes of 5.4,
3.5, 3.6, and 2.6 \nwm2sr at 3.6, 4.5, 5.8 and 8 \um respectively.

\subsubsection{mid-IR}

Elbaz et al (2002) used ISOCAM observations of almost 1,000
galaxies at 15 \um to estimate the contribution to the CIB at that
wavelength from sources down to 50 $\mu$Jy. They obtain that these
galaxies produce $2.4 \pm 0.5$ \nwm2sr and their contribution may
saturate at fluxes $\lsim 30-50 \mu$Jy.

Spitzer/MIPS 24 \um channel has best mid-IR resolution and its
confusion limit corresponds to fainter fluxes. The MIPS FWHM at 24
\um corresponds to 6$^{\prime \prime}$. Papovich et al (2004)
present the number counts of $\simeq 5\times 10^4$ sources from
the 24 \um Spitzer deep surveys. They show that the counts probe a
previously undetected population of very luminous galaxies at high
$z$. The data were obtained in five fields from Spitzer
characterization and GTO (Guaranteed Time Observations)
observations. The largest and shallowest of the fields (Bootes)
subtended ca. 9 deg$^2$ with an exposure of 87 secs per pixel and
the smallest and deepest (ELAIS) was ca. 130 arcmin$^2$ exposed
for just under one hour in each pixel. The counts out to 60 \um
give $\sim 2$ \nwm2sr and Papovich et al (2004) estimate that
extrapolating to sources below that limit gives around 3 \nwm2sr
in total.

\begin{figure}[h]
\centering \leavevmode \epsfxsize=1. \columnwidth
\epsfbox{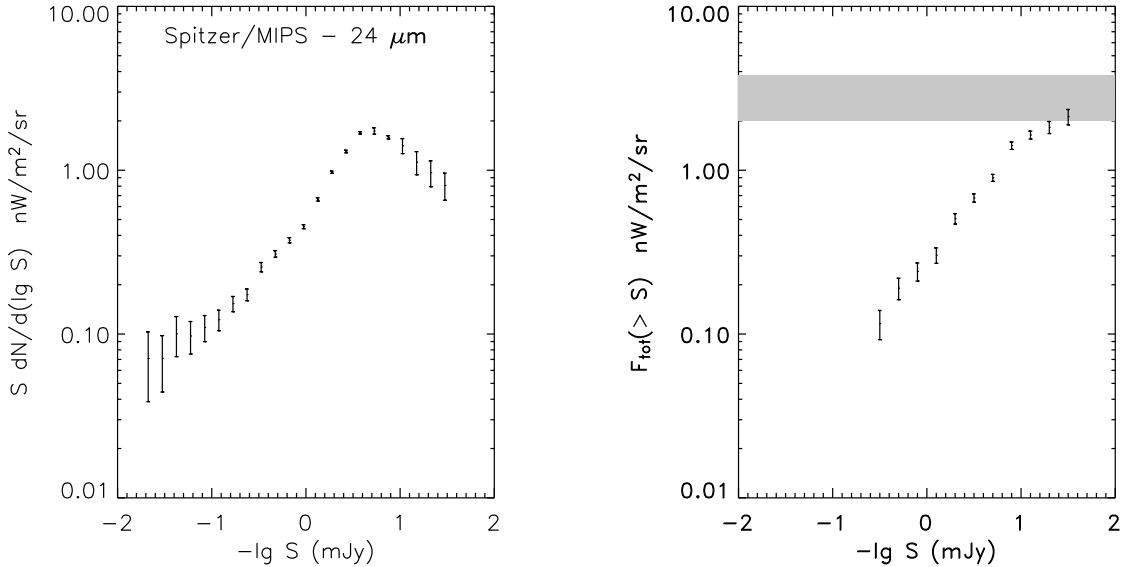} \vspace{0.5cm} \caption[]{ \scriptsize{{\bf
Left}: differential flux distribution in \nwm2sr from
the MIPS counts at 24 \um (Papovich et al 2004)\\
{\bf Right}: cumulative flux contribution as function of the
source flux from galaxies shown in the left panel. Shaded area
corresponds to the total flux with its uncertainty including the
extrapolation from sources fainter than $60 \mu$Jy (Papovich et al
2004). } } \label{spitzer_mips_24}
\end{figure}

\subsubsection{Far-IR and sub-mm}

The number counts at mid- and far-IR indicate that the IR-luminous
galaxies evolved more rapidly than their optical counterparts and
make a substantial contribution to the star formation at higher
$z$ (Elbaz et al 1999).

A deep survey (FIRBACK) was performed by ISOPHOT at 170 \um (Lemke
et al 1996) out to the depth of 120 mJy where galaxy counts can no
longer be fitted with a Euclidean slope. This is confirmed with
the MIPS observations of the Chandra Deep Field (Dole et al 2004)
which indicate a {\rm lower} limit on the CIB at 170 \um of 1.4
\nwm2sr corresponding to about $\sim 10$\% of the CIB. Dole et al
(2004) used early observations with the Spitzer MIPS instrument to
measure galaxy counts at 70 and 160 \um down to 15 and 50 mJy
respectively. Their counts are consistent with the 60 \um IRAS
counts (Lonsdale et al 1990) and the FIRBACK 170 \um survey (Dole
et al 2001). From evolutionary modeling they suggest that most of
the observed faintest sources  lie at $z\sim 0.7$ with a tail out
to $z\sim 2$. Integrating the flux from the observed sources leads
to the total flux of 0.95 and 1.4 \nwm2sr at 70 and 1 60 \um
respectively. As Fig. \ref{cib_dc} shows this corresponds to a
small part of the total CIB levels at these wavelengths, so most
of the CIB at 70 and 160 \um must come from still fainter (and
more distant) sources.

The FIR background can be resolved into individual sources with
detections by the SCUBA instrument at the James Maxwell Telescope
(Smail et al 1997, Blain et al 2002). The total flux at 850 \um
from the the SCUBA sources is estimated to be $0.5\pm 0.2$ \nwm2sr
(Smail et al 1997).

Serjeant et al (2004) presented statistical detections of galaxies
in the Spitzer Early Release Observations through a stacking
analysis of their reanalyzed SCUBA 8 mJy survey maps of Scott
et al (2002) combined with a Spitzer identification of their positions.
Integrating the sub-mm fluxes of the Spitzer populations they find
that the 5.8 \um galaxies contribute only $0.12 \pm 0.05$ \nwm2sr
at 850 \um, but at the same time contribute $2.4 \pm 0.7$ \nwm2sr
at 450 \um, or almost the entire CIB at that wavelength. This
contribution is shown with filled circle at 450 \um in Fig.
\ref{cib_dc}.

\subsection{CIB fluctuations from clustering of ordinary galaxies}

In surveys where the beam is small, galaxies can be resolved and
removed down to some limiting magnitude allowing, in principle, to
probe the CIB from fainter and typically more distant systems. At
$z$=1, the angular scale of $1\dasec$ subtends comoving scale
15$h^{-1}$Kpc, for the WMAP cosmological parameters, so the
identified galaxies can be excised almost completely in surveys
with arcsec scale resolution. In practice, the precise value of
the limiting magnitude depends on the instrument noise levels,
foreground emissions, etc.

\begin{figure}[h]
\centering \leavevmode \epsfxsize=0.75 \columnwidth
\epsfbox{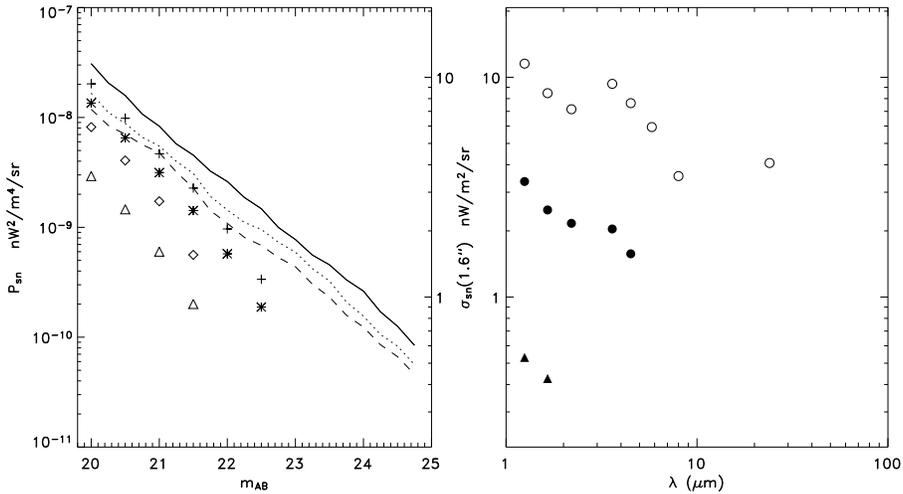} \vspace{0.5cm} \caption[]{ \scriptsize{ {\bf
Left:} Shot noise power spectrum (left vertical axis), evaluated
according to eq. \ref{power_shotnoise}, from galaxies fainter than
the AB magnitude shown in the horizontal axis. It was computed for
galaxy counts summarized in Sec.6.6. Left vertical axis shows the
corresponding value of $\sigma_{\rm sn}$ for a beam of $\simeq
1.6^{\prime\prime}$ in radius. The lines show the shot noise from
HDF galaxy counts by Madau \& Pozzetti (2000): solid, dotted and
dashed lines correspond to  to J, H and K bands respectively.
Symbols show the shot noise from galaxies in the IRAC Spitzer
counts data from Fazio et al (2004): plus signs, asterisks,
diamonds and triangles correspond 3.6, 4.5, 5.8 and 8 \um
respectively. {\bf Right:} $\sigma_{\rm sn}$ vs $\lambda$
evaluated from existing deep galaxy counts data. Open circles
correspond to galaxies fainter than $m_{AB}=20$, filled circles to
$m_{AB} > 22$ and filled triangles to $m_{AB} > 25$.} }
\label{sigma_conf}
\end{figure}

Small beam also means that the shot-noise component of the
fluctuations discussed in Sec. 3.2.2 may be important. The left
panel of Fig. \ref{sigma_conf} plots the shot noise power spectrum
(left vertical axis) and the value of $\sigma_{\rm sn}$ (right
vertical axis) given by eqs. \ref{power_shotnoise},\ref{shotnoise}
vs the AB magnitude down to which galaxies have been removed. The
right panel shows $\sigma_{\rm sn}$ vs the wavelength for galaxies
fainter than $m_{AB}= 20$ (open circles), 22 (filled circles), and
25 (filled triangles). The shot noise was evaluated from galaxy
counts data summarized in Sec. 6.6. In the figure we assumed the
beam of $\omega_{\rm beam} = 2.1 \times 10^{-10}$ sr or, in case
of a spherical beam, a radius of $\sim 1.6^{\prime\prime}$. We did
not attempt to calculate the (probably substantial) statistical
uncertainties in the shot noise components because the various
systematic, cosmic variance, etc effects in the different galaxy
surveys are difficult to quantify (which would probably make such
error bars misleading), but the figure should give a reliable
approximation for the magnitude of the expected shot noise.

\begin{figure}[h]
\centering \leavevmode \epsfxsize=0.75 \columnwidth
\epsfbox{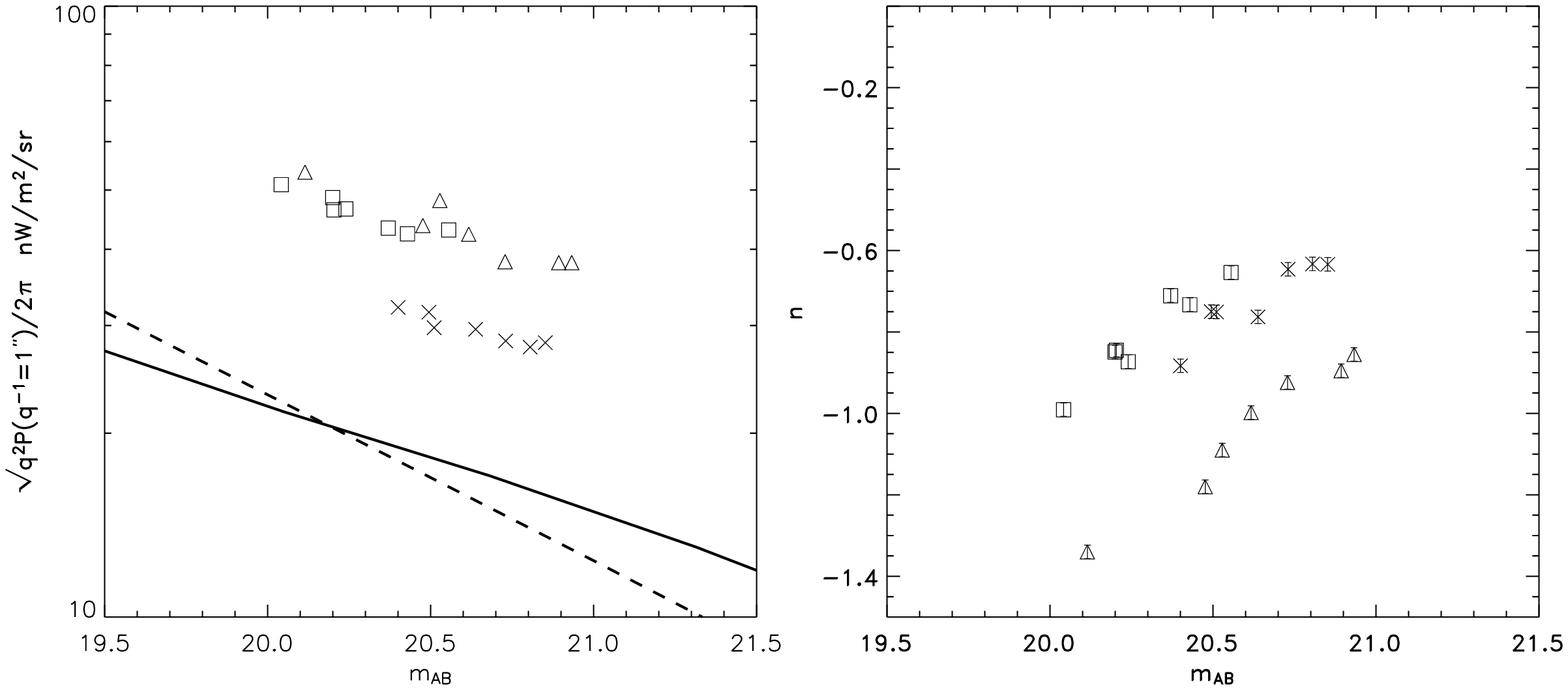} \caption[]{ \scriptsize{ {\bf Left}: Amplitude
of the RMS CIB flux fluctuation at $q^{-1}=1$ arcsec from Fig.
\ref{komsc} vs AB clipping magnitude beyond which galaxies were
excised from the field. Open triangles, open squares and crosses
correspond to the 2MASS J,H,K bands respectively. Dashed and solid
lines show the fluctuation from passively evolving galaxies,
including the shot-noise component, for J and K bands
respectively. {\bf Right}: Power law index $n$ for the CIB power
spectrum $P(q) \propto q^n$ from the deep 2MASS data shown in Fig.
\ref{komsc} is plotted vs the limiting clipping magnitude. Same
symbol notation as in the left panel. } }
\label{komsc_interpretation}
\end{figure}

The left panel of Fig. \ref{komsc_interpretation} shows the
amplitude of the detected CIB fluctuation at $q^{-1}=1$ arcsec
from the analysis of the deep 2MASS data in J, H, K bands
(Kashlinsky et al 2002, Odenwald et al 2003). It is plotted vs the
AB magnitude above which galaxies galaxies were clipped out for
the three 2MASS bands. Although the present deep galaxy counts
data have potentially appreciable uncertainties for evaluating the
shot noise amplitude,  comparison with Fig. \ref{sigma_conf} shows
that the detected CIB fluctuations are substantially higher than
the shot-noise fluctuation from galaxies remaining in the field.
The largest angular scales probed by the analysis were $\pi/q \sim
1.5^\prime$ which at $z$=1 corresponds to $\simeq$1$h^{-1}$Mpc.
Hence the data are probing the range of scales where galaxy
clustering is in the non-linear regime. The lines show the
amplitude for passively evolving galaxies with synthetic spectra
from Jimenez \& Kashlinsky (1999) for J and K bands. They are
normalized to the observations of the galaxy luminosity densities
in Fig. \ref{figlumden}, the observed power spectrum of galaxy
clustering, include the shot-noise component from Fig.
\ref{sigma_conf}, and assume no biasing. The high numbers for the
near-IR fluctuations are consistent with other findings
(Kashlinsky \& Odenwald 2000, Matsumoto et al 2000, 2003) after
accounting for the beam difference and the fact that in the
large-beam DIRBE and IRTS studies no galaxies were removed.

The figure illustrates that the detected CIB anisotropy levels are
large compared to simple no-evolution assumptions. They may imply
either significant evolution of galaxy populations at early cosmic
times, or strong biasing of faint (more distant) galaxy
population. The latter would then have to be wavelength dependent
because the discrepancy is different in the various bands.
Alternatively, Maggliochetti et al (2002) ascribe the difference
to contribution from Population III stars at $z\sim 10-20$; this
contribution is discussed in the next section.

The slope of the detected power spectrum of the CIB from
Kashlinsky et al (2002) depends on the clipping magnitude cutoff
as shown in the right panel of Fig. \ref{komsc_interpretation}. If
progressively fainter galaxies are removed the slope of the power
spectrum flattens. This is consistent with fainter galaxies being
at higher $z$ when the clustering pattern was less evolved and the
slope reflects evolution of galaxy clustering with time. The slope
of this dependence is similar in all three bands. The power
spectrum of the present-day galaxy clustering on non-linear scales
has the spectral index $n\simeq -1.3$ (Groth \& Peebles 1977,
Maddox et al 1990), so it would appear that the deep 2MASS J band
patch with the lowest $m_{\rm cut}$ probes galaxies out to smaller
$z$. In CDM models the non-linear power spectrum of galaxy
clustering evolves toward steeper slope (higher $|n|$) at low $z$
consistent with the trend in Fig. \ref{komsc_interpretation}b.

\subsection{Cumulative flux from galaxy counts vs CIB
measurements}

\begin{deluxetable}{c c | c}
\tabletypesize{\scriptsize}
\rotate
\tablecaption{Integrated CIB flux from observed NIR galaxy counts}
\startdata
$\lambda$ & Flux  (\nwm2sr ) & Comments$^a$ \\
1.25 \um $^1$ & $9.71^{+3.00}_{-1.90}$ & Saturates at $m_{\rm AB} \gsim 24-25$\\
1.65\um $^1$ & $9.02^{+2.62}_{-1.68}$ & Saturates at $m_{\rm AB} \gsim 24-25$ \\
2.2 \um $^1$& $7.92^{+2.04}_{-1.21}$ & Saturates at $m_{\rm AB} \gsim 22-23$ \\
3.6 \um $^2$& $5.27 \pm 1.02$ & Saturates at $m_{\rm AB} \gsim 22-23$\\
4.5 \um $^2$&  $3.95 \pm 0.77$ & Saturates at $m_{\rm AB} \gsim 22-23$\\
5.8\um $^2$  & $\gsim 2.73 \pm 0.22$ & Possibly saturates at $m_{\rm AB} \gsim 22-23$, but data too uncertain. $^b$ \\
8 \um $^2$ & $\gsim 2.46 \pm 0.21$ & Possibly saturates at $m_{\rm AB} \gsim 22-23$, but data too uncertain. $^b$\\
\hline
 & & \\
1.25 \um $\leq \lambda \leq 8$ \um &  $11.4 \pm 4.6$  & NIR (bolometric) flux \\
1.25 \um $\leq \lambda \leq 4.5$ \um  & $9.7 \pm 3.2$ & -- \\
 & & \\
\hline
15 \um $^3$ & $2.4\pm 0.5$  & Saturates at $S \lsim 0.05$ mJy. $^b$\\
24 \um $^4$ &  $2.7_{-0.7}^{+1.1}$ & Saturates at $S \lsim 0.05$ mJy. $^b$\\
70 \um $^5$ &  $>0.95$ & Does not saturate out to $S\gsim $15 mJy. $^c$\\
170 \um $^5$ & $>1.4$ & Does not saturate out to $S\gsim $180 mJy. $^c$\\
450 \um $^6$ & $2.4 \pm 0.7$ & Sub-mm fluxes of Spitzer sources at 24 \um account for most of the CIB\\
850 \um $^7$ & $0.5 \pm 0.2$ & Sources out to $S\simeq 0.5-1$ mJy account for most of the CIB\\
\enddata
\tablecomments{ References: $^1$ Madau \& Pozzetti (2000), $^2$ Fazio
et al (2004), $^3$ Elbaz et al (2002), $^4$ Papovich et al (2004),
$^5$ Dole et al (2004), $^6$ Serjeant et al (2004), $^7$ Blain et
al (1999) \\
$^a$ Saturation point is defined as the magnitude where
$\frac{1}{F(<m)} \frac{dF}{dm} \leq 0.2 $ \\
$^b$ CIB not measured directly at that wavelength\\
$^c$ Counts are not measured to sufficiently faint limits. }
\label{galcounts}
\end{deluxetable}

At many IR wavelengths the counts from ordinary galaxies are deep
enough to make a 'reasonable' guess about their total contribution
to the CIB. This would indicate whether other sources of radiation
existed that produced the observed CIB flux and its fluctuations.
Table \ref{galcounts} summarizes the total fluxes observed
directly by the deepest currently available galaxy populations and
one can compare them with the CIB measurements summarized in Fig.
\ref{cib_dc}.

At mid-IR there are no direct CIB measurements to compare, but
indirect upper limits are close to the lower limits from the
counts so probably no surprises are expected. At far-IR most of
the CIB has been resolved at 450 and 850 \um with the flux from
the observed counts not yet saturating. At NIR the counts from the
observed 'ordinary' galaxy populations saturate at levels a factor
of $\sim 2-3$ {\rm lower} than the claimed levels of the mean CIB
measurements and the fluctuations analysis. Furthermore, the CIB
fluctuations results from the deep 2MASS data suggest that the
excess arises in  galaxy populations with $m_{\rm AB} \gsim 21$
and likely comes from early times.

\begin{figure}[h]
\centering \leavevmode \epsfxsize=0.75 \columnwidth
\epsfbox{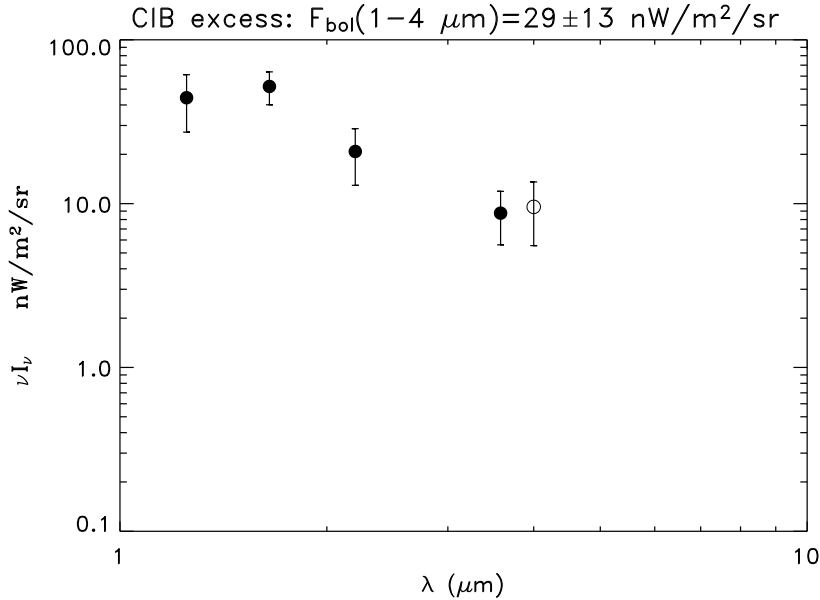} \vspace{0.5cm} \caption[]{ \scriptsize{ CIB
excess at near-IR: filled circles correspond to CIB measurements
minus the cumulative flux from the observed galaxy populations. As
discussed in the text, the various measurements are consistent,
but for the purposes of this plot, the CIB mean levels were taken
from Cambresy et al (2001) at J band, and from the IRTS
measurements at 1.63 to 4 \um. Filled circles show wavelengths
where galaxy counts can be summed directly. Open circle
corresponds to extrapolation of the observed fluxes from galaxy
counts given in Table \ref{galcounts} to $\lambda=4$ \um, the
maximal wavelengths of the IRTS measurements. } } \label{excess}
\end{figure}

Fig. \ref{excess} plots the NIR CIB excess left over after galaxy
counts contributions from Table \ref{galcounts} are subtracted
from the observed CIB levels shown in Fig. \ref{cib_dc}. The last
point is plotted at 4 \um, where IRTS data are still available and
galaxy counts were extrapolated to this wavelengths using the
numbers in Table \ref{galcounts}. Integrating over the points in
the figure leads to the total bolometric flux for the near-IR CIB
excess between 1 and 4 \um of $29.4 \pm 13.0$ \nwm2sr . Ignoring
the last point (open circle) in the figures leads to $26.4 \pm
12.2$ \nwm2sr .

Zodiacal light at these wavelengths would come from reflected
Solar emission and observations suggest a slope $\propto
\lambda^{-2}$ (Kelsall et al 1998 and Fig. \ref{foregrounds}).
Assuming that zodiacal light errors are proportional to the
zodiacal flux, the excess may be attributable to the residual zodi
emission if the points at 1.25 and 4 \um can be discarded.
Comparison with the zodiacal light surface brightness shown in
Fig. \ref{foregrounds} suggests that in this case, the relative
errors in the zodiacal emission modeling would have to have
amplitude of $\gsim $ 40--50 \% (see Kelsall et al 1998 and Fig.
\ref{foregrounds}) at the near-IR bands. They would also have to
be the same in all the different CIB measurements with different
instruments and different zodiacal models used. Galactic cirrus
emission would at these wavelengths contribute emission with a
much smaller and a roughly flat SED (Leinert et al 1998).

\section{Population III}

The epoch of the first stars, when Population III stars formed, is
now emerging as the next cosmological frontier. It is not clear
what these stars' properties were, when they formed or how long
their era lasted before leading to the stars and galaxies we see
today. On the other hand, if the excess near-IR CIB is real, as
many independent measurements suggest, it may well be attributable
to emissions by these stars and would make the CIB and its
structure a unique probe directly into the epoch and efficiency of
Population III formation and evolution. This would provide a
powerful application of the CIB-related science to one of the most
outstanding remaining questions in the standard cosmological
picture.

Recent measurements of quasar spectra reveal the Gunn-Peterson
trough at $\lambda_{\rm rest} \leq 1216 \AA$ at $z\gsim 6-7$
indicating the location of the reionization epoch (Becker et al
2001, Djorgovski et al 2001). These quasars already show the
presence of metals suggesting that the first metal free stars had
to form at still earlier epochs and enrich the IGM by $z\sim 7$.
At epochs corresponding to $z\gsim 7$ the Universe had to contain
significant amounts of neutral hydrogen leading to absorption of
all radiation at wavelengths shorter than the Lyman limit.
Interestingly, the J band filter probes wavelengths longer than
the Lyman limit at $z\lsim 14$ and the K band at $z \lsim 20$, so
much of the near-IR CIB can probe the Population III era even if
reionization has not yet occurred.

\subsection{What were the first stars?}

Population III stars are the first stars to form in the metal-free
Universe and so far remained an entirely hypothetical (and
theoretical) class of objects since introduced by Rees in the late
seventies (Rees 1980 and references cited therein). Much of what
can be attributed to consequences of that early and possibly brief
era depends on the the mass Population III stars had and we start
this section with a brief discussion of the theoretical spectrum
of possibilities and the current ideas.

Should the first stars have been different from modern day stellar
objects? We know from observations that star formation takes
places in gas clouds much more massive than $M_{\rm Chandra}
=(\frac{\hbar c}{Gm_p^2})^{3/2} m_p \simeq 1.44 M_\odot$, a
mass-scale of a typical self-gravitating object supported by
nuclear fusion, and must be caused by gravitational collapse
(stellar material is much more dense than the clouds in which they
form) and fragmentation (stars are much smaller in mass than the
parental clouds). Thus the gravitational collapse processes must
lead to final fragments that are just in the right mass range to
support nuclear burning.

Hoyle (1953) provided a plausible and elegant answer to this that
became known as ``opacity limited fragmentation theory". It goes
as follows: the cooling time is much shorter in galactic mass gas
clouds than the dynamical collapse time and the ratio of the two
decreases with increasing gas density (Rees \& Ostriker 1977).
Gravitational collapse thus proceeds isothermally leading to a
decrease in the Jeans mass inside the cloud, $M_{\rm Jeans}
\propto T^{3/2}\rho^{-1/2}$. The cloud becomes susceptible to
fragmentation into masses of $\sim M_{\rm Jeans}$, which
themselves collapse and fragment into smaller pieces. This process
continues until the density increases enough so that the optical
depth across the fragment is sufficiently high to trap the
radiation released by the collapse. This requires $\tau \propto
\kappa \rho^{2/3}M^{1/3}$ of order unity (for absorption; for
scattering this happens at higher $\tau$). For clouds of solar
metallicity the collapse proceeds at $T\sim 10$K and fragmentation
stops at $\sim 0.01 M_\odot$ (Low \& Lynden-Bell 1977, Silk 1977).
For metal free gas clouds the collapse proceeds at higher T
($\gsim 10^4$K in the absence of hydrogen molecules, Dalgarno \&
McCray 1972). Thus the early thinking in the pre-CDM era was that
the first stars would have to be much more massive. On the other
hand, Rees (1976) has shown, from very general arguments of
maximal cooling efficiency, that the minimal mass is roughly
independent of $T$ and is $M_{\rm min} \simeq M_{\rm Ch}
(k_BT/m_pc^2)^{1/4}$. I.e. in metal free gas the cooling is very
inefficient so $T$ is high, but at the same time - because opacity
is generated by the same coolants - $\tau$ is smaller and the
fragmentation stops at higher densities and smaller $M_{\rm
Jeans}$.

However, a consensus based on recent numerical investigations
within the framework of the $\Lambda$CDM models is now emerging
that fragmentation of the first collapsing clouds at redshifts
$z_*\sim 10-30$ was very inefficient and that the first metal-free
stars after all were very massive objects with mass $\gsim 100
M_\odot$ (\cite{abel,bromm}, see recent review by Bromm \& Larson
2004). Such stars would form in small mini-halos and live only a
few million years, much less than the age of the Universe ($\simeq
2\times10^8$  years at $z=20$), making their direct detection
still more difficult. On the other hand the net radiation produced
by these massive stars could give substantial contributions to the
total diffuse background light (\cite{mjr}) and since their light
is red-shifted much of that contribution will be today in the
infrared bands (\cite{bond}). This could make them significant
contributors to the near-IR CIB.

Such massive, metal-free stars will be dominated by radiation
pressure and would radiate close to the Eddington limit: $L \simeq
L_{\rm Edd} = \frac{4 \pi G m_p c}{\sigma_T} M \simeq 1.3 \times
10^{38} M/M_\odot$ erg/sec, where $\sigma_T$ is the cross section
due to electron (Thompson) scattering. The energy spectrum for
emission from these stars will be quite featureless and close to
that of a black body at $T \sim 10^{4.8-5}$K (Schaerer 2003,
Tumlinson et al 2003). Unlike their metal rich counterparts
Population III stars are not expected to have significant
mass-loss during their lifetime (Baraffe et al 2001). If
sufficiently massive ($\gsim 240 M_\odot$, Bond et al 1984, Heger
et al 2003) such stars would also avoid SN explosions and collapse
directly to black holes in which case their numbers can
significantly exceed that required to produce the metallicities
observed in Population II stars.  The lifetime of these stars will
be independent of mass: $t_L\simeq \epsilon M c^2/L \simeq 3
\times 10^6$ years, where $\epsilon =0.007$ is the efficiency of
the hydrogen burning. These numbers are in good agreement with
detailed computations (e.g. Schaerer 2002).

\begin{figure}[h]
\centering \leavevmode \epsfxsize=0.7 \columnwidth
\epsfbox{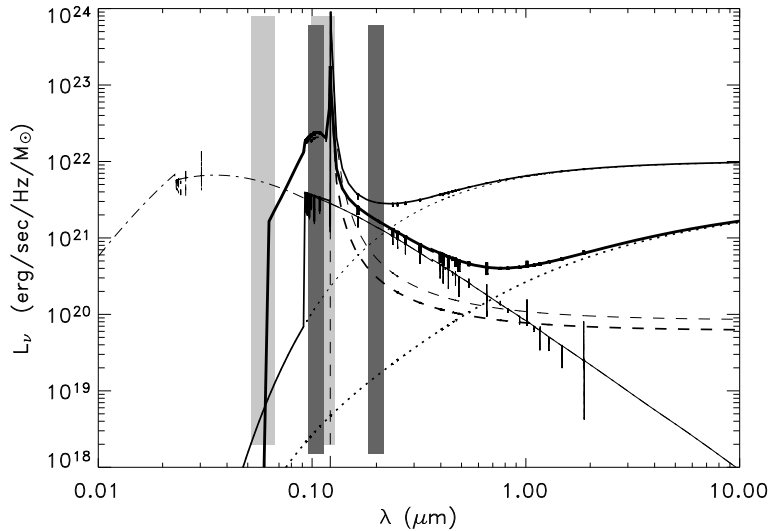} \vspace{0.5cm} \caption[]{ \scriptsize{ An
example of the Population III SED produced by a 1000 $M_\odot$
metals-free star from Santos et al (2002). Light and dark shaded
area show the range probed respectively by the J and K band DIRBE
filter at $z=20$ and 10. Dashed-dotted line shows the spectrum of
a 300 $M_\odot$ metal free star; thin solid line shows the part of
that SED longward of the Lyman limit. Dotted lines show the
free-free emission from the surrounding medium and dashed lines
show the nebulae emission; thick and thin lines of each type
correspond to the two extremes photon propagation considered by
Santos et al (2002). Thick solid lines correspond to the total SED
produced by the Population III emission in the two extreme cases.}
} \label{pop3santos}
\end{figure}

From the WMAP large-scale polarization measurements (Kogut et al
2003) it is known that the Universe's optical depth to
recombination is $\tau \simeq 0.2$ requiring the re-ionization to
occur by $z_*\simeq 20$. Numerical calculations suggest that when
a Population III star forms it is surrounded by a gaseous nebula
from the gas of the host halo which was not incorporated into the
stars (Bromm et al 1999). The nebula and the IGM would remain
neutral in the absence of Population III. In general, the
resultant SED would thus be made up of three components: 1) direct
stellar emission not absorbed by Lyman continuum; 2) Lyman
emission of the absorbed part of stellar SED, 3) free-free
emission from the IGM. Consequently, Santos et al (2002) consider
two extremes for the reprocessing of the ionizing radiation from
Population III: 1) Each Population III star is surrounded by a
dense nebula with all the reprocessing of ionizing radiation
taking place there, and 2) the nebula is optically thin to the
ionizing photons and all the reprocessing takes place in the IGM.
Fig. \ref{pop3santos} shows the SED from Santos et al (2002)
produced by a 1000 $M_\odot$ metal-free star for the two extreme
cases of photon propagation. Superimposed are also the J and K
filters redshifted to $z=20$ and 10.

There are several intuitive reasons why Population III produce
significant both CIB levels and fluctuations:

$\bullet$ Each unit mass of
Population III (if made of massive stars) emits $\gsim 10^5$ more
light than normal stars.

$\bullet$ A higher fraction of the total
luminosity would be redshifted into the NIR bands from hot objects
at these redshifts than from $z\sim 2-3$. E.g. the K band
redshifted to $z\simeq 20$ corresponds to rest wavelength of 1000
\AA, roughly the Lyman limit wavelength.

$\bullet$ Massive stars (assuming Population III were indeed
massive stars) radiate with a higher mean radiative efficiency,
$\epsilon =0.007$, than the present day stellar populations.

$\bullet$ Because their era was presumably brief, Population III
epochs contain less projected volume than the ordinary galaxy
populations spanning the epochs of Population I and I stars.
Hence, the larger relative fluctuations.

$\bullet$ Biasing is higher for Population III because they form
out of rarer regions which leads to the amplified correlations.

\subsection{Isotropic component of CIB}

It was noted by several authors that the near-IR CIB signals such
as e.g. detected in the 2MASS long integration data (Kashlinsky et
al 2003) may come from the Population III stars (Magliocchetti,
Salvaterra \& Ferrara 2003,
 Santos, Bromm \& Kamionkowski 2002,
Salvaterra \& Ferrara 2003, Kashlinsky et al 1999, Cooray et al
2004, Kashlinsky et al 2004). The argument can be shown to be
(almost) model-independent provided Population III stars were
indeed very massive.

Our discussion in Sections 7.2 through 7.4 follows Kashlinsky et
al (2004):

For a flat Universe the co-moving volume per unit solid angle
contained in the cosmic time interval $dt$ is $dV = c (1+z)^{-1}
d_L^2 dt$, where $d_L$ is the luminosity distance. Each Population
III star will produce flux $Lb_{\nu^\prime}(1+z)/4\pi d_L^2$,
where $b_\nu$ is the fraction of the total energy spectrum emitted
per unit frequency and $\nu^\prime = \nu(1+z)$ is the rest-frame
frequency. (The Population III SED is normalized so that $\int
b_\nu d\nu=1$). The co-moving mass density in these stars is $\int
M n(M,t)dM =\Omega_{\rm baryon} \frac{3H_0^2}{8 \pi G} f_*$, where
$f_*$ is the fraction of the total baryonic mass in the Universe
locked in Population III stars at time $t$. The net flux per unit
frequency from a population of such stars with mass function
$n(M)dM$ is given by:
\begin{equation}
\frac{d}{dt} I_\nu = \frac{\int L n(M,t) dM}{4\pi d_L^2} \;(1+z)
\; \langle{b_{\nu^\prime}}\rangle \;\frac{dV}{dt} \\=
\frac{c}{4\pi} \langle b_{\nu^\prime} \rangle \langle \frac{L}{M}
\rangle f_* \rho_{\rm baryonic} \label{didt}
\end{equation}
Here $\langle b_\nu \rangle \equiv \int L n(M,t) b_\nu dM/\int L
n(M,t) dM$  denotes the mean Population III SED averaged over
their initial mass function and $\langle \frac{L}{M} \rangle
\equiv \int L n(M,t) dM/\int M n(M,t) dM$. For Gaussian density
field $f_* \sim 5\times10^{-2} - 3\times 10^{-3}$ if on average
Population III formed in 2--3 sigma regions (see later).  Provided
that $L/M=$ constant, this result does not depend on the details
of the initial mass function of Population III stars (cf. Fig. 5
of Salvaterra \& Ferrara 2003). (Note that for the present day
stellar populations their mass-to-light ratio depends strongly on
stellar mass and $\langle \frac{L}{M} \rangle$ is much smaller
than the Population III value of $4\pi G m_p c/\sigma_T \simeq
3.3\times 10^4 L_\odot/M_\odot$ leading to substantially smaller
net fluxes). Assuming no significant mass loss (Baraffe et al
2001) during their lifetime $t_L$ for Population III stars, these
stars will produce CIB of amplitude:
\begin{equation}
\nu I_\nu = \frac{3}{8 \pi} \; \frac{1}{4\pi R_H^2} \;
\frac{c^5}{G} \; \epsilon \Omega_{\rm baryon} \frac{1}{t_L} \int
f_* \langle \nu^\prime b_{\nu^\prime} \rangle \frac{dt}{1+z}
\label{cib_bol}
\end{equation}

Note that in eq. \ref{cib_bol}, $c^5/G$ is the maximal luminosity
that can be achieved by gravitational processes. It enters there
because ultimately the nuclear burning of stars is caused by
gravity as they evolve in gravitational equilibrium with the
(radiation) pressure.  Eq. \ref{cib_bol}, has a simple meaning
illustrating its relative model-independence: the cumulative
bolometric flux produced is $L_{\rm max} = c^5/G \simeq
10^{26}L_\odot$ distributed over the surface of the Hubble radius,
$4\pi R_H^2$, times the model dependent, but more-or-less
understood, dimensionless factors. The spectral distribution of
the produced diffuse background will be determined by
$\langle\overline{b_\nu}\rangle\equiv \int f_*(1+z)^{-1} b_\nu
dt/\int f_*(1+z)^{-1}dt$ i.e. the mean SED averaged over the
Population III era redshifts. The value of $L_{\rm max}/4\pi
R_H^2$ is $\simeq 3\times10^8 \; {\rm nW/m^2/sr}$ so even with
small values of $\Omega_{\rm baryon}, \epsilon, f_*$, the net flux
from Population III stars could be indeed substantial.

The first stars had to form in the rare regions of the density
field that reached the turn-around over-density while the bulk of
the matter was still in linear regime (density contrast $<1$).
They would then turn-around and, if certain conditions are met,
collapse and form the first generation of stars. For Gaussian
density field, as is the case with the CDM models, this would
specify the value of $f_*$ at each time. It is perhaps a bit
unexpected that the WMAP polarization measurements indicated that
first stars started forming at $z_* = 20^{+10}_{-9}$, but the most
straightforward explanation is that the earliest stars formed out
of very high peaks of the density field leading to smaller $f_*$.

Once they turn around and begin to collapse the primordial clouds
will quickly heat up to their virial temperature, $T_{\rm vir}$,
and only efficient cooling will enable isothermal collapse when
gas pressure is smaller than gravity (or $T< T_{\rm vir}$) . The
details of what determines collapse and formation of Population
III stars are not yet clear, but molecular cooling seems critical
in the initial stages of metal-free collapse. In general $H_2$
cooling is effective at $T\gsim 400$ K (Santos et al 2002), but
some simulations suggest rotational $H_2$ cooling can effectively
dissipate binding energy of the cloud only if $T > 2000$ K
(Miralda-Escude 2003). The numbers that follow have thus been
evaluated for $T_{\rm vir} =$ 400K and 2000K. The relation between
comoving scale $r$, the mass contained by that radius and the
virial temperature at the onset of the collapse is:
\begin{equation}
M = 3.6\times 10^{11} \frac{\Omega}{0.3}
\left(\frac{r}{1h^{-1}{\rm Mpc}}\right)^3 h^{-1} M_\odot \;\; ;
\;\; T_{\rm vir} = 36 \left(\frac{\Omega}{0.3}\right)^{1/3}
\left(\frac{M}{M_\odot}\right)^{2/3} (1+z) \; {\rm K}
\label{tvirvsr}
\end{equation}

In the spherical linear approximation, the mass that at some early
epoch, $z_i$, had density contrast ($\delta_{\rm col}=1.68 $ times
the growth factor between $z_i$ and $z$) will collapse at redshift
$z$. For Gaussian density field, the regions that collapse
correspond to $\eta =\delta_{\rm col} /[\langle \delta_M^2 \rangle
]^{\frac{1}{2}} >1$ standard deviations the value of $f_* \simeq
{\rm erfc}(\eta/\sqrt{2})$. The value of the RMS density contrast,
$[\langle \delta_M^2 \rangle ]^{\frac{1}{2}}$ at redshift $z$ can
be evaluated given the power spectrum of the assumed model
($\Lambda$CDM) normalized to large-scale CMB data and can be
estimated from Fig. \ref{p3} and eqs. \ref{tvirvsr}. Population
III stars had to form out of $\eta \sim 2-3$ sigma rare
fluctuations which for Gaussian distribution correspond to $f_*$
varying between $\sim 5\times 10^{-2}$ and $\sim 3\times 10^{-3}$;
departures from spherical symmetry would accelerate the collapse
thereby decreasing $\eta$ and increasing $f_*$. Because $\eta$ is
a decreasing function of decreasing $z$, and $f_*$ is a (rapidly)
increasing function of decreasing $\eta$, the average
$\overline{f_*}$ will be dominated by the late times of Population
III evolution.

\begin{figure}[h]
\centering \leavevmode \epsfxsize=0.75 \columnwidth
\epsfbox{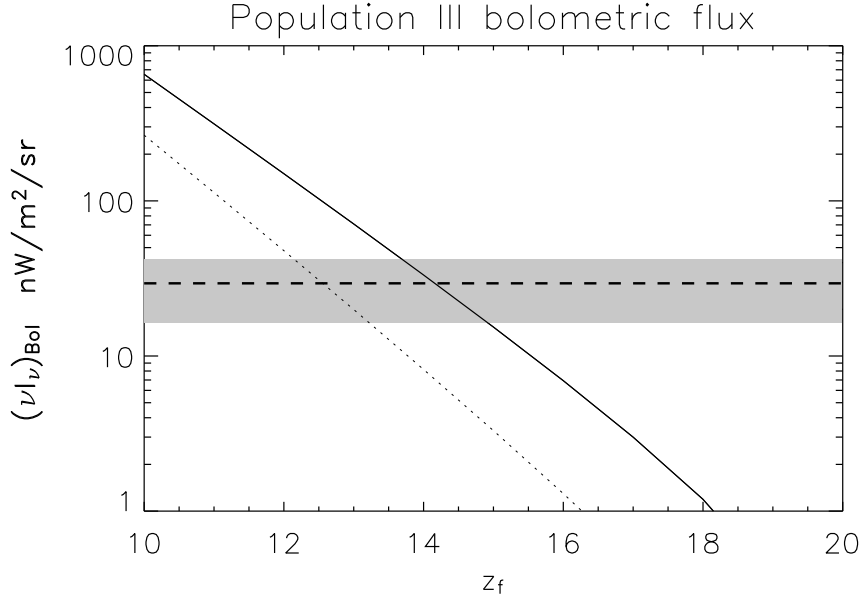} \vspace{0.5cm} \caption[]{
\scriptsize{Bolometric flux produced from the Population III era
lasting until the redshift $z_f$ shown along the horizontal axis.
Solid line assumes Population III stars form in collapsing regions
with $T_{\rm vir} = 400$ K and dashed line corresponds to $T_{\rm
vir} = 400$ K. Shaded area shows the bolometric flux of the CIB
excess at near-IR evaluated from Fig. \ref{excess} with its
uncertainty.} } \label{pop3bol}
\end{figure}

Fig. \ref{pop3bol} shows the total bolometric flux evaluated
according to eq. \ref{cib_bol} evaluated for $\Lambda$CDM model
for $T_{\rm vir}=400$ K(solid line) and $T_{\rm vir} = 2000$ K
plotted vs the redshift marking the end of the putative Population
III era, $z_f$. The lines in the figure assume that efficiency of
Population III formation is regulated only by parameter $f_*$. The
prediction for the total bolometric flux produced by Population
III is fairly robust. On the other hand, the way it is distributed
among the various bands or its energy spectrum depends on the
details of how Lyman-$\alpha$ ionizing photons are absorbed and
re-emitted along the line-of-sight. Santos et al (2002) model the
redistribution of the total flux produced from the Population III
era and produce good fits to the observed near-IR CIB at J and K
bands. Clearly, in order to account for the near-IR CIB excess
measured at 1--4 \um, the diffuse bolometric flux from Population
III stars must exceed the observed excess; any additional flux
will then have to be redistributed to different wavelengths. This
can be achieved if the Population III stars were massive and their
era lasted at least until $z_f \sim 15$. Interestingly, the
existence of the CIB excess at J band suggests that if this excess
originated from the Population III era, the latter should have
extended to $z\lsim 14$.

This discussion suggests that Population III era could have
produced CIB at levels comparable to those in Fig. \ref{excess},
but it could also have produced substantially lower (or higher)
fluxes, while the measurements of the CIB mean levels can be
significantly affected by the systematics and be mistaken for the
various residual errors. On the other hand, as is shown below,
Population III stars, whose emission arises at epochs when the
spatial spectrum of galaxy clustering is not yet evolved, should
have produced a unique and measurable signature via the CIB
fluctuations. It is that signature, both its spectrum in the
angular and energy frequency domain, that could provide the
ultimate insight into the Population III epochs (Cooray et al
2004, Kashlinsky et al 2004).

\subsection{Contribution to anisotropies in CIB}

In order to evaluate the amplitude and spectrum of the CIB from
Population III era we start with the density field of the
$\Lambda$CDM model. For reference, a 1 arcminute angular scale
subtends comoving scale $\simeq 1.5 h^{-1}$Mpc between $z=6$ and
$z=15$ for WMAP cosmological parameters and equation of state with
$w=-1$. As Fig. \ref{p3} shows, on sub-arcminute scales the
density field is in quasi-linear to non-linear regime (density
contrast $\gsim 0.2$) for the $\Lambda$CDM model. There the
spectrum due to clustering evolution was modified significantly
over that of the linear $\Lambda$CDM spectrum, and the
fluctuations amplitude has increased from its primordial value
especially since the effective spectral index on these scales for
$\Lambda$CDM model was $n \lsim -2$.

At $z=20$ the Universe is $\sim 2 \times 10^8$ years old which is
much larger than the age of the individual Population III stars,
$t_L \sim 3 \times 10^6$ years. If Population III era lasted for
only one generation forming at $z_*$, the CIB fluctuations will be
$\delta_{\rm CIB} \sim \sqrt{\pi}\Delta(qd_A^{-1};z)|_{z_*}$,
where $\Delta(k)$ is given by eq. \ref{Delta3}. In what follows we
define $t_*$ as the time-length of the period in which Population
III stars were the dominant luminosity sources in the Universe.

Several points are worth noting about eq. \ref{limber_del2} when
applied to Population III era: 1) For given amplitude of the mean
CIB levels from Population III, the value of $\Delta(k)$ is
inversely proportional to $\sqrt{t_*}$ and thus measures the
duration of the Population III era. The density perturbations grow
with decreasing $z$, thus most of the contribution to the integral
in eq. \ref{limber} comes from the low end of $z$ over some range
of $b_\nu$. At some wavelengths the overall dependence on $t_*$
wins out at some (shortest NIR) wavelengths and the longer the
Population III phase, the smaller are the relative fluctuations of
the CIB from them. 2) In the Harrison-Zeldovich regime of the
power spectrum, $P_3 \propto k$, one would have $\delta_{\rm CIB}
\propto q^{1.5}$. 3) The transition to the Harrison-Zeldovich
regime occurs in the linear regime at all relevant redshifts and
happens at the co-moving scale equal to the horizon scale at the
matter-radiation equality. All this would allow probing of the
Population III era, its duration, and the primordial power
spectrum at high redshifts on scales that are currently not probed
well. Interestingly, short duration of the Population III will
lead to smaller CIB flux, but larger relative fluctuations and
vice versa.

In addition to the small angular scale increase due to non-linear
gravitational evolution, the fluctuations in the clustering
distribution of Population III systems will be amplified because,
within the framework of the $\Lambda$CDM model these systems had
to form out of rare peaks of the primordial density field (Kaiser
1984, Jensen \& Szalay 1986, Politzer \& Wise 1996, Kashlinsky
1987, 1991, 1998). This leads to biased (enhanced) 2-point
correlation function of the Population III regions, $\xi_b$ over
that of the underlying density field, $\xi(r) =\frac{1}{2\pi^2}
\int P_{\Lambda{\rm CDM}}(k) k^2 j_0(kr) dk$. In the $\Lambda$CDM
model the Population III stars at $z\sim 10-20$ will form in
regions having $\eta>1$. In this limit the biasing factor given by
eq. 29 of Kashlinsky (1998) reduces to check:
\begin{equation}
\xi_b=\frac{1}{2} [ \exp\left( \frac{\eta^4}{\delta_{\rm col}^2}
\xi\right) -1 ]
\label{xi_biased}
\end{equation}
In the limit of $\xi \rightarrow 0$ this reduces to the more
familiar form of $\xi_b(r) \simeq \frac{\eta^4}{2 \delta_{\rm
col}^2} \xi(r)$. We adopt eq. \ref{xi_biased} in evaluating the
numbers below.

There will also be shot-noise fluctuations due to individual
Population III systems entering the beam. The relative magnitude
of these fluctuations will be $N_{\rm beam}^{-1/2}$, where $N_{\rm
beam}$ is the number of the Population III systems within the
beam. This component may be important at very small angular
scales, where $N_{\rm beam} \sim 1 $, and will contribute to the
power spectrum: $P_{\rm SN} = \frac{1}{n_2}$, where $n_2 = c \int
n_* d_L^2 (1+z)^{-1} dt$ is the projected angular number density
of Population III systems. The detection of the shot-noise
component in the CIB power spectrum at small angular scales will
give a direct measure of both the duration of the Population III
era and constrain the makeup and masses of the Population III
systems. Calculations show that unless the Population III era was
very short, the shot noise correction to the CIB would be small on
scales greater than a few acrseconds (Kashlinsky et al 2004).

Because of the Lyman absorption by the surrounding matter prior to
reionization, essentially all the flux emitted by Population III
stars at rest wavelength less than 912 \AA\ will be absorbed by
the Lyman continuum absorption in the surrounding medium (Gunn \&
Peterson 1965, Yoshii \& Peterson 1994, Haiman 2002, Santos 2004)
and the contribution to the CIB will be cut-off at $\lambda < 912
(1+z_f)$ \AA . At longer wavelengths the flux, including emission
from the nebulae around each star, will propagate without
significant attenuation. Population III stars emit as black bodies
at $\log T \simeq 4.8-5$ (Schaerer 2002) and the Lyman limit at
$z\sim 20$ is shifted to $\sim 2 \mu$m at $z=0$. The 2MASS J band
filter contains emission out to 1.4 \um\, and represents the
shortest wavelength where excess in the CIB over that from
ordinary galaxies has been measured (Kashlinsky \& Odenwald 2000,
Cambresy et al 2001, Kashlinsky et al 2002, Odenwald et al 2003).
If the measured excesses in CIB fluctuations
 and isotropic component at J band are indeed attributable
to Population III, their era must have lasted until $z \lsim 14$.

Fig. \ref{del_pop3} shows the resultant CIB fluctuations from
Population III stars from 1.25 to 8 \um. The
fluctuations are normalized to reproduce the CIB excess at 2.2 \um
of 10 \nwm2sr . All Population III stars were assumed to start
forming at $z=20$, but differently colored lines correspond to
different values of $z_f$, the redshift of the end of Population
III era. Because of the Lyman continuum absorption there would be
no appreciable CIB fluctuations at J band if $z_f \lsim 14$ (no
blue lines) and the contribution at J band will come from $z \sim
13$, rather than $z=20$. Taken at face value, the presence of
excess CIB fluctuations in J band indicates that the Population
III stars were possibly the dominant sources of luminosity until $z_f
\lsim 14$.

\subsection{Can CIB anisotropies from Population III be measured?}

Both DIRBE and 2MASS data indicate CIB anisotropies at amplitudes
larger than the contribution from ordinary galaxy populations
(Kashlinsky \& Odenwald 2000, Kashlinsky et al 2002, Odenwald et
al 2003) and are consistent with significant contributions due to
Population III. However, because the measured signal contains
contributions from remaining galaxies (all galaxies for DIRBE and
$K\geq 19$ galaxies for 2MASS), it is difficult to isolate the
contribution from the Population III stars.

\begin{figure}[h]
\centering \leavevmode \epsfxsize=1. \columnwidth
\epsfbox{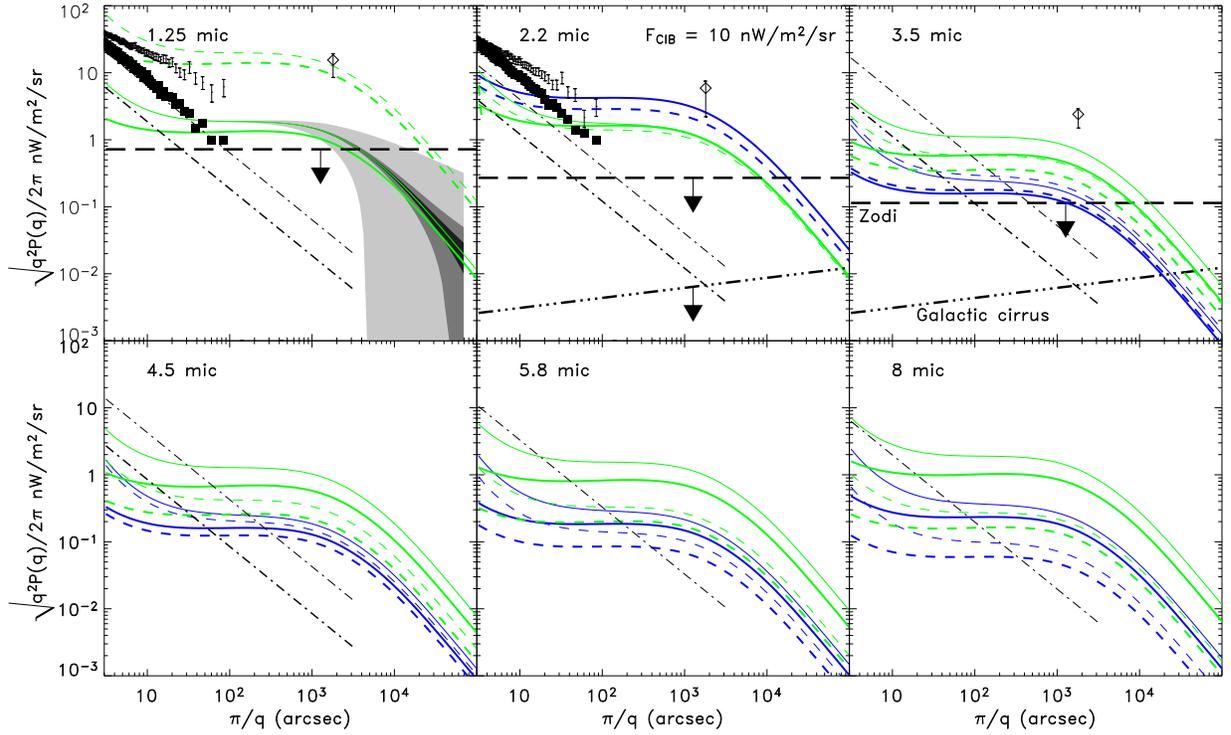} \caption[]{ \scriptsize{ CIB fluctuations,
$\sqrt{q^2P_2/2\pi}$, from Population III stars forming at
$z_*=20$ are shown with colored lines. Blue and green lines
correspond to Population III era ending at $z_f = 15$ and 10
respectively. Thick and thin colored lines correspond to the two
extremes of Population III SED modeling after Santos et al (2002)
and shown in Fig. \ref{pop3santos}. Solid and dashed colored lines
correspond Population III forming in haloes with $T_{\rm vir}\geq
400$K and $T_{\rm vir}\geq 2000$K. Filled squares at 1.25 and 2.2
\um show residual atmospheric fluctuations from ground based 2MASS
measurements after 1 hour of exposure from Odenwald et al (2003).
Thick long dashes are adopted from Kashlinsky et al (2004) and
denote the upper limit on zodiacal light fluctuations from
\cite{abraham} scaled to the corresponding band. Thick
dashed-triple-dotted line denotes cirrus fluctuations: these are
upper limits at 2.2 micron and an estimate from \cite{kiss} at 3.5
micron. Dot-dashed lines correspond to shot noise from galaxies
fainter than AB magnitude = 22 (thickest), and $m_{\rm AB}=20$
(thin). Diamonds with error bars show the CIB fluctuations at
$\sim 0.7^\circ$ from the COBE DIRBE fluctuations analysis
\cite{ko}. Note that because of the large DIRBE beam, these
results include contribution from all galaxies as well as other
sources such as Population III. The K band CIB fluctuation from
deep integration 2MASS data is shown at the largest scale
accessible to those data with $\times$; the 2MASS data shown in
the figure were taken for the the patches for which galaxies were
removed brighter than $K \simeq 19$ (Kashlinsky, Odenwald et al
2002, Odenwald, Kashlinsky et al 2003). The cosmic variance
1-sigma uncertainty is shown with shaded regions for 1.25 \um. The
darkest shade corresponds to a total of 1,000 deg$^2$ observed,
and the lighter shade corresponds to a total of 100 deg$^2$. } }
\label{del_pop3}
\end{figure}

Population III stars, if massive, should have left a unique and
measurable signature in the near-IR CIB anisotropies over angular
scales from $\sim 1$ arcminute to several degrees as Fig.
\ref{del_pop3} shows. Detection of these fluctuations depends on
the identification and removal of various foreground emission (and
noise) contributions: atmosphere (for ground-based measurements),
zodiacal light from the Solar system, Galactic cirrus emission,
and instrument noise.

Kashlinsky et al (2004) discuss observability of the Population
III CIB fluctuations vs the confusion arising from the various
foregrounds. The confusion may arise from 1) Galactic cirrus
emission, 2) galaxies with ordinary stellar populations not
removed from the data, 3) zodiacal light fluctuation, and 4)
atmospheric fluctuations from ground measurements. They estimated
the contributions from the various foregrounds to the fluctuations
from 1 to 3.5 \um. The numbers from their discussion are plotted
in the top panels of Fig. \ref{del_pop3}.

Below is brief discussion of the various contaminants when
searching for the CIB fluctuations from the Population III era:

$\bullet$ {\it Atmosphere}: Filled squares in Fig. \ref{del_pop3}
show the residual atmospheric fluctuations at 1.25 and 2.2 \um\ on
sub-arcminute scales after one hour of integration from one of the
deep 2MASS fields measurements (Odenwald, Kashlinsky et al 2003).
Atmospheric gradients become important on larger angular scales
($\gsim 0.05-0.1^\circ$) and their effects can be highly variable
on a wide range of time scales (e.g. Adams \& Skrutskie 1995,
http://pegasus.phast.umass.edu/adams/airglowpage.html) making
detection of arcminute and degree scale CIB fluctuations difficult
in ground based observations. Yet these problems can be completely
avoided with space--based experiments.

$\bullet$ {\it Ordinary galaxies} must be eliminated to
sufficiently faint levels so that the remaining fluctuations in
their cumulative emission are sufficiently small. The
dashed-dotted lines in Fig. \ref{del_pop3} show the shot-noise
fluctuations estimated from the observed deep counts from 1 to 8
\um. Thin lines correspond to galaxies brighter than $m_{\rm
AB}=20$ identified and removed and thick lines when the same is
possible out to $m_{\rm AB}=22$. One can see that ideally one
would have to identify galaxies out to $m_{\rm AB}\gsim 23-24$ in
order to measure the CIB fluctuations from Population III era
across the entire range of scales.

$\bullet$ {\it Zodiacal light} emission from interplanetary dust
(IPD) is the brightest foreground at most IR wavelengths over most
of the sky. There are some structures in this emission associated
with particular asteroid families, comets, and an earth--resonant
ring, but these structures tend to be confined to low ecliptic
latitudes or otherwise localized. The main IPD cloud is generally
modeled with a smooth density distribution. Observationally,
intensity fluctuations of the main IPD cloud have been limited to
$<0.2$\%
 at 25 $\mu$m \cite{abraham}. Extrapolating this limit to other
DIRBE wavelengths using the observed mean high--latitude zodiacal
light spectrum from COBE DIRBE (Kelsall et al 1998), Kashlinsky,
Arendt et al (2004) arrive at the limits shown in Figure
\ref{del_pop3}. Because the Earth is moving with respect to
(orbiting within) the IPD cloud, the zodiacal light varies over
time. Likewise, any zodiacal light fluctuations will not remain
fixed in celestial coordinates. Therefore repeated observations of
a field on timescales of weeks to months should be able to
distinguish and reject any zodiacal light fluctuations from the
invariant Galactic and CIB fluctuations. The upper limit on
zodiacal light fluctuations can be evaluated also at the bottom
panels of Fig. \ref{del_pop3} using extrapolations of the DIRBE-
measured zodiacal light spectrum from Fig. \ref{foregrounds}. The
zodiacal light is expected to be very smooth at all the near-IR
bands.

$\bullet$ {\it Galactic cirrus}. Intensity fluctuations of the
Galactic foregrounds are perhaps the most difficult to distinguish
from those of the CIB. Stellar emission may exhibit structure from
binaries, clusters and associations, and from large scale tidal
streams ripped from past and present dwarf galaxy satellites of
the Milky Way. At long IR wavelengths, stellar emission is
minimized by virtue of being far out on the Rayleigh--Jeans tail
of the stellar spectrum (apart from certain rare classes of dusty
stars). At near-IR wavelengths stellar emission is important, but
with sufficient sensitivity and angular resolution most Galactic
stellar emission, and related structure, can be resolved and
removed. IR emission from the ISM (cirrus) is intrinsically
diffuse and cannot be resolved. Cirrus emission is known to extend
to wavelengths as short as 3 $\mu$m (Arendt et al 1998).
Statistically, the structure of the cirrus emission can be modeled
with power--law distributions. Using the mean cirrus spectrum,
measurements made for the cirrus fluctuations in the far-IR with
ISO (Kiss et al 2003) were scaled all the way to 3.5 $\mu$m by
Kashlinsky, Arendt et al (2004), providing the estimated
fluctuation contribution from cirrus that is shown in Figure
\ref{del_pop3}. The extrapolation to shorter wavelengths is highly
uncertain, because cirrus (diffuse ISM) emission has not been
detected at these wavelengths, and the effects of extinction may
become more significant than those of emission, but cirrus
contribution is generally expected to be several orders of
magnitude lower than the CIB fluctuations (Arendt et al 1998). We
do not show the expected cirrus fluctuations at the wavelengths
corresponding to the bottom panels of Fig. \ref{del_pop3} as the
simple extrapolations of the cirrus SED from Fig.
\ref{foregrounds} may be inadequate due to the presence of PAH
emission at these wavelengths.

Figure \ref{del_pop3} shows that CIB fluctuations from Population
III would be the dominant source of diffuse light fluctuations on
arcminute and degree scales even if Population III stars epoch was
briefer, and their diffuse flux smaller, than the current CIB
numbers suggest. Their angular power spectrum should be very
different from other sources of diffuse emission and its
measurement thus presents a way to actually discover Population
III and measure the duration of their era and their spatial
distribution. The latter would provide direct information on
primordial power spectrum on scales and at epochs that are not
easily attainable by conventional surveys. This measurement would
be imperative to make and is feasible with the present day space
technology.

\begin{figure}[h]
\centering \leavevmode \epsfxsize=1. \columnwidth
\epsfbox{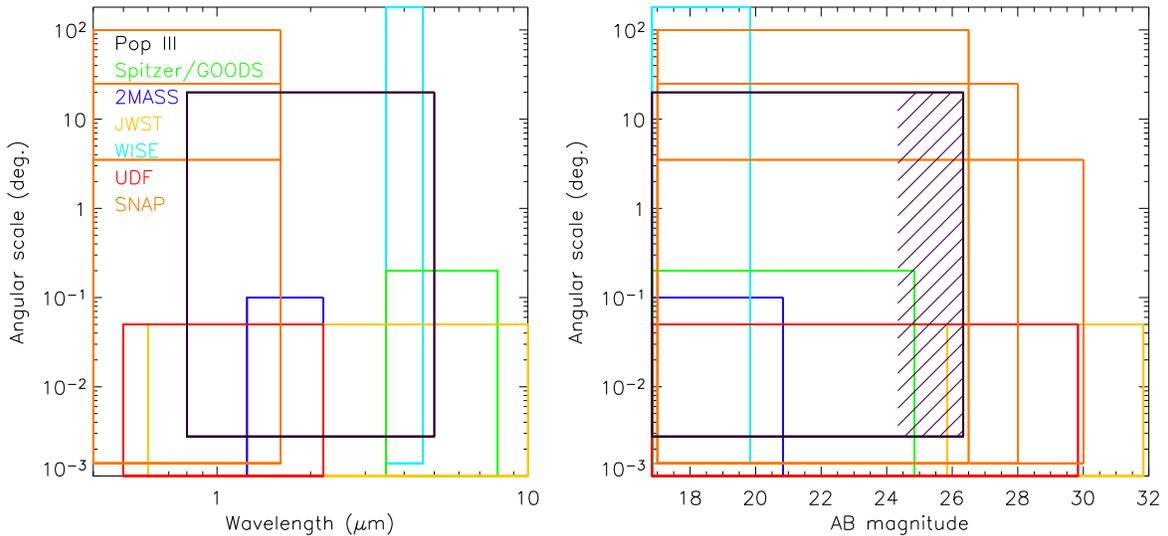} \caption[]{ \scriptsize{Black rectangular box
shows the regions of the parameters required for a space mission
to measure the CIB anisotropies from Population III. Cross-hatched
region in the right panel displays the magnitude range where
Population III stars are expected to dominate CIB anisotropies.
Other colors correspond to the currently existing datasets or
planned NASA missions which can be utilized for this purpose: red
is the Hubble Deep Field from HST
(http://www.stsci.edu/ftp/science/hdf/hdf.html), dark blue is for
the deep 2MASS data \cite{komsc,okmsc}, green is the NASA's {\it
Spitzer} GOODS project (http://www.stsci.edu/ftp/science/goods/),
light blue corresponds to the planned NASA WISE MIDEX mission
(http://www.astro.ucla.edu/~wright/WISE), yellow corresponds to
the field-of-view of the NASA James Webb Space Telescope planned
for the end of the decade (http://jwst.gsfc.nasa.gov) and orange
color to the three surveys planned by SNAP (http://snap.lbl.gov).
} } \label{future}
\end{figure}

The previous discussion suggests that CIB fluctuations from the
Population III era are observable, but because of the large-scale
atmospheric gradients, space observations are required for
detection. The following is needed:

1) In order to make certain that the signal is not contaminated by
distant ordinary galaxies with normal stellar populations, one
would need to conduct a deep enough survey in order to identify
and eliminate ordinary galaxies from the field. In practice, as
Fig. \ref{cib_dc} indicates, going to $m_{\rm AB}\simeq 24$ would
be sufficient.

2) A direct signature of Population III signal is that there
should be a Gunn-Peterson Lyman-break feature in the spectral
energy distribution of the CIB fluctuations, i.e. at wavelengths
$\leq 912 (1+z_f)^{-1}$ \AA. That spectral drop would provide an
indication of the epoch corresponding to the end of the Population
III era. At longer wavelengths the spectral energy distribution of
the CIB fluctuations would probe the history of energy emission
and its re-distribution by the concurrent IGM. At wavelengths
$\gsim 10 \mic$ zodiacal light fluctuations may become dominant.
Hence one would want $0.9 \mic \lsim \lambda \lsim 5-10 \mic$.

3) On angular scales from a few arcminutes to $\sim 10^\circ$
Population III would produce CIB anisotropies with a distinct and
measurable angular spectrum, but its measurement will be limited
by the cosmic or sampling variance. Calculations show that for
reliable results on scales up to $\theta$ one would need to covers
area a few times larger (Kashlinsky et al 2004). Shaded regions in
the upper left panel of Fig. \ref{del_pop3} show that in order to
get reliable results on scales up to $\theta\gsim 1^\circ$ one
would need to observe areas of $\gsim 10^\circ$ across.

Fig. \ref{future} shows the parameters a space-based survey should
have in order to probe Population III CIB anisotropies and
compares it with the currently planned space missions. The left
panel shows the projection on the (angular scale)--wavelength
plane; the right panel shows the projection  onto the (angular
scale)--sensitivity plane. Clearly, such a Population III
experiment needs to have a wide field-of-view (FOV) and at the
same time have good angular resolution (preferably up to a few
arcseconds). The sensitivity requirements were taken to reach the
ability to identify and eliminate from the data ordinary galaxies
that at all the near-IR wavelengths contribute $\sigma_{\rm sn}
\lsim 1$ \nwm2sr ; Fig. \ref{sigma_conf} shows that this is
reachable at $m_{\rm AB} \gsim 24$ which with modern technologies
and a $\sim 0.5$ m mirror is achievable from space in $\sim 1$ hr
integrations. None of the currently planned surveys cover
perfectly the required range of parameters, although a combination
of various missions instruments may do a part of the job. Cooray
et al (2004) propose a rocket experiment similar to an earlier
experiment by Xu et al (2002) with an FOV of a few degrees and
$\sim$ arcmin resolution, whereas Kashlinsky et al (2004) argue
for a dedicated space mission based on the existing detector
technologies.

\section{Snapshot of the future}

It is important to resolve as much of the CIB as possible into
individual contributors and isolate the contributions to the total
mean flux and anisotropies by the cosmic epoch. It is also
important to isolate the epochs from which luminous sources
contribute to the part of the CIB which remains unresolvable by
contemporary instruments and techniques. Significant progress in
these directions is expected to happen with the current (Spitzer)
and planned space missions which should cover larger scales, have
better angular resolution and go deeper across the entire range of
the IR spectrum. In this section we give a brief overview of the
IR missions and their potential for CIB studies.

\subsection*{Spitzer}

The Spitzer telescope (initially known as SIRTF - Space InfraRed
Space Facility) has been launched aboard a Boeing Delta II rocket
in August 2003. The telescope carries three instruments: InfraRed
Array Camera (IRAC), Multiband Imaging Photometer (MIPS) and
InfraRed Spectrograph (IRS). IRAC is a four-channel camera that
provides simultaneous 5.12$^\prime \times 5.12^\prime$ images. The
pixel scale is 1.2$^{\prime\prime}$ in all IRAC bands. The
Multiband Imaging Photometer for SIRTF (MIPS) is designed to
provide very deep imaging and mapping at 24, 70, and 160 \um. In
integrations of 2000 seconds, it reaches 5$-\sigma$ detection
limits at these wavelengths
 of 0.2, 0.5, and 8 mJy, respectively. The pixel scale is
 2.4$^{\prime\prime}$ at 24 \um, 4.5$^{\prime\prime}$ at
 70 \um\ and 15$^{\prime\prime}$ at 160 \um.

On the observational side the following surveys have already been
approved by the Spitzer Legacy
\footnote{http://ssc.spitzer.caltech.edu/legacy/} and Director's
Discretionary \footnote{http://ssc.spitzer.caltech.edu/fls/}
programs:

\noindent $\bullet$ {\bf First Look Survey}: The extragalactic
component of the First Look Survey (FLS) is intended to reach the
1-sigma sensitivities of $\simeq 10\mu$J with IRAC and 1 to 27 mJy
with MIPS over a 4 sq. deg. and a 1 sq. deg. field. A smaller 0.25
sq. deg. region will be covered with 10 times the nominal
integration time. Although the survey is very shallow, the 160
\um\ data are already expected to be confusion limited even for
this shallow survey.

\noindent $\bullet$ {\bf SWIRE}: The SWIRE project (Lonsdale et al
2004) will cover 7 fields between 5 and 15 sq. deg. in size with
sensitivities that are several times better than the FLS. A stated
goal of the SWIRE project is to investigate galaxy formation in
the $0.5 < z < 2.5$ range.

\noindent $\bullet$ {\bf GOODS}: The GOODS project
\footnote{http://www.stsci.edu/science/goods} will observe 300
square arcmin divided into two fields: HDF North and Chandra Deep
Field South. Mean exposure time per position is 25 hours per band
with IRAC. At 24 \um, MIPS will obtain 10 hour exposures, if on
orbit experience shows that this will be a distinct advantage over
planned Guaranteed Time Observations (GTO). MIPS GTO proposals for
these regions at 70 and 160 \um\ are sufficiently deep, that the
GOODS project will not duplicate these observations. If it will
prove useful, the GOODS observations will also include an
ultra-deep field in the HDF-N region with up to 100 hours of
integration. The GOODS team intends to use the resolved galaxy
counts, which should include $L^*$ galaxies to redshifts to $z$ =
5, to establish the best lower limits on the 3.6 -- 24 \um\ mean
CIB.

On the data analysis side of the long term programs, the National
Science Foundation has approved:

\noindent  $\bullet$ {\bf LIBRA}, or Looking for Infrared
Background Radiation Anisotropies,
\footnote{http://haiti.gsfc.nasa.gov/kashlinsky/LIBRA} is an
NSF-supported 5 year project to measure CIB anisotropies from
early epochs. The integrated light of all galaxies in the deepest
counts to date fails to match the observed mean level of the CIB,
indicating a significant high-redshift contribution to the CIB.
The project will use the Spitzer and deep 2MASS data at 1 - 160
\um\ to measure the spatial fluctuations of this residual high-$z$
portion of the CIB. Analysis of these spatial fluctuations will
provide information on the luminosity of the early universe, and
on the developing structure of galaxies and clusters of galaxies
at early epochs.

\subsection*{Astro-F}

Japan's Astro-F or the Infrared Imaging Surveyor (IRIS) is an
infrared astronomy mission scheduled for launch in 2005 into a
sun-synchronous polar orbit. IRIS employs a 70 cm telescope cooled
to 6 K using liquid helium. ASTRO-F is designed for advanced
surveys with two observation modes. The first is a general survey,
one rotation in one orbital period, which is used for the all sky
survey. The second is a pointing mode for imaging and
spectroscopic observations of a limited region of the sky. Two
focal-plane instruments are installed. One is Far-Infrared
Surveyor (FIS) and the other is Infrared Camera (IRC). The FIS is
a photometer optimized for all-sky survey with far-infrared
arrays, and is expected to produce catalogs of infrared sources
with much better sensitivity and higher angular resolution than
the IRAS. The FIS can be operated as an imager or a
Fourier-transform spectrometer in the pointing mode. The IRC is a
three-channel camera that covers the wavelength bands from 2\um to
25\um, and has the capability to perform low-resolution
spectroscopy with prisms/grisms on filter wheels. The field of
view of the IRC is 10 arcmin and the spatial resolution is
approximately 2 arcsec. Large format arrays are used to attain the
deep survey with wide field and high angular resolution. IRC
observations are carried out only in pointing mode.

The detection limits will be 1 - 100 $\mu$Jy in the near-mid
infrared and 10-100 mJy in the far infrared. It is planned to
conduct a large area ($\sim$4 square degree) survey at the K and L
bands for CIB studies.

\subsection*{WISE}

WISE (Wide-field Infrared Survey Explorer)
\footnote{http://wise.ssl.berkeley.edu/} has been selected as
NASA's next Medium-Class Explorer (MIDEX) mission with a tentative
launch date in 2008. It will have four bands at 3.5, 4.6, 12 and
23 \um with $\simeq 3^{\prime\prime}$ pixels. During its 6 month
mission WISE satellite will provide an all-sky coverage about
1,000 more sensitive than IRAS. It should detect ULIRGs out to
$z\sim 3$ and bright $L_*$ IR galaxies out to $z\sim 1$.

\subsection*{Herschel and Planck}

The European Space Agency's Herschel Space Observatory
\footnote{http://astro.esa.int/herschel/} (formerly called Far
Infrared and Submillimetre Telescope, or FIRST), with an
anticipated launch in 2007, will be the first space observatory
covering the full far infrared and sub-millimeter waveband, and
its passively cooled telescope will have a 3.5 meter diameter
mirror. It will be located at L2 and will provide photometry and
spectroscopy in the 57 to 670 \um range (Pilbratt 2004). The
Observatory will have 3 instruments: The Photodetector Array
Camera and Spectrometer (PACS), the Spectral and Photometric
Imaging REceiver (SPIRE) instrument, and the Heterodyne Instrument
for the Far Infrared (HIFI) instrument. SPIRE is made of two
sub-instruments: a three-band imaging photometer operating at 250,
360 and 520 \um, and an imaging Fourier Transform Spectrometer
(FTS) covering 200-670 \um. The field of view of the photometer
will be $4^\prime \times 8^\prime$, observed simultaneously in the
three spectral bands. An internal beam steering mirror allows
spatial modulation of the telescope beam and will be used to
jiggle the field of view in order to produce fully-sampled images.
Observations can also be made by scanning the telescope without
chopping. In addition to pointed imaging and spectroscopy of
distant galaxies, it is planned that the telescope will conduct
deep confusion limited surveys of $\gsim 100$ deg$^2$ of the sky.

Planck satellite \footnote{http://www.rssd.esa.int/Planck} is
designed mainly for CMB measurements and will be launched together
with Herschel. Its High Frequency Instrument (HFI) has bands at
350, 550, 850 and 1380 \um with resolution of 10 to 5 arcminutes
and will be useful for CIB studies. Its all sky survey will
probably be too shallow to constrain galaxy evolution, but will
give good measurements of the bright end of galaxy counts. With
its scanning strategy Planck will survey some high Galactic
latitude areas much deeper (a factor $\sim 3$ in flux) and they
could be useful for studies of the far-IR to sub-mm CIB
fluctuations at arcminute to degree angular scales.

\subsection*{JWST}

Successor to the Hubble Space Telescope, the James Webb Space
Telescope \footnote{http://www.jwst.nasa.gov} is a large,
infrared-optimized space telescope scheduled for launch in August,
2011 (Sabelhaus \& Decker 2004). JWST will have a large mirror,
6.5 meters in diameter. It will reside in an L2 Lissajous orbit.
It will have two instruments of direct relevance to CIB science:
the NIR camera (NIRCAM) and the MIR Instrument (MIRI). The NIRCAM
will cover wavelengths from 0.6 to 5 \um with 0.03 arcsec/pixel
resolution from 0.6 to 2.3 \um and 0.064 arcsec/pixel at 2.3 to 5
\um. The MIRI will provide the JWST with imaging and spectroscopy
at wavelengths from 5 through 27 microns with $\sim 0.2$
arcsec/pixel. The FOV will be $\simeq 1.5^\prime \times
1.5^\prime$. The telescope will be able to detect Lyman break
galaxies out to $z\gsim 10$ and supernovae out to $z\gsim 20$
(Gardner et al 2004). Its science goals in cosmology will be to
identify the end of dark ages and the assembly of galaxies at
observer bands from visible to mid-IR.

\subsection*{SNAP}

The SuperNovae Acceleration Probe (SNAP), or the Joint Dark Energy
Mission, \footnote{http://snap.lbl.gov} is planned as a combined
DOE and NASA space mission designed to probe the equation-of-state
of the Universe with high $z$ supernovae out to $z\sim 2$
(Aldering et al 2004). It is scheduled tentatively for launch
early in the next decade. SNAP will cover wavelengths from visible
to 1.7 \um and will be useful for NIR CIB studies. Panoramic, wide
and deep surveys are currently planned to cover up to $\sim 10^4,
1000$ and 15 deg$^2$ and will go to AB magnitudes of $\sim 26.5,
28$ and 30 respectively. Fig. \ref{future} shows that these SNAP
surveys at the longest SNAP wavebands should be very useful in
uncovering the potential high-$z$ contribution to the CIB
fluctuations at $\lsim 1.5 $\um including from Population III.

\subsection*{SCUBA-2, SOFIA, ALMA}

SCUBA-2
\footnote{http://www.jach.hawaii.edu/JACpublic/JCMT/scuba/scuba2/}
will replace the current SCUBA on the James Clerk Maxwell
Telescope in 2006. It will have a total of $\sim 10,000$
bolometers and offer simultaneous imaging of a 50 arcmin$^2$ FOV
at 450 and 850 \um mapping the sky up to 1,000 times faster than
the current SCUBA array (Audley et al 2004).

The Stratospheric Observatory for Infrared Astronomy (SOFIA)
\footnote{http://sofia.arc.nasa.gov/Sofia/} is under development
by NASA, and is expected to have its initial science operations in
early 2007.  This observatory, with its 2.5 m telescope, operates
at 41,000 to 45,000 feet aboard a Boeing 747 aircraft,  providing
access to much of the far-IR spectrum.   It will provide
complementary capabilities to the Spitzer Space Telescope,
permitting higher spatial resolution, and hence lower confusion
limits at far infrared wavelengths. The High Angular Resolution
Widefield Camera (HAWC) provides imaging capability at 50, 90,
160, and 215 \um. Surveys are planned  to detect high redshift
galaxies (D. A. Harper, private communication). The instrument
will reach the confusion limit in about 40 hours, so such
observations will be carried out over very limited areas. The
instrument will be useful to follow up observations of high
redshift sources detected  by Spitzer or other surveys.

The Atacama Large Millimeter Array (ALMA)
\footnote{http://alma.nrao.edu and
http://www.eso.org/projects/alma/} in Chile is planned as a
synthesis radio telescope built in international collaboration
between North America and Europe which will operate at millimeter
and sub-mm wavelengths. It will reach resolution of $\lsim
1^{\prime\prime}$ over fields of several square arcmin.

\subsection*{SAFIR, SPIRIT, SPECS}

Planned for the end of the next decade these prospective NASA
missions will achieve both high angular resolution (arcsecond to
sub-arcsecond) and high sensitivity ($\sim \mu$Jy) necessary to
resolve and characterize most of the sources comprising the cosmic
infrared background at far-IR and sub-mm wavelengths.

Recommended by the National Academy of Sciences Decadal Review
\footnote{Astronomy and Astrophysics in the New Millennium 2001,
National Research Council\\
http://www.nap.edu/books/0309070317/html/} SAFIR
\footnote{http://safir.jpl.nasa.gov/} stands for a Single Aperture
Far-Infrared observatory (Lester et al 2004). It is a large
cryogenic space telescope scheduled for launch around 2015 to
2020. The SAFIR telescope will operate between 20 \um and 1 mm and
will be cooled to about 5 K. The combination of large mirror size
and cold temperature will make SAFIR significantly more sensitive
than the Spitzer and Herschel instruments with sensitivity limited
only by the irreducible noise of photons in the astrophysical
backgrounds.

In the Community Plan for Far-IR/Submillimiter Space Astronomy
\footnote{in Proceedings ``New Concepts for
Far-Infrared/Submillimiter Space Astronomy", eds. D.J. Benford and
D. Leisawitz (Washington, DC: NASA), NASA CP-2003-212233, pp.
XV-XXV (2003)} it is envisaged that SAFIR will be followed by a
kilometer baseline FIR interferometer designed to provide both
wide field-of-view imaging and spectroscopy. This interferometer
is widely known as the Sub-millimeter Probe of the Evolution of
Cosmic Structure (SPECS). SAFIR and SPECS were selected for study
under the ROSS/Vision Mission study program, and the Space
Infrared Interferometer Telescope (SPIRIT), a science pathfinder
for SPECS, was selected for study under the NASA ROSS Origins
Science mission concept study program. These will be the first
instruments that would achieve sub-arcsecond resolution at
FIR/sub-mm wavelengths and provide the $\mu$Jy-level sensitivity
over a field-of-view of $\sim$ 1 arcmin$^2$ (Leisawitz 2004). All
three observatories, in addition to high sensitivity and
resolution, will have spectral resolution of the order of $R\sim
1000$.

\section{Concluding remarks}

Cosmic Infrared Background presents a complementary, and sometimes
the only, way to detect cumulative emission from galaxies at all
cosmic times, including from objects inaccessible to current or
future telescopic studies. Thanks to better detector sensitivities
and new measurement techniques it is now possible to identify and
remove foreground emissions at IR wavelengths and begin to measure
the CIB. Because of the difficulties in identifying the mean level
of the CIB in the presence of strong foreground emissions, it is
often useful to attempt to measure the CIB fluctuations spectrum.
The latter also allows to isolate the contributions to the CIB
from progressively fainter galaxies and, on average, earlier
cosmic epochs from surveys with $\sim$ arcsec angular resolution.

We have reviewed how mean levels of the CIB and, both the
amplitude and the angular spectrum of, its anisotropies relate to
the properties of the underlying galaxy populations. The various
foreground contributions and the latest measurements of the mean
CIB and its fluctuations were summarized from the near-IR to
sub-mm. There are mutually consistent detections of the mean
levels and fluctuations of the CIB in the near-IR, low upper
limits in the mid-IR and detections of the mean CIB in the far-IR
longward of 100 \um.

There are two broad classes of contributors to the CIB: from
'ordinary' (metal rich) stellar populations and from objects from
the Population III era. The data on the present-day galaxy
luminosity density present an important normalization point for
interpretation of the CIB measurements. We showed that by
constructing realistic SEDs for the stellar component one can get
a fair agreements between the different luminosity density
measurements at various wavelengths. This implies that it is
unlikely that significant flux is unaccounted for in galaxy
measurements at different bands and with different instruments and
methods, and the results are consistent with the independently
derived measurements of the present-day stellar density parameter,
$\Omega_*$. The same holds true at longer wavelengths where
emission from dust is the dominant contributor.

Total flux from observed ordinary galaxy populations gives a lower
limit on their contribution to the CIB and can be estimated from a
variety of galaxy counts data from the near-IR to sub-mm bands. We
summarized such estimates based on various most recent
measurements and conclude that 1) at far-IR wavelengths the
observed background is most likely accounted for by the observed
galaxy populations, which are probably located at $z\gsim 1-2$; 2)
at mid-IR wavelengths the total fluxes saturate at levels which
are just below the best current upper limits on the CIB and
probably account for at least a large fraction of the CIB; 3) at
near-IR wavelengths the total contributions from the observed
galaxy population saturate at levels significantly lower than the
claimed detections of the CIB mean levels and anisotropies. The
latter detections, if taken at face value, would indicate
substantial contributions from much earlier epochs than probed by
the observed faintest galaxy populations.

A plausible candidate to explain the excess in the near-IR CIB
would be emissions from Population III era objects. Their expected
contributions should have left a unique and observable signature
in the spectrum of the CIB anisotropies if Population III were
massive stars as is currently expected. The contributions from the
various foreground emissions are such that this signature can, and
should be, observed with future space-based instruments and
properly tuned observations.

Finally, the current (Spitzer) and planned space missions and new
instruments in the IR and sub-mm are expected to bring a wealth of
high quality observational data. This data can and, no doubt, will
be used for further progress in the CIB studies and in identifying
the history of light emission from collapsed objects in the early
Universe.

\section*{Acknowledgments}

I am grateful to my collaborators on the CIB-related topics who
have contributed much to the results discussed in this review. In
alphabetical order they are: Rick Arendt, Roc Cutri, Jon Gardner,
Mike Hauser, Raul Jimenez, John Mather, Harvey Moseley, Sten
Odenwald, Mike Skrutskie. In addition to them, over the years I
have benefited from many useful discussions and correspondence
with Dominic Benford, Eli Dwek, Dale Fixsen, Tom Kelsall, Alex
Kutyrev, Toshio Matsumoto, Bernard Pagel. I am grateful to the
following people who have supplied some of the data shown in the
figures above: Rick Arendt (IRAC galaxy counts), Dale Fixsen
(FIRAS spectral response), Tom Jarrett (2MASS star counts), Drs.
Iwamuro and Maihara (Subaru deep counts), Dr Toshio Matsumoto
(IRTS results), Lucia Pozzetti and Piero Madau (HDF galaxy
counts), Mike Santos and Mark Kamionkowski (Population III SED
spectra). I thank David Leisawitz for fruitful discussions on the
future sub-mm space missions and Alan Sweigart on mysteries of
stellar evolution. Finally, I am grateful to Rick Arendt for
careful reading of and comments on the various drafts of this
manuscript, to Dale Fixsen for likewise stimulating comments on
the manuscript, and to the Editor, Marc Kamionkowski, for patience
and encouragement in this, what turned out to be a
longer-than-planned, project. I acknowledge useful comments from
an anonymous referee. This work was supported by the National
Science Foundation grant No. AST-0406587.

\end{document}